\documentclass[aps,prd,preprint,tightenlines,floatfix,showpacs,
groupedaddress]{revtex4}

\usepackage{subfigure}
\usepackage{graphicx}

\begin{document}
\preprint{UFIFT-HET-05-23}

\vspace{1cm}

\date{October 24, 2005}

\title{The Inner Caustics of Cold Dark Matter Halos}

\author{Aravind Natarajan$^b$ and Pierre Sikivie$^{a,b}$}

\affiliation{
$^a$ School of Natural Sciences,~Institute for Advanced Study,
~Princeton,~NJ 08540\\
$^b$ Institute for Fundamental Theory,~
Department of Physics,~University of Florida,
~Gainesville,~FL 32611-8440\\}

\begin{abstract} 

We prove that a flow of cold collisionless particles from all 
directions in and out of a region necessarily forms a caustic.  A 
corollary is that, in cold dark matter cosmology,  galactic halos 
have inner caustics in addition to the more obvious outer caustics.  
The outer caustics are fold catastrophes located on topological 
spheres surrounding the galaxy.  To obtain the catastrophe structure 
of the inner caustics, we simulate the infall of cold collisionless
particles in a fixed gravitational potential.  The structure 
of inner caustics depends on the angular momentum distribution 
of the infalling particles.  We confirm a previous result that 
the inner caustic is a ``tricusp ring" when the initial velocity 
field is dominated by net overall rotation.  A tricusp ring is a 
closed tube whose cross section is a section of an elliptic umbilic
catastrophe.  However, tidal torque theory predicts that the initial 
velocity field is irrotational.  For irrotational initial velocity 
fields, we find the inner caustic to have a tent-like structure
which we describe in detail in terms of the known catastrophes.  
We also show how the tent caustic transforms into a tricusp ring 
when a rotational component is added to the initial velocity field.

\vspace{0.34cm}
\noindent
\end{abstract}
\maketitle

\section{Introduction\label{one}}

It is generally believed at present \cite{WMAP} that approximately 23\% 
of the energy density of the universe is in the form of ``cold dark
matter" (CDM).  The CDM particles must be non-baryonic, collisionless, 
and of small primordial velocity dispersion.  Particle physicists have 
put forth several candidates with the required properties.  Two among
these, the axion and the neutralino, have the distinction of having 
been originally postulated for purely particle physics reasons. The
axion solves the ``strong CP problem", whereas the neutralino is a
prediction of supersymmetric extensions of the Standard Model.  The
primordial velocity dispersions of these two candidates are very 
small, of order $3 \cdot 10^{-17}~c$ for axions and $10^{-12}~c$
for neutralinos.

A central problem in dark matter studies is the question of how CDM is
distributed in the halos of galaxies, and in particular in the halos 
of spiral galaxies such as our own Milky Way.  Indeed knowledge of 
this distribution is essential for understanding galactic dynamics 
and for predicting signals in direct and indirect searches for dark 
matter on Earth.

Three main approaches towards determining the dark matter distribution 
in galactic halos have been put forth.  The first assumes that galactic
halos are isothermal \cite{LB}. This assumption is highly predictive and, 
in fact, some of the predictions of the isothermal model agree with 
observation, to wit the flatness of rotation curves and the existence of 
core radii \cite{Paolo}.  However, even if galactic halos were thermalized 
in the past, they do not remain so because surrounding dark matter keeps 
falling onto them \cite{ipser,rob}. There is no mechanism which thermalizes 
the flows of axions or neutralinos which fall late onto the galaxy.  The 
second approach is to carry out $N$-body simulations of the halos \cite{Nsim}.  
This does presumably give a correct description if the number $N$ of
particles in the simulations is sufficiently large. However present
simulations have only $N \sim 10^7$  and, as a result, suffer from
2-body relaxation \cite{2bo} and from inadequate sampling of phase space
\cite{sing}.

The third approach exploits the fact that the CDM particles lie on a 3-dim. 
hypersurface in 6-dim. phase space \cite{ipser,sing,Trem}. This implies that 
the velocity distribution of dark matter particles at every point in physical 
space is discrete and that there are surfaces in physical space, called caustics, 
where the density of dark matter particles is very large.  We have recently 
argued  \cite{rob} that these discrete flows and caustics are a robust 
prediction of cold dark matter cosmology.  The reader may wish to consult 
ref. \cite{rob} for background information and a list of references. 

In the present paper we investigate the catastrophe structure of the caustic 
formed in a flow of cold collisionless dark matter particles falling in and 
out of a galactic gravitational potential well.  We call the caustics thus 
formed the ``inner caustics" of the galactic halo.  Galactic halos also have 
``outer caustics".  An outer caustic occurs near where an outflow of dark 
matter turns around before falling back in \cite{FG,Bert}.  The catastrophe 
structure of outer caustics is simple.  They are {\it fold} ($A_2$) 
catastrophes \cite{cat} located on topological spheres surrounding the 
galaxy.  We will see  below that the catastrophe structure of inner caustics 
is relatively more complicated.

We start off below (Section II) by proving that there always is  
an inner caustic, i.e. that it is impossible for cold collisionless 
particles to flow from all directions in and out of a region without 
forming a caustic.  The proof indicates that the inner caustic occurs 
near where the particles with the most angular momentum are at their 
distance of closest approach to a central point of the region.  This 
suggests that the structure of an inner caustic depends mainly on the 
angular momentum distribution of the infalling particles.

In previous work \cite{sing}, it was shown by analytical methods that 
the inner caustic is a tricusp ring when the initial velocity field 
$\vec{v}(\vec{r})$ of the infalling particles is dominated by net 
overall rotation.  A tricusp ring is a closed tube whose cross section 
is a section of an {\it elliptic umbilic} catastrophe \cite{cat}. We 
confirm this result in Section IV.   Fig.~\ref{axial-b} shows an axially 
symmetric tricusp ring as it appears in our simulations.

However, as will be shown below (Section III), the leading theory 
for the origin of the angular momentum of galaxies, namely tidal 
torque theory \cite{Pee,Dor,Whi}, predicts that the initial velocity 
field of dark matter particles is a pure gradient.  A pure gradient 
field is of course irrotational ($\vec{\nabla} \times \vec{v} = 0$).  
It can nonetheless carry net angular momentum 
$\vec{L} = \int d^3 r~\rho(\vec{r})~\vec{r} \times \vec{v}(\vec{r})$
because the density $\rho$ is inhomogeneous.

Motivated by tidal torque theory, we want to determine the structure 
of the inner caustics of galactic halos when the initial velocity 
field is irrotational and, more generally, when the initial velocity 
field is not dominated by net overall rotation.  We address this 
issue by simulating the infall of cold collisionless particles
in a fixed gravitational potential well.  The inner caustic is 
revealed by finding the locus of points where the Jacobian of 
the map between the initial and final positions of the particles
vanishes, i.e. where the map is singular.

We consider all initial velocity fields which are linear in position
$\vec{r}$, i.e. fields of the type $\vec{v}=M\,\vec{r}$ where $M$ is 
a $3\times 3$ real matrix.  We set $Tr(M)=0$ and more generally 
ignore the radial component $v_r=\hat{r}\cdot\vec{v}\,$ because it 
does not contribute to angular momentum and hence does not have 
much influence upon the inner caustics.  Tidal torque theory predicts 
$M^T=M$, but we consider the more general case in which $M$ has 
both symmetric and anti-symmetric parts.  

The outline of our paper is as follows.  In Section~\ref{two}, we
introduce the formalism of zero velocity dispersion flows and caustics,
and give an existence proof of inner caustics.  In Section~\ref{three}, 
we discuss our numerical techniques and the initial conditions 
expected from tidal torque theory.  In Section IV, we simulate 
flows which are dominated by net overall rotation, confirm that 
the inner caustic is a tricusp ring in that case, and demonstrate 
the stability of the tricusp ring under small perturbations.  In 
Section~\ref{five} we simulate flows which are not dominated by 
overall rotation and derive the structure of inner caustics for
those cases.  In Section VI, we summarize our conclusions.

In this paper, unless stated otherwise, the words ``circle" and 
``sphere" will be used in their topological, rather than geometrical, 
sense.  So an ellipse will be called a circle, etc. 

\section{Existence proof of inner caustics\label{two}}

Consider a flow of cold collisionless particles falling from all 
directions in and out of the gravitational potential well of a 
galaxy.  That the particles are collisionless means that they
move under the influence of purely gravitational forces.  We 
assume in this paper that Newtonian gravity applies, and so 
the particles obey the equation of motion
\begin{equation}
{d^2 \vec{r} \over dt^2}~=~-~\vec{\nabla} \Phi(\vec{r},t)
\label{Newt}
\end{equation}
where $\Phi(\vec{r},t)$ is the gravitational potential of the 
galaxy.  Note however that the existence proof of inner caustics 
given below would hold equally well in General Relativity.  That 
the particles are cold means that they have negligible velocity 
dispersion.  We set the velocity dispersion equal to zero.  However,
the presence of a small velocity dispersion does not affect the 
existence of caustics, and does not change their structure.  It 
only cuts off the divergence of the dark matter density at the 
location of the caustics.

As was mentioned in the Introduction, CDM particles lie on a 
3-dim. hypersurface in 6-dim. phase space.  Because the number 
of particles is huge - of order $10^{84}$ axions and/or $10^{68}$
neutralinos per galactic halo - the particles can be labeled
by three continuous parameters $\vec{\alpha} = (\alpha_1, \alpha_2,
\alpha_3)$ which are chosen arbitrarily.  For example, one may 
choose $\vec{\alpha}$ to be the position $\vec{q}$ of the particle 
at some early initial time $t_{\rm in}$, before shell crossings 
have occurred.  Other parametrizations may be more convenient 
however, depending on the problem at hand.

Let $\vec{x}(\vec{\alpha};t)$ be the position of the particle 
labeled $\vec{\alpha}$, at time $t$.  After shell crossings 
have occurred, there will in general be particles with different 
values of $\vec{\alpha}$ at the same location $\vec{r}$.  Let
$\vec{\alpha}_j(\vec{r},t)$, with $j = 1 ... n(\vec{r},t)$, be the 
solutions of $\vec{r} = \vec{x}(\vec{\alpha};t)$.  Thus $n(\vec{r}, t)$
is the number of distinct flows at location $\vec{r}$ at time $t$.
$n$ is always a positive odd integer because $n = 1$ to start with 
(before shell crossings have occurred) and the number of solutions 
of $\vec{r} = \vec{x}(\vec{\alpha};t)$ can only change by two at 
a time.  Let ${d^3 N \over d \alpha^3}(\vec{\alpha})$ be the 
number density of particles in the chosen parameter space.  The 
mass density of particles in physical space is then \cite{sing}
\begin{equation}
d(\vec{r}, t) = m \sum_{j=1}^{n(\vec{r},t)}
{d^3 N \over d\alpha^3}(\vec{\alpha})
{1 \over |D(\vec{\alpha}, t)|}~ 
\bigg|_{\vec{\alpha} = \vec{\alpha}_j(\vec{r},t)} 
\label{dmden}
\end{equation}
where $m$ is the particle mass and 
\begin{equation}
D(\vec{\alpha}, t) = 
det \left({\partial \vec{x}(\vec{\alpha}, t) \over 
\partial \vec{\alpha}} \right)~~~\ .
\label{jacob}
\end{equation}
The magnitude of $D$ is the Jacobian of the map $\vec{\alpha}
\rightarrow \vec{x}$.  Caustics occur where $D=0$, {\it i.e.} 
where the map is singular.  Note that, although $D$ is not 
reparametrization invariant, the density $d$ and the caustic 
condition $D=0$ are reparametrization invariant.  Caustics lie 
generically on 2-dim. surfaces because, in general, the condition 
$D=0$ defines a surface in physical space.  Only in special cases 
does the condition $D=0$ define an isolated line or point.  Hence 
isolated line caustics and isolated point caustics are degenerate 
cases; they are unstable towards becoming caustic surfaces.  Finally 
note that the map $\vec{\alpha} \rightarrow \vec{x}$ is singular 
where the number of flows $n$ changes.  So caustics lie generically 
at the boundaries between regions which have different numbers of 
flows.  On one side of a caustic surface are two more flows than 
on the other.

We now show that a continous flow of CDM particles falling in and 
out of a gravitational potential well cannot be free of caustics.   
Let us define a geometrical sphere of radius $R$ surrounding 
the potential well.  The precise value of $R$ and the precise 
location of the sphere's center do not matter.  Let us label 
the particles by $\vec{\alpha} = (\theta, \phi, \tau)$ where 
$\tau$ is the time when the particle crosses the sphere on 
its way into the well, and $\theta$ and $\phi$ are the polar 
coordinates of its crossing point on the sphere.  
$\vec{x}(\theta, \phi, \tau; t)$ is the position of the particle 
as a function of time $t$.  We will show that, at any $t$, 
\begin{equation}
D = det~{\partial (x,y,z) \over \partial (\theta, \phi, \tau)}
= {\partial \vec{x} \over \partial \tau} \cdot 
\left({\partial \vec{x} \over \partial \phi} \times
{\partial \vec{x} \over \partial \theta}\right)
\label{dotcross}
\end{equation}
vanishes at at least one point inside the sphere.  The flow 
is considered at an arbitrary fixed time $t$.  We will suppress 
the label $t$ henceforth.

For each $(\theta, \phi)$, the initial crossing time parameter 
$\tau$ has some range: 
$\tau_{\rm out}(\theta, \phi) < \tau < \tau_{\rm in}(\theta, \phi)$ 
where $\tau_{\rm in}~(\tau_{\rm out})$ is the initial crossing time 
of particles presently crossing the sphere on the way in (out).
[The careful reader may notice that $\tau_{\rm in}(\theta,\phi )=t$,
but this does not play a special role in what's to follow.]
Let us choose the origin $\vec{x} = 0$ at the center of the sphere 
and consider the function of three variables 
$r(\theta, \phi, \tau) = \sqrt{\vec{x}(\theta, \phi, \tau)
\cdot \vec{x}(\theta, \phi, \tau)}$.  We have 
$~{\partial r \over \partial \tau}
|_{\theta, \phi, \tau_{\rm out}(\theta,\phi)}~<~0~$ and 
$~{\partial r \over \partial \tau}
|_{\theta, \phi, \tau_{\rm in}(\theta,\phi)}~>~0$.
Hence for all $(\theta, \phi)$ there exists $\tau_0(\theta, \phi)$ 
such that 
\begin{equation}
r(\theta, \phi, \tau_0(\theta, \phi))
= {\rm min}~r(\theta, \phi, \tau) \equiv r_{\rm min}(\theta, \phi)~~~\ ,
\label{min}
\end{equation}
where the minimum is over $\tau \in [\tau_{\rm out}(\theta, \phi),
\tau_{\rm in}(\theta, \phi)]$ for fixed $(\theta, \phi)$. 
$r_{\rm min}(\theta, \phi)$ is the closest distance to the origin 
among all particles labeled $(\theta, \phi)$.  Let us first 
assume that $r_{\rm min}(\theta, \phi) \neq 0$ for some 
$(\theta,\phi)$ .  We will return later to the opposite case, 
where $r_{\rm min}(\theta, \phi) = 0$ for all $(\theta, \phi)$.

We have
\begin{equation}
{\partial r \over \partial\tau} 
\bigg|_{\theta, \phi, \tau_0(\theta, \phi)} = 
{\vec{x} \over r} \cdot {\partial \vec{x} \over \partial\tau}
\bigg|_{\theta, \phi, \tau_0(\theta, \phi)} = 0 ~~~\ 
\label{partau}
\end{equation}
for all $(\theta, \phi)$ such that $r_{\rm min}(\theta, \phi) \neq 0$.
Now, $r_{\rm min}(\theta, \phi)$ has a maximum value over the sphere 
$S_2 = \{(\theta, \phi)\}$.  Let $(\theta_0, \phi_0)$ be such that
\begin{equation} 
r_{\rm min}(\theta_0, \phi_0) = 
{\rm max}~r_{\rm min}(\theta, \phi)~~~\ .
\label{max}
\end{equation}
We have then
\begin{equation}
{\partial r \over \partial\theta} \bigg|_{\vec{\alpha}_0}~=~
{\vec{x} \over r} \cdot {\partial \vec{x} \over \partial\theta}
\bigg|_{\vec{\alpha}_0} = 0~~,~~
{\partial r \over \partial\phi} \bigg|_{\vec{\alpha}_0}~=~
{\vec{x} \over r} \cdot {\partial \vec{x} \over \partial\phi}
\bigg|_{\vec{\alpha}_0} = 0
\label{parthetaphi}
\end{equation}
with $\vec{\alpha}_0 \equiv (\theta_0, \phi_0, \tau_0(\theta_0, \phi_0))$.
Note that $\vec{x}_0 \equiv \vec{x}(\vec{\alpha}_0) \neq 0$.
Eqs. (\ref{partau}) and (\ref{parthetaphi}) imply that 
${\partial \vec{x} \over \partial\theta}|_{\vec{\alpha}_0}$, 
${\partial \vec{x} \over \partial\phi}|_{\vec{\alpha}_0}$ and 
${\partial \vec{x} \over \partial\tau}|_{\vec{\alpha}_0}$ are all 
perpendicular to $\vec{x}_0$.  Hence those three vectors are 
linearly dependent, and therefore $D(\vec{\alpha}_0) = 0$.  This 
implies that $\vec{x}_0$ is the location of a caustic.  Note that 
$\vec{x}_0$ depends on the choice of origin.  If we move the origin 
about, $\vec{x}_0$ will move too.  Thus the inner caustic is in 
general spatially extended.  This is as expected since caustics lie 
generically on surfaces.
 
Next we consider what happens in the special case where
$r_{\rm min}(\theta, \phi) = 0$ for all $(\theta, \phi)$. 
Then $\vec{x}(\theta, \phi, \tau_0(\theta, \phi)) = 0$ for all 
$(\theta, \phi)$.  Therefore, for $\tau$ near $\tau_0(\theta,\phi)$,
\begin{equation}
\vec{x}(\theta, \phi, \tau) = 
\vec{\nu}(\theta, \phi)~(\tau -\tau_0(\theta, \phi))
+ O((\tau - \tau_0(\theta, \phi))^2)
\label{taylor}
\end{equation}
where $\vec{\nu}(\theta,\phi) = 
{\partial \vec{x} \over \partial \tau}
\bigg|_{\theta, \phi, \tau_0(\theta,\phi)}$.  We may reparametrize
$\theta^\prime = \theta,~\phi^\prime = \phi,
~\tau^\prime = \tau - \tau_0(\theta, \phi)$, and rename
$(\theta^\prime,\phi^\prime,\tau^\prime) \rightarrow
(\theta,\phi,\tau)$.  In this parametrization, 
$\vec{x}(\theta,\phi,\tau) = \vec{\nu}(\theta,\phi)\tau + O(\tau^2)$
for small $\tau$.  Hence
\begin{equation}
D(\theta,\phi,\tau) = \vec{\nu}(\theta,\phi)\cdot
\left({\partial \vec{\nu} \over \partial \theta} \times
{\partial \vec{\nu} \over \partial \phi} \right) \tau^2~~~\ .
\label{spec}
\end{equation}
Since $D=0$ at $\tau =0$, the origin is the location of a caustic.  
This completes the proof.  Note that the case $\vec{x}_0 =0$ is 
special since the whole inner caustic has collapsed to a point.

To conclude this section, let us remark that in the situation 
discussed here, where CDM particles fall in and out of a 
gravitational potential well, the number of participating 
flows inside the sphere with radius $R$ is everywhere an even 
integer.  Indeed just inside the $r = R$ surface there are two 
flows, one going in and one going out, and (as remarked earlier) 
the number of flows can only change by two at a time.  The reader 
may wonder how this fits with the statement that the number of 
flows is everywhere odd.  The resolution of this little puzzle 
is that, in actual realizations such as a galactic halo, there 
is always an odd number of additional flows present which are 
not participating in the in and out flow under consideration. 

\section{Initial conditions and numerical techniques\label{three}}

In this section we describe how the simulations are done.  A central 
issue is the initial conditions we give to the particles.  In subsection 
A we show that in the tidal torque theory for the origin of the angular
momentum of spiral galaxies the flow of CDM particles is irrotational to
all orders in perturbation theory, i.e. to all orders in an expansion in
powers of the density perturbations.  In subsection B, we present the
initial conditions used in the simulations.  They are a generalization of
the prediction of tidal torque theory to allow for rotational flow as well
as irrotational flow.  In subsection C, we describe the steps carried out
in doing the simulations.

\subsection{Irrotational flow in tidal torque theory}

Net rotation is a striking property of isolated spiral galaxies.
Yet it is not clear at present that the origin of this net rotation 
is well understood \cite{Pri}.  The leading hypothesis is that 
net rotation of spiral galaxies is the result of torque applied 
by the tidal gravitational forces of neighboring density perturbations 
in the very early stages of structure formation \cite{Pee,Dor,Whi}.  
This hypothesis is called ``tidal torque theory".  It is assumed in 
this context that general relativisitic effects are unimportant, and 
that Newtonian gravity applies.

In CDM cosmology, density perturbations enter the non-linear regime 
when shell crossings occur and caustics form.  Before that time, or 
wherever shell crossings have not occurred yet, there is a single 
flow at every physical location $\vec{r}$.  Let us call
$\vec{v}(\vec{r},t)$ the velocity of that primordial flow.  
In the formalism of Section II, $\vec{v}(\vec{r},t)$ is obtained 
by eliminating $\vec{\alpha}$ from $\vec{r} = \vec{x}(\vec{\alpha},t)$ 
and $\vec{v} = {\partial \vec{x} \over \partial t} (\vec{\alpha},t)$.
For collisionless particles
\begin{equation}
{d \vec{v} \over d t} (\vec{r},t) = 
{\partial \vec{v} \over \partial t}(\vec{r},t) 
+ (\vec{v}(\vec{r},t) \cdot \vec{\nabla}) \vec{v}(\vec{r},t)
= - \vec{\nabla} \Phi(\vec{r},t)
\label{folmot}
\end{equation}
where $\Phi(\vec{r},t)$ is the gravitational potential.  
Eq. (\ref{folmot}) neglects relativistic effects such as the 
gravitomagnetic force of General Relativity.  Eq. (\ref{folmot}) 
implies
\begin{eqnarray}
{\partial \over \partial t} (\vec{\nabla} \times \vec{v}) &=&
- \vec{\nabla} \times [(\vec{v} \cdot \vec{\nabla}) \vec{v}] \nonumber\\
&=& - \hat{i} \epsilon_{ijk} [(\partial_j v_l)(\partial_l v_k)
+ v_l \partial_l \partial_j v_k]~~ \ .
\label{curlv}
\end{eqnarray}
If $\vec{v} = \vec{\nabla} \chi$, both terms on the RHS of
Eq. (\ref{curlv}) vanish.  Hence a pure gradient initial velocity 
field remains pure gradient at all times.

In zeroth order of perturbation theory, the flow is given by 
Hubble's law: $\vec{v}(\vec{r},t) = H(t) \vec{r}$ where 
$H(t) = {\dot{a} \over a}$ and $a(t)$ is the scale factor.
That velocity field is certainly a pure gradient.  In first 
order, the particle trajectories are given by \cite{Zel}
\begin{equation}
\vec{x}(\vec{q},t) = a(t) [\vec{q} 
- b(t) \vec{\nabla}_q \Phi(\vec{q})]
\label{Zelap}
\end{equation}
where $\vec{q}$ is the particle's position at a very early time, 
and $\Phi(\vec{q})$ is the gravitational potential.  [The latter 
is time independent in first order perturbation theory when expressed 
in terms of the co-moving coordinate $\vec{q}$, i.e. 
$\Phi (\vec{r} = a(t) \vec{q}, t) = \Phi(\vec{q})$.]  In the 
matter dominated era, both $a(t)$ and $b(t)$ are proportional 
to $t^{2 \over 3}$.  As expected, the velocity field implied by
Eq. (\ref{Zelap}) is irrotational.  Our remark implies that the 
velocity field remains irrotational to all orders of perturbation 
theory.

\subsection{The linear initial velocity field approximation}

Eq. (\ref{Zelap}) implies
\begin{equation}
\vec{v}(\vec{r},t) = H(t)\vec{r} 
- a(t) {db \over dt}(t)~ 
\vec{\nabla}_q \Phi(\vec{q})|_{\vec{q} = {1 \over a(t)}\vec{r}}~~\ .
\label{Zelvel}
\end{equation}
Following ref. \cite{Whi}, let us choose $\vec{q} = 0$ at a 
minimum of $\Phi$ and expand $\Phi$ in Taylor series up to 
second order in the $q$'s.  The velocity field is then a 
linear function of position:
\begin{equation}
\vec{v} (\vec{r}) = M~\vec{r}
\label{lin}
\end{equation}
where $M$ is a symmetric ($M^{\rm T} = M$) $3 \times 3$ matrix.
When a shell of particles surrounding a galaxy approaches turnaround, 
its initial Hubble expansion has been canceled by the attractive 
gravitational force of the galaxy.  For such a shell the trace of 
$M$ is approximately zero.

In Sections IV and V we numerically integrate the equations of 
motion of particles falling in and out of the galactic gravitational 
potential well.  The particles are given initial conditions on a 
geometrical turnaround sphere as follows.  The particle labeled 
($\theta,\phi,\tau$) starts at time 
$\tau$ at location 
\begin{equation}
\vec{r}_{\rm in}(\theta,\phi) = R~\vec{n}(\theta,\phi)
\label{inpos}
\end{equation}
where $\vec{n}(\theta,\phi)$ is the unit vector in the direction 
defined by polar angle $\theta$ and azimuth $\phi$.  $R$ is the 
radius of the turnaround sphere.  The particles are assigned 
velocities of the form of Eq. (\ref{lin}) with $Tr~M = 0$.  We 
write $M = S + A$ with $S^{\rm T} = S$ and $A^{\rm T} = - A$.
Tidal torque theory predicts $A = 0$.  However, as tidal torque
theory may not be the final word on the origin of the angular
momentum of galaxies, we will study the inner caustics for
$A \neq 0$ as well as $A = 0$.

We believe the above set of initial conditions is sufficiently 
general for our purpose of studying the structure of inner caustics.  
Indeed, the structure of inner caustics is determined by the 
distribution of distances of closest approach of the particles 
falling in.  This in turn is determined by the distribution of 
angular momenta on the turnaround sphere, or equivalently by the 
tangential components of the initial velocity field on the 
turnaround sphere.  Caustics are stable under deformations.  
So to find the structure of inner caustics it is not necessary 
to use the most general initial conditions, but only initial 
conditions which are sufficiently representative of those that 
occur in reality.  We expect that to include higher order terms 
in the Taylor expansion of the initial velocity field as a 
function of position will only deform, without changing their 
essential structure, the inner caustics found when assuming the 
initial velocity field is of the form of Eq. (\ref{lin}).  We verify 
this expectation explicitly in the case of the tricusp caustic ring 
when we study its stability in Section IV.

We choose coordinate axes such that $S$ is diagonal:
\begin{equation}
M = {1 \over R} \left(
\begin{array}{ccc}
g_1 & - c_3 & c_2 \\
c_3 & g_2 & - c_1 \\
- c_2 & c_1 & -g_1-g_2 
\end{array}
\right)~~~\ .
\label{M}
\end{equation}
$g_1$ and $g_2$ parametrize the symmetric part of $M$ which 
yields the gradient part of $\vec{v}$, whereas $c_1$, $c_2$
and $c_3$ parametrize the anti-symmetric part of $M$ which yields 
the curl part of $\vec{v}$ and describes a rigid rotation of 
angular velocity $\vec{\omega} = {\vec{c} \over R}$.  In terms 
of the five parameters, the components of the initial velocity 
field tangent to the turnaround sphere are 
\begin{eqnarray}
v_\phi(\theta,\phi) &=& \vec{v} \cdot \hat{\phi}\nonumber\\
&=& (g_2 - g_1) \sin \theta ~ \sin \phi ~ \cos \phi
- \cos \theta (c_1 \cos \phi + c_2 \sin \phi) + c_3 \sin\theta\nonumber\\
v_\theta(\theta,\phi) &=& \vec{v} \cdot \hat{\theta} \nonumber\\
&=& \sin \theta ~ \cos \theta ~ [g_1 (1 + \cos^2 \phi)
+ g_2 (1 + \sin^2 \phi)] - c_1 \sin \phi + c_2 \cos \phi~~~~\ .
\label{vin}
\end{eqnarray}
The radial component $\vec{v} \cdot \hat{r}$ of the initial 
velocity field does not contribute to the angular momentum, 
and is set equal to zero.  We verify in Section IV that the 
inclusion of radial initial velocities on the turnaround 
sphere has very little effect on the position of tricusp
caustic rings, and no effect on their structure.

To conclude this subsection, we discuss the symmetry properties 
of our initial velocity field.  Almost always we take the
gravitational potential to be spherically symmetric, in which 
case the symmetry properties of the initial velocity field are 
those of the subsequent evolution as well. In the irrotational 
case ($c_1 = c_2 = c_3 =0$), the initial velocity field is 
reflection symmetric about the $x=0$, $y=0$ and $z=0$ planes.  
Moreover it is axially symmetric when two of the three eigenvalues 
($g_1$, $g_2$, and $g_3 = -g_1 - g_2$) are equal.  Most often
we will chose the axes such that $g_1 \leq g_2 \leq g_3$. The 
parameter space is then $g_1 \leq 0$ and 
$g_1 \leq g_2 \leq - {1 \over 2} g_1$. When $g_1 = g_2$, the 
initial velocity field is axially symmetric about the $z$ axis.  
When $g_2 = - {1 \over 2} g_1$, it is symmetric about the $x$ axis. 

In the case of pure rotation ($g_1 = g_2 =0$), we may choose 
axes such that $\vec{c} = c \hat{z}$.  The initial velocity 
field is always axially symmetric in this case.  When $g_1$, 
$g_2$, $c_1$, $c_2$ and $c_3$ are all different from zero, the 
initial velocity field has no symmetry.  When axial symmetry 
about the $z$ axis is imposed, $c_1 = c_2 = 0$ and $g_1 = g_2$.

\subsection{How the simulations are done}

We simulate a single flow of zero velocity dispersion falling in 
and out of a time independent gravitational potential $\Phi(r)$, 
which is specified below.  The initial conditions are 
Eqs. (\ref{inpos},\ref{vin}) plus $v_r = \vec{v}\cdot\hat{r} = 0$.  
We solve the equation of motion, Eq. (\ref{Newt}), numerically to 
obtain for all $(\theta,\phi)$ the trajectory 
$\vec{x}(\theta,\phi,\tau;t)$ of the particle which 
started at position $(\theta,\phi)$ on the turnaround sphere, at 
time $\tau$.  Since neither the potential $\Phi$ nor the initial 
conditions are time-varying, the simulated flows are stationary, 
i.e. $\vec{x}(\theta,\phi,\tau;t) = \vec{x}(\theta,\phi,t-\tau)$.  
The simulation of non-stationary flows would be straightforward 
but considerably more memory intensive and time consuming, without 
being more revealing of the structure of inner caustics.   The 
only change with respect to simulations of stationary flows would 
be that the caustics deform in a time dependent way.  This would 
not teach us anything new about the structure of inner caustics.  

From $\vec{x}(\theta,\phi,t^\prime = t-\tau)$ we calculate 
$D = det~\frac{\partial(x,y,z)}{\partial(\theta,\phi, t^\prime)}$.  
We then plot the points where $D=0$.  The set of these points
is the inner caustic surface.

Unless stated otherwise, $\Phi$ is the gravitational potential  
produced by the matter density profile:
\begin{equation}
\rho(r) = {v_{\rm rot}^2 \over 4 \pi G (r^2 + a^2)}~~~\ ,
\label{d_profile}
\end{equation}
which implies an asymptotically flat rotation curve with rotation 
velocity $v_{\rm rot}$.  $a$ is the core radius.  We will refer
to the density profile of Eq.~(\ref{d_profile}) as the ``isothermal"
profile although it is only an appoximation to an exact isothermal 
profile.  The equation of motion is then
\begin{equation} 
\frac{d^2\vec r}{dt^2}=
-\frac{v_{\rm rot}^2}{r} 
\left[1-\frac{a}{r}\tan^{-1}\left(\frac{r}{a}\right)\right]\hat{r}~~~\ .
\label{acc}
\end{equation} 
We use everywhere $R$, the radius of the turnaround sphere, as the 
unit of distance and $v_{\rm rot}$ as the unit of velocity.  So our 
unit of time is ${R \over v_{\rm rot}}$, which is of order the infall 
time.  The core radius $a$ is always set equal to 0.0285.

The five parameters $g_1,~g_2,~c_1,~c_2$ and $c_3$, when expressed in
units of $v_{\rm rot}$, are related to and of order the dimensionless
angular momentum $j$ defined in \cite{STW}.  It was estimated in
that paper that the average of $j$ over the turnaround sphere is
approximately 0.2 in the case of the Milky Way.  This sets the overall
scale for the values of $g_1~...~c_3$ we are interested in and which 
are used in our simulations.  Note that it is the relative values of 
these five parameters that determines the structure of inner caustics.  
The overall scale of the parameters merely determines the overall size 
of the inner caustics relative to $R$.

Since we simulate the infall of a single cold flow in a fixed 
potential, the particle resolution is not a critical issue.  We 
chose a resolution of one particle per degree interval in $\theta$ 
and $\phi$, and a time step of $10^{-4}$.

\section{The tricusp ring\label{four}}

It was found in ref. \cite{sing} that the inner caustic is a 
``tricusp ring" when the initial velocity distribution is dominated 
by a rotational component.  A ``tricusp ring" is a closed tube 
whose cross section has the tricusp shape characteristic of the 
elliptic umbilic ($D_{-4}$) catastrophe.  In this section, we perform 
simulations of flows with initial velocity fields dominated by a 
rotational component, show that the inner caustic is indeed a 
tricusp ring and that this structure is stable under perturbations.

\subsection{Axially symmetric case}

\begin{figure}[!ht]

\subfigure[]{\label{infall-a}\includegraphics[width=2.1in,height=2.1in]
{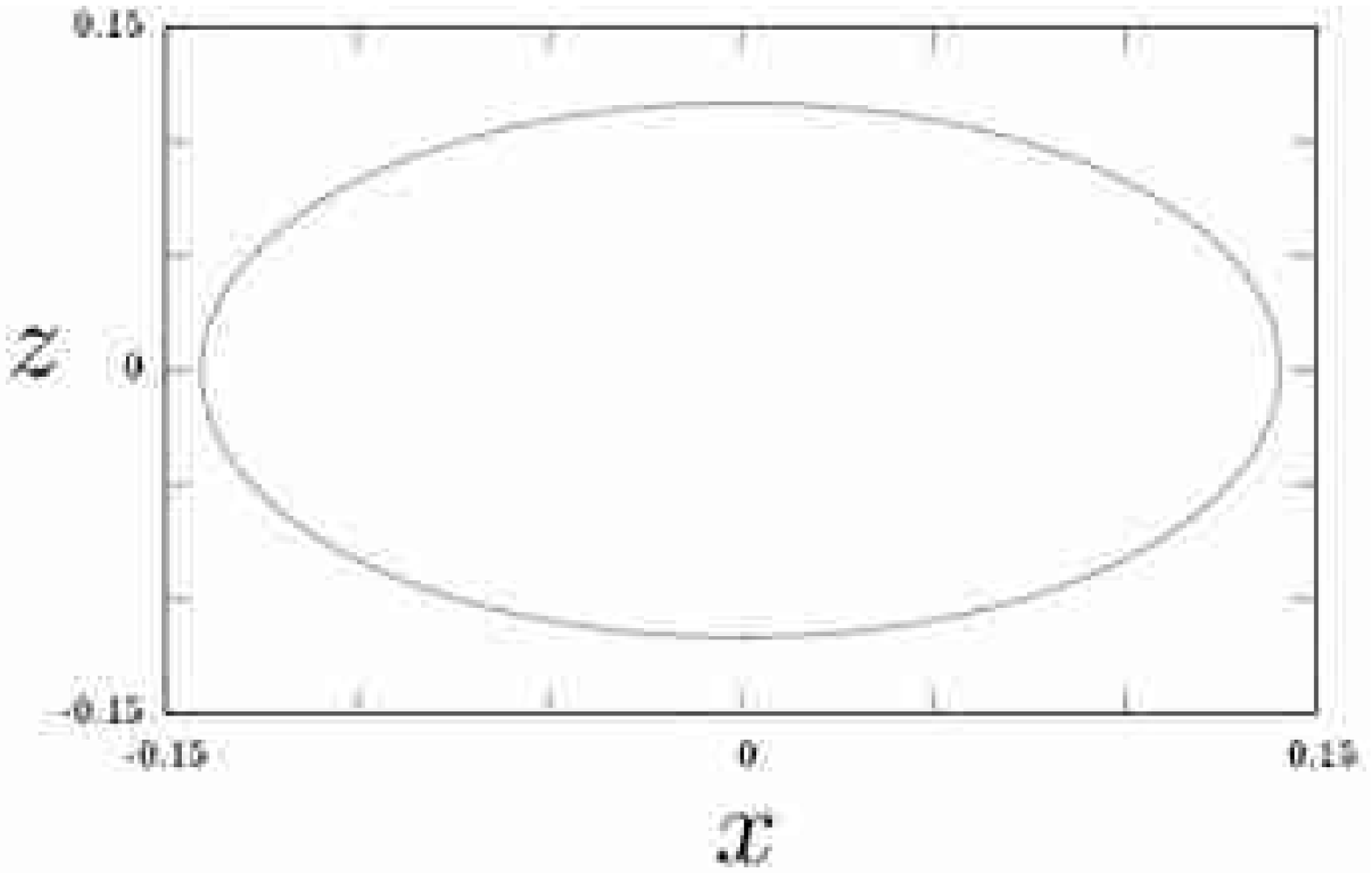}}
\hspace{.5in}
\subfigure[]{\label{infall-b}\includegraphics[width=2.1in,height=2.1in]
{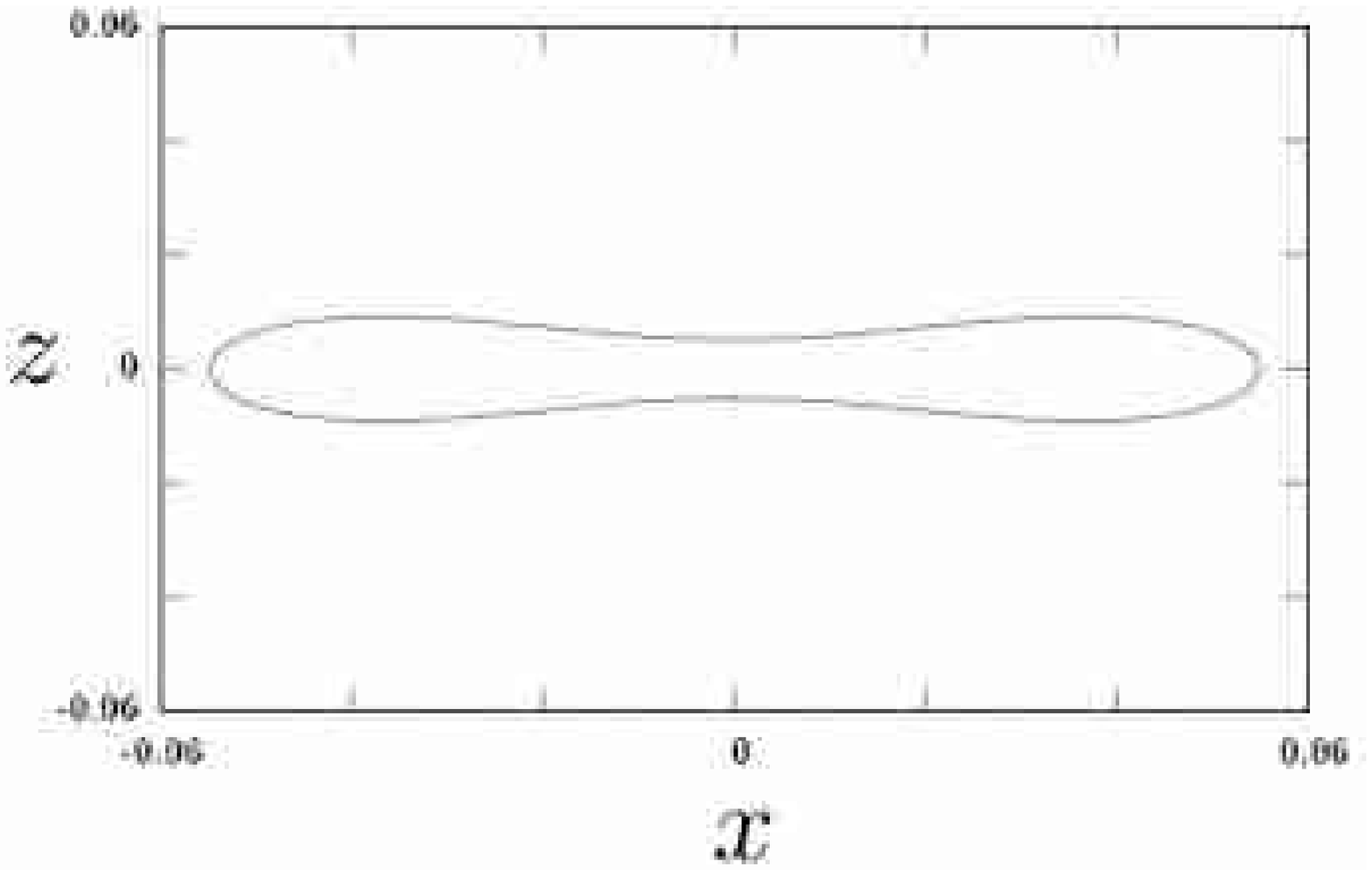}}\\
\subfigure[]{\label{infall-c}\includegraphics[width=2.1in,height=2.1in]
{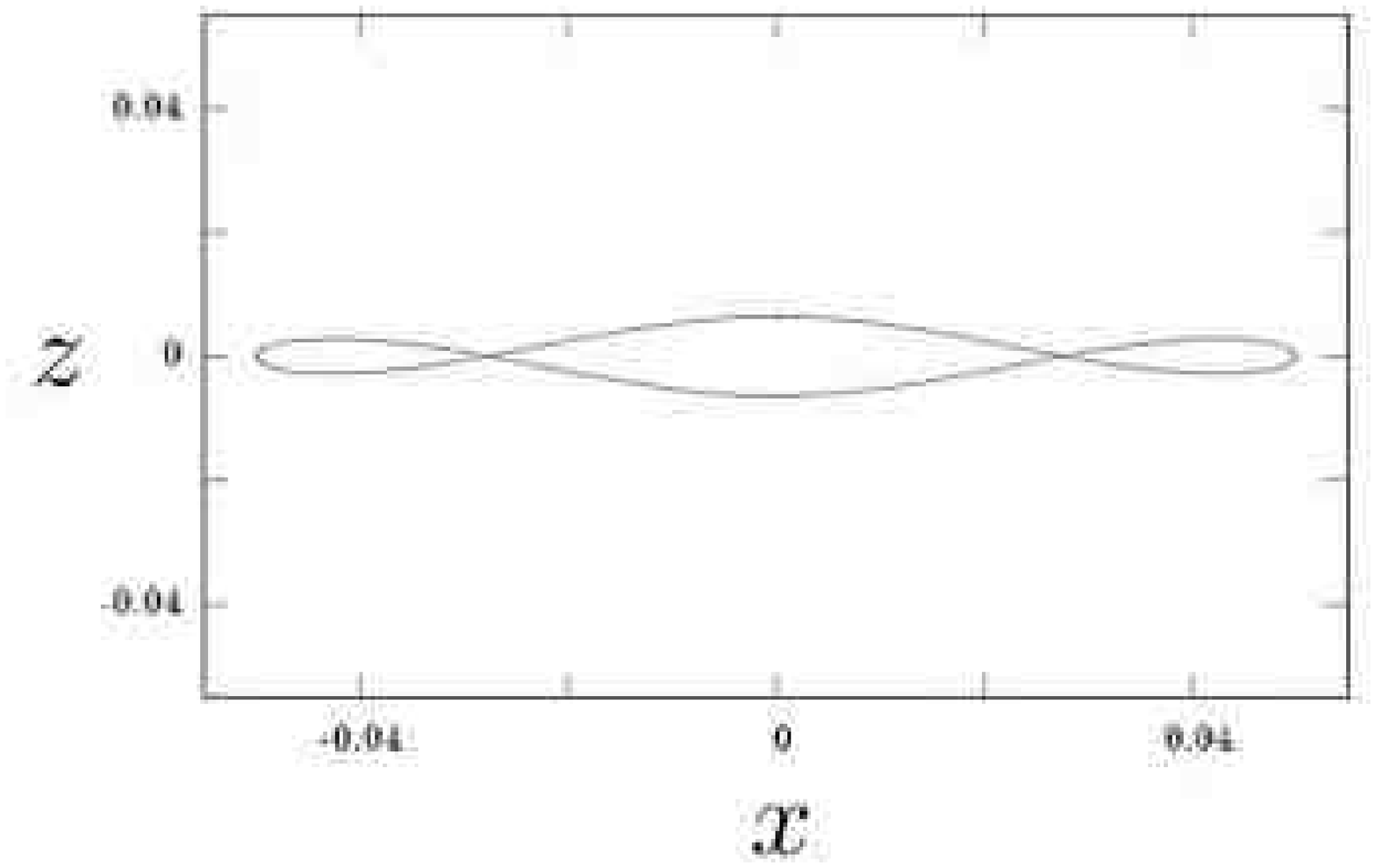}}
\hspace{.5in}
\subfigure[]{\label{infall-d}\includegraphics[width=2.1in,height=2.1in]
{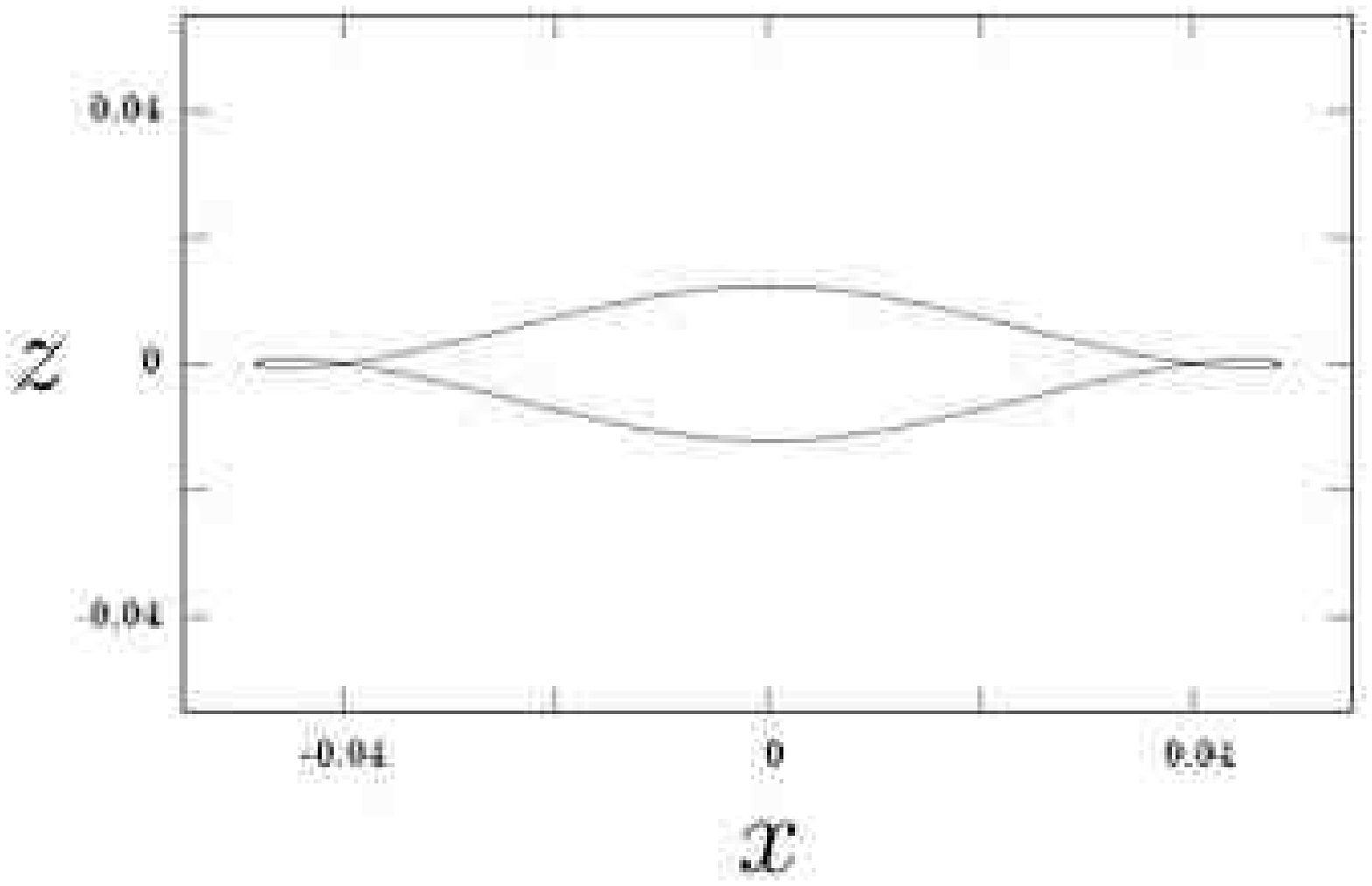}}\\
\subfigure[]{\label{infall-e}\includegraphics[width=2.1in,height=2.1in]
{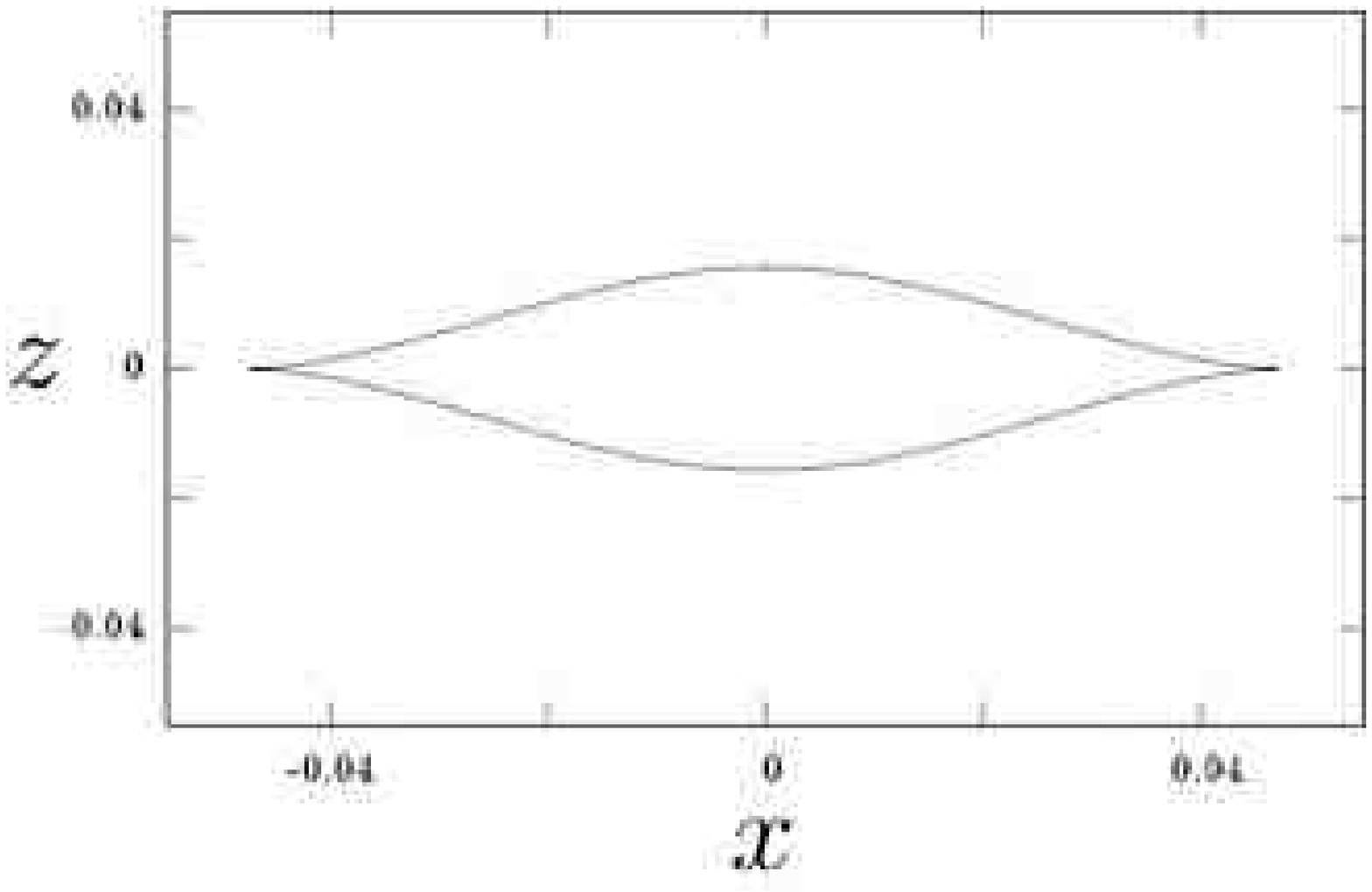}}
\hspace{.5in}
\subfigure[]{\label{infall-f}\includegraphics[width=2.1in,height=2.1in]
{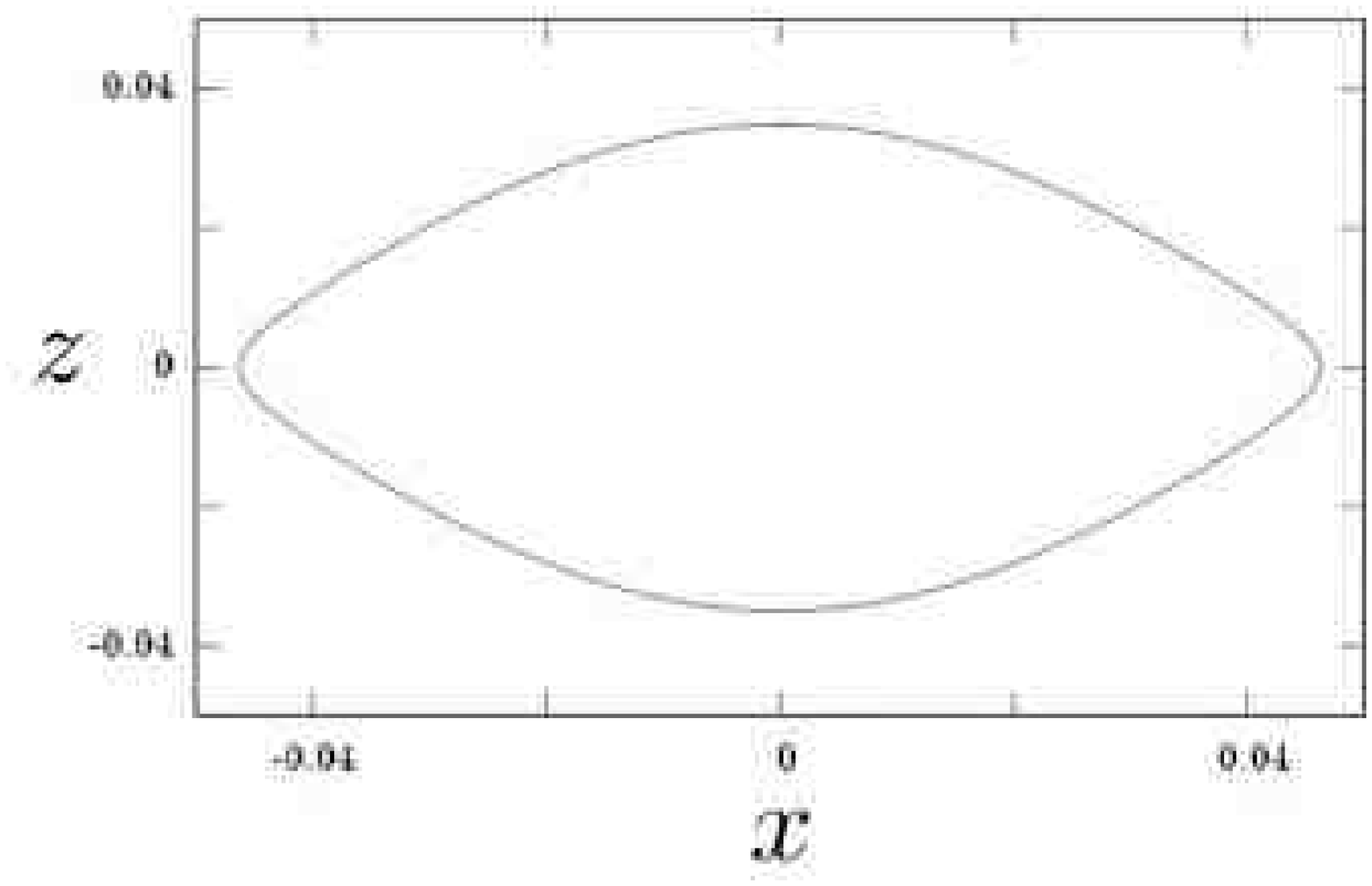}}

\caption{Infall of a cold collisionless shell for the axially symmetric 
initial velocity field $\vec v=-0.1\,\sin\theta\;\hat\phi$.  The six frames 
show the shell in cross section at times $t^\prime$ =  1.206, 1.258, 1.263, 
1.265, 1.267, 1.275. The continuous infall of many such shells produces the 
caustic shown in Figs.~\ref{axial-a} and~\ref{axial-b}.  $x$ and $z$ are in 
units of the outer turnaround radius $R$.  Note that here, and in many 
subsequent figures, the $x$ and $z$ scales vary from frame to frame.
\label{infall}}
\end{figure}

Consider the simple case where the initial velocity field is a 
linear function of the coordinates and is purely rotational, i.e.
when Eq. (\ref{lin}) holds with an anti-symmetric matrix $M$.
The initial velocity field is then axially symmetric.  Let us 
choose coordinates such that $\vec{v} = c_3~\sin\theta~\hat{\phi}$.
Fig. \ref{infall} shows the resulting infall of a single shell in 
$xz$ cross section, at successive times, for $c_3 = - 0.1$.

As the shell falls in, deviations from spherical symmetry appear due to 
the presence of angular momentum. The particles at the poles have zero 
angular momentum and fall in faster than the particles near the equator
~[\ref{infall-b}].  They are the first to cross the $x y$ plane~
[\ref{infall-c}]. The shell completes the process of turning itself 
inside out in Fig.~\ref{infall-e}, forming a crease. Finally the 
shell increases in size again and regains an approximately spherical 
shape~[\ref{infall-f}].

\begin{figure} 
\subfigure[]{\label{axial-a}\includegraphics[width=3.2in,height=3.2in]
{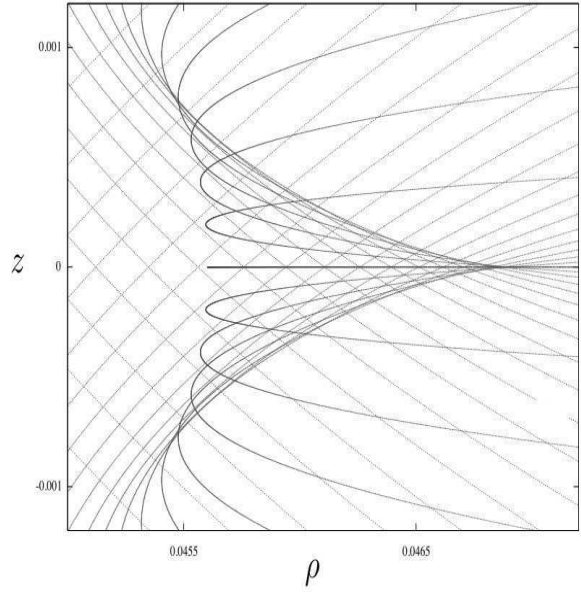}}
\subfigure[]{\label{axial-b}\includegraphics[width=4in,height=3in]
{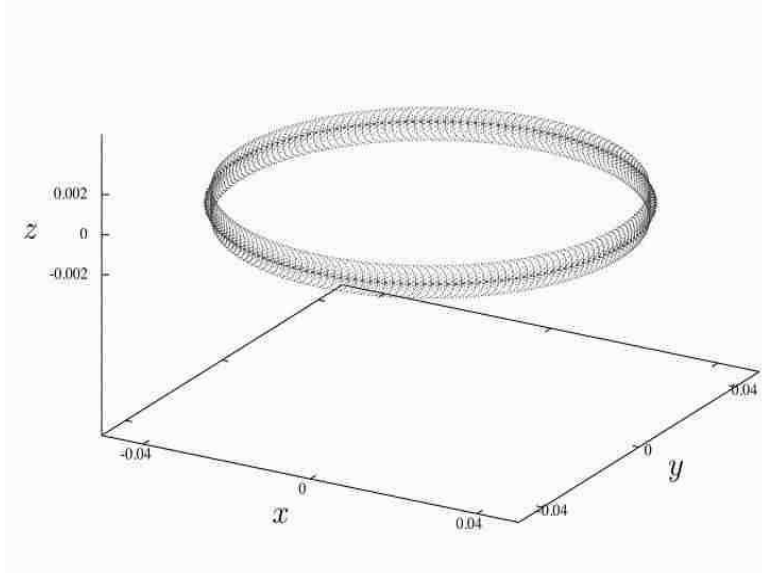}}

\caption{(a) Dark matter flows near the inner caustic for the initial 
velocity field $\vec v=-0.1\,\sin\theta\;\hat\phi$.  The inner 
caustic is a ring whose cross section has three cusps, one of which 
points away from the galactic center.  The three cusps are clearly 
visible in the figure.  There are four flows everywhere inside the 
caustic surface, and two everywhere outside. (b) The axially symmetric
tricusp ring in three dimensions. Note that here, and in many subsequent 
figures, the caustic is stretched in the $z$ direction for greater 
clarity.\label{axial}}
\end{figure}

The inner caustic occurs at and near the location of the crease in
Fig.~\ref{infall}.  Fig. \ref{axial-a} shows the flow near the crease
in $\rho z$ cross section where $\rho\equiv\sqrt{x^2+y^2}$. The figure 
shows that the $\rho z$ plane is divided into two regions, one with 
two flows at each point and the other with four flows at each point. 
The boundary that separates these two regions is the caustic. The 
dark matter density is infinite there in the limit of zero velocity 
dispersion. Fig.~\ref{axial-b} shows the caustic in three dimensions.
It is indeed the ``tricusp ring" described in ref. \cite{sing}. 

\subsection{Perturbing the initial velocity field}

\subsubsection{Breaking axial symmetry}

\begin{figure}
\resizebox{3.5in}{3.5in}{\includegraphics{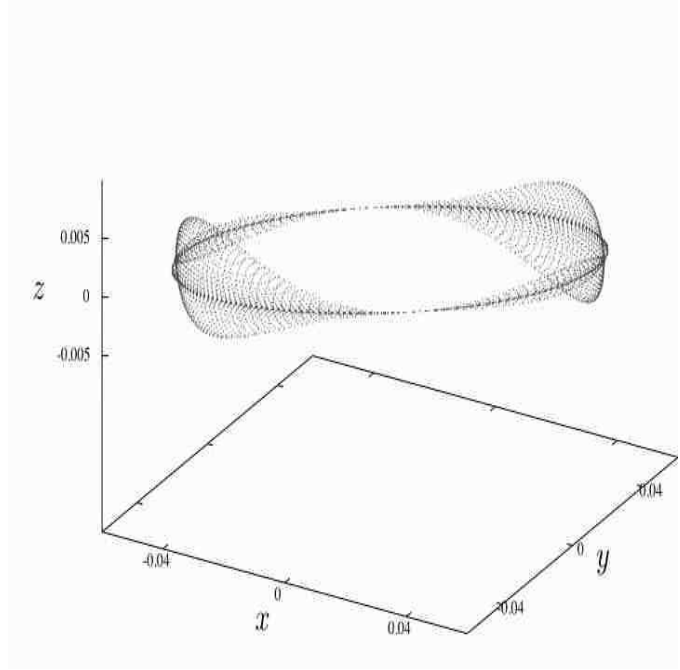}}
\caption{Tricusp ring for the initial velocity field of Eq.~(\ref{vin})
with $c_3 = - 0.1,~g_1 = - 0.033,~g_2 = 0.0267$.  It has reflection 
symmetry about the $x y$, $x z$ and $y z$ planes.
\label{ell_umb}}
\end{figure}

We introduce departures from axial symmetry by adding gradient terms, 
proportional to $g_1$ and $g_2$.  Fig.~\ref{ell_umb} shows the inner 
caustic for the initial velocity field of Eqs. (\ref{vin}) with 
$c_3 = - 0.1,~g_1 = - 0.033,~g_2 = 0.0267$.  It is again a tricusp 
ring but the cross section is now $\phi$-dependent.  The tricusp shrinks 
to a point four times along the ring.  In the neighborhood of each such 
point, the catastrophe is the full \emph{elliptic umbilic}$(D_{-4})$
\cite{cat}. 

\subsubsection{A random perturbation}

\begin{figure}
\resizebox{3.5in}{3.5in}{\includegraphics{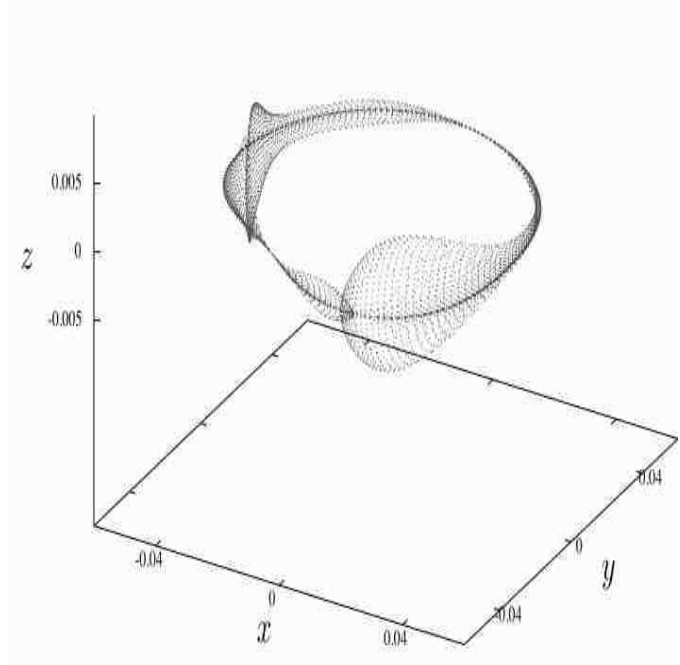}}
\caption{Tricusp ring for the initial velocity field given 
in Eq. (\ref{rand}).\label{perturb}}
\end{figure}

We added randomly chosen perturbations to the previously discussed 
axially symmetric initial velocity field.  The inner caustic 
shown in Fig.~\ref{perturb} is for the initial velocity field
\begin{eqnarray}
\vec{v} &=& - 0.1~\sin\theta~\hat{\phi}\nonumber\\
&-& 0.01~\sin\theta~\left[\sin(2\phi) + 0.5 \sin(3\phi) 
+ 0.25 \sin(4\phi)\right] \hat{\phi}
\label{rand}
\end{eqnarray}
Although the caustic is deformed from what it was in Fig. \ref{axial}, 
its structure remains a ring whose cross section is everywhere a tricusp.

\subsubsection{The effect of radial velocities}

We added radial velocities to the previously discussed axially 
symmetric initial velocity field.  We find that the radial 
velocity components result in only relatively small changes in the 
dimensions of the tricusp ring.  For the initial velocity field
\begin{equation}
\vec{v} = c_3~\sin\theta~(\hat{\phi} + \hat{r})
\label{rad}
\end{equation}
with $c_3 = - 0.1$, the tricusp ring radius was decreased by 0.28\%,
compared to what it was for the original initial velocity field  
($\vec{v} = c_3~\sin\theta~\hat{\phi}$), and the transverse 
dimensions of the tricusp ring were reduced by 11\% and 14\% in 
the directions perpendicular and parallel to the plane of the ring.  
For $c_3 = - 0.3$, the radius was reduced by 1.7\% and the transverse 
dimensions by 69\% and 73\% respectively.

Let us explain why radial velocities on the turnaround sphere have 
only a small effect on the inner caustics.  The inner caustics are 
determined by the distribution of distances $r_{\rm min}$ of closest 
approach to the galactic center of the infalling particles.  The 
distance of closest approach is determined by angular momentum 
conservation:  $\ell = r_{\rm min} v_{\rm max}$, where $\ell$ is 
specific angular momentum and $v_{\rm max}$ is the speed at the 
moment of closest approach.  The latter is determined by energy 
conservation
\begin{equation}
{1 \over 2} v_{\rm max}^2 = 
{1 \over 2}(v_\phi^2 + v_\theta^2 + v_r^2) + \Phi(R) - 
\Phi(r_{\rm min})~~~\ .
\label{encon}
\end{equation}
The main contribution to $v_{\rm max}$ is from the gravitational 
potential energy released while the particle falls in.  The initial
velocity components provide only corrections to $v_{\rm max}$ which 
are second order in $v_\phi, v_\theta$ and $v_r$.  Since $\ell$ does 
not depend on $v_r$ at all, radial velocities produce only second 
order corrections to the distances of closest approach.

\subsection{Modifying the gravitational potential}

In this subsection we verify that, when the initial velocity 
field is dominated by a rotational component, the inner caustic 
is a tricusp ring independently of the choice of gravitational 
potential.
 
\subsubsection{The NFW density profile}

We carried out simulations of the infall of collisionless particles
in the gravitational potential produced by the density profile of 
Navarro, Frenk and White~\cite{nfw} :
\begin{equation}
\rho(r)=\frac{\rho_s}{\frac{r}{r_s}
{\left[1+\frac{r}{r_s}\right]}^2}~~~\ .
\label{nfw_profile}
\end{equation}
The scale length $r_s$ was chosen to be 25 kpc. $\rho_s$ was determined
by requiring that the rotational velocity at galactocentric distance
$r_\odot$ = 8.5 kpc is 220 km/s.  The acceleration of a particle orbiting 
in the potential produced by the NFW density profile is then:
\begin{eqnarray}
\vec a(r)=-\frac{(220~{\rm km/s})^2\;x^{2}_\odot}{r_\odot\;x^2}
\left[\frac{\ln(1+x)-\frac{x}{1+x}}{\ln(1+x_\odot)-\frac{x_\odot}
{1+x_\odot}}\right]\;\hat r
\end{eqnarray}
where $x = r/r_s$ and $x_\odot = r_\odot/r_s$. 

\begin{figure}
\resizebox{3.5in}{3.5in}{\includegraphics{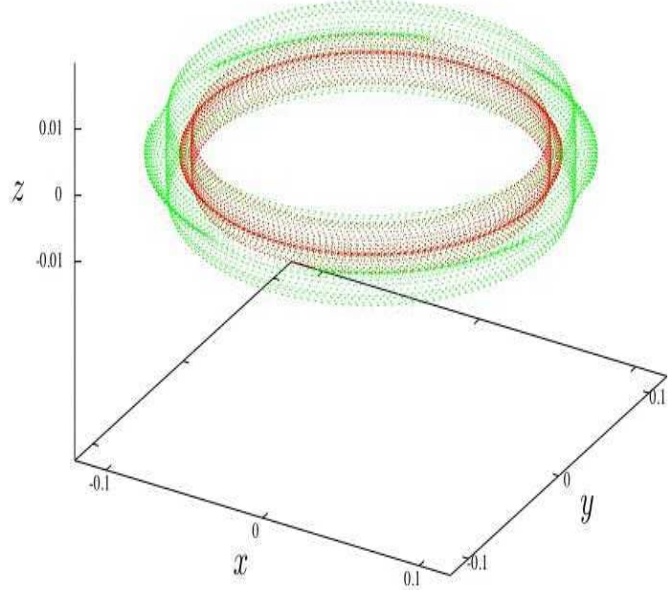}}
\caption{Tricusp rings caused by the same initial velocity field but 
different gravitational potentials.  The larger ring (green) was 
obtained using the NFW profile, Eq.~(\ref{nfw_profile}), while the 
smaller ring (red) was obtained using the density profile of 
Eq.~(\ref{d_profile}). \label{nfwiso}}
\end{figure}

Fig.~\ref{nfwiso} shows the result of two simulations plotted on the same 
figure. The outer caustic ring is obtained using the density profile of 
Eq.~(\ref{nfw_profile}), while the inner caustic ring is obtained using 
the density profile of Eq.~(\ref{d_profile}) with $v_{\rm rot} = 220$ km/s 
and $a = 4.96$ kpc.  In both cases, the turnaround radius $R = 174$ kpc and 
the initial velocity field $\vec{v} = 0.2~\sin\theta~\hat{\phi}$.  As always, 
the coordinates $x, y, z$ are in units of $R$. The inner caustic is a tricusp 
ring in each case, but with different dimensions.  The ring caustic produced 
by the NFW profile has larger radius than that produced by the isothermal 
profile because the NFW gravitational potential is shallower than the 
isothermal one at the location of the caustic 
($r_{\rm caustic} \simeq 16$ kpc).  Since $\ell = r_{\rm min}~v_{\rm max}$ 
is the same, $r_{\rm min}$ is larger in the NFW case because $v_{\rm max}$ 
is smaller.

\subsubsection{Breaking spherical symmetry}

\begin{figure}
\resizebox{3.5in}{3.5in}{\includegraphics{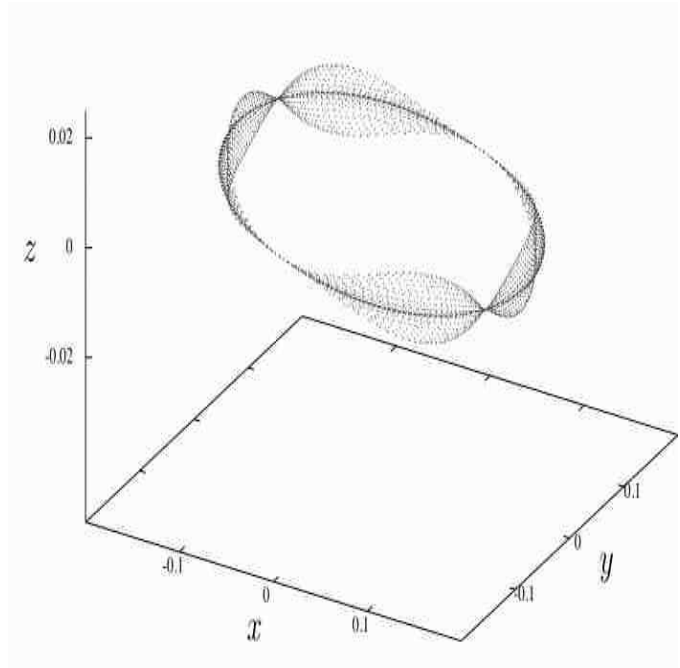}}
\caption{Tricusp ring for the initial velocity field 
$\vec{v} = 0.2~\sin\theta~\hat{\phi}$ and the non spherically 
symmetric gravitational potential of Eq. (\ref{nsph}).\label{triax}}
\end{figure}

We also simulated the infall of collisionless particles in a 
non spherically symmetric gravitational potential.  For the 
latter, we chose the triaxial form:
\begin{equation}
\Phi(r) = -v_{\rm rot}^2 \ln\left(\frac{R}
{\sqrt{{\left(\frac{x}{a_1}\right)}^2 + 
{\left(\frac{y}{a_2}\right)}^2 + 
{\left(\frac{z}{a_3}\right)}^2}}\right)
\label{nsph}
\end{equation}
where $a_1$, $a_2$ and $a_3$ are dimensionless numbers. 
Fig.~\ref{triax} shows the inner caustic for the case where
$a_1=0.95$, $a_2=1.0$ and $a_3=1.05$, and the initial velocity 
field $\vec{v} = 0.2~\sin\theta~\hat{\phi}$.  It is again a 
tricusp ring.  Its axial symmetry is lost due to the absence 
of axial symmetry in the potential.  The tricusp ring still 
has reflection symmetry about the $xy$, $yz$ and $xz$ planes.  
As in Fig.~\ref{ell_umb} the tricusp shrinks to a point four 
times along the ring.

\section{General structure of inner caustics\label{five}}

In this section, we describe the structure of inner caustics when 
the initial velocity field is not dominated by a rotational component.  
In subsection A, we discuss the axially symmetric case, whereas the 
non axially symmetric case is discussed in subsection B.  In each 
subsection, we simulate first irrotational (i.e. pure gradient) 
initial velocity fields.  As was mentioned in Section III, 
irrotational initial velocity fields are predicted by tidal torque 
theory.  We will find the inner caustics produced by irrotational 
velocity fields to have a definite structure which we refer to as 
the ``tent caustic".  After describing the tent caustic, we add a 
rotational component to the initial velocity field and see how the 
tent caustic deforms into a tricusp ring.

\subsection{The axially symmetric case} 

The initial velocity field of Eqs.~(\ref{vin}) is symmetric about 
the $z$ axis when $c_1 = c_2 = 0$ and $g_1 = g_2$.  Then
\begin{equation}
\vec{v} = {3 \over 2}~g_1~\sin(2\theta)~\hat{\theta}
+ c_3~\sin\theta~\hat{\phi}~~~\ . 
\label{axsym}
\end{equation}
We first simulate the flow and obtain the inner caustic 
in the irrotational case, $c_3 = 0$.  Next we see how the 
flow and inner caustic are modified when $c_3 \neq 0$.

\subsubsection{Infall of a cold collisionless shell}

Figs. \ref{inf} and \ref{case1} show the infall of a cold
collisionless shell whose initial velocity field is given
by Eq.~(\ref{axsym}) with $c_3 = 0$ and $g_1 = - 0.0333$.
Since $c_3 = 0$, each particle stays in the plane containing 
the $z$ axis and its initial position on the turnaround sphere.  
Figs. \ref{inf} and \ref{case1} show the particles in the 
$y = 0$ plane.  The angular momentum vanishes at $\theta=0$ 
and $\theta=\pi/2$ where $\theta$ is the polar coordinate 
of the particle at its initial position.  Hence, the particles 
labeled $\theta = 0$ and $\theta = \pi/2$ follow radial orbits. 
The angular momentum increases in magnitude from $\theta=\pi/2$, 
reaches a maximum at $\theta=\pi/4$ and returns to zero at 
$\theta=0$. The sign of the angular momentum does not change 
during this interval. 

The shell starts out as shown in Fig.~\ref{inf-a}. As the 
shell falls in, the particles at $\theta\ne 0,\pi/2$ move 
towards the poles. These particles feel an angular momentum 
barrier and fall in more slowly than the particles at 
$\theta=0$.  This results in the formation of a loop in 
Fig. \ref{inf-c}.  The formation of the loop implies a 
cusp caustic on the $z$ axis.  The particles labeled $\theta=0$ 
and $\theta=\pi$ have crossed the $z=0$ plane and the particles 
labeled $\theta=\pi/2$ have crossed the $x=0$ plane in 
Fig.~\ref{inf-g}. The shell then takes the form shown in 
Fig.~\ref{inf-h}.  The further evolution is shown in 
Fig.~\ref{case1}.  The loop that is present near the 
$z = 0$ plane in Fig.~\ref{case1-b} disappears through 
the sequence of Figs.~\ref{case1-c} -~\ref{case1-g}.  The 
disappearance of the loop implies the existence of a cusp 
caustic in the $z = 0$ plane as well.  In Fig.~\ref{case1-h} 
the shell has regained an approximately spherical form and is 
expanding to its original size.

For larger values of $|g_1|$ the early evolution is qualitatively 
the same as in Fig.~\ref{inf}, but the late evolution is 
qualitatively different from Fig.~\ref{case1}.  Fig.~\ref{case2}
shows the late evolution of a cold collisionless shell with 
the initial velocity field of Eq.~(\ref{axsym}) with $c_3 = 0$ 
and $g_1 = - 0.0667$.  The loop which is present near the $z = 0$ 
plane in Fig.~\ref{case2-b} disappears by a more complicated 
sequence [\ref{case2-c} -~\ref{case2-g}] than was the case in
Fig.~\ref{case1}.  In Fig.~\ref{case2} the particles near 
$\theta=\pi/2$ cross the $z=0$ plane before the sphere turns 
itself inside out. This crossover produces additional structure, 
and a more complicated caustic, than for the $g_1 = - 0.0333$ case.  
The critical value of $|g_1|$, below which the qualitative evolution 
is that of Fig.~\ref{case1}, and above which that of Fig.~\ref{case2}, 
is $g_{1*} \simeq 0.05$.

\subsubsection{Caustic Structure}

\begin{figure}

\subfigure[Caustic structure.]{\label{caust2-a}\resizebox{4in}{4in}
{\includegraphics{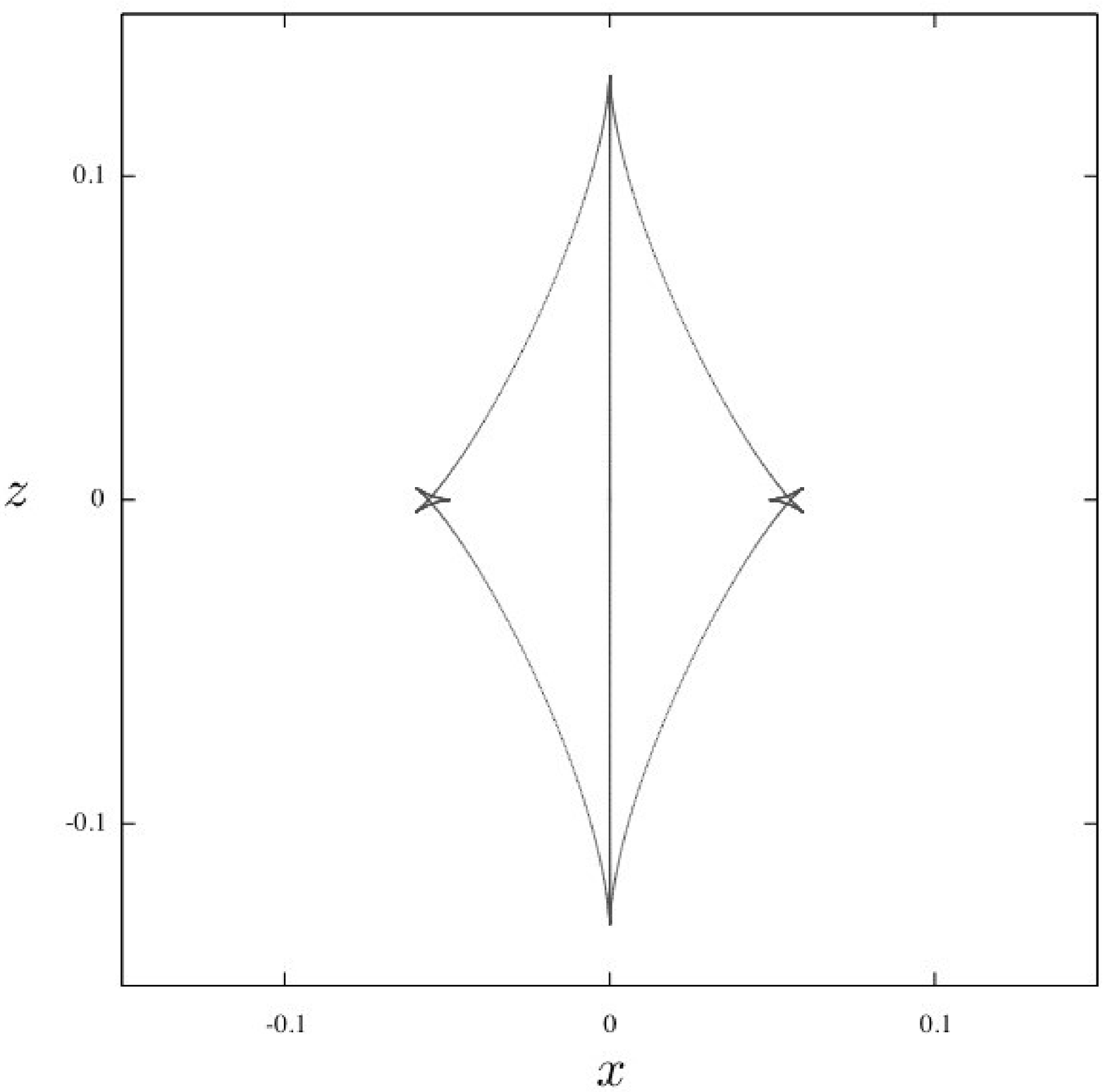}}}
\subfigure[Dark matter flows near the butterfly caustic.]
{\label{caust2-b}\resizebox{4in}{3in}
{\includegraphics{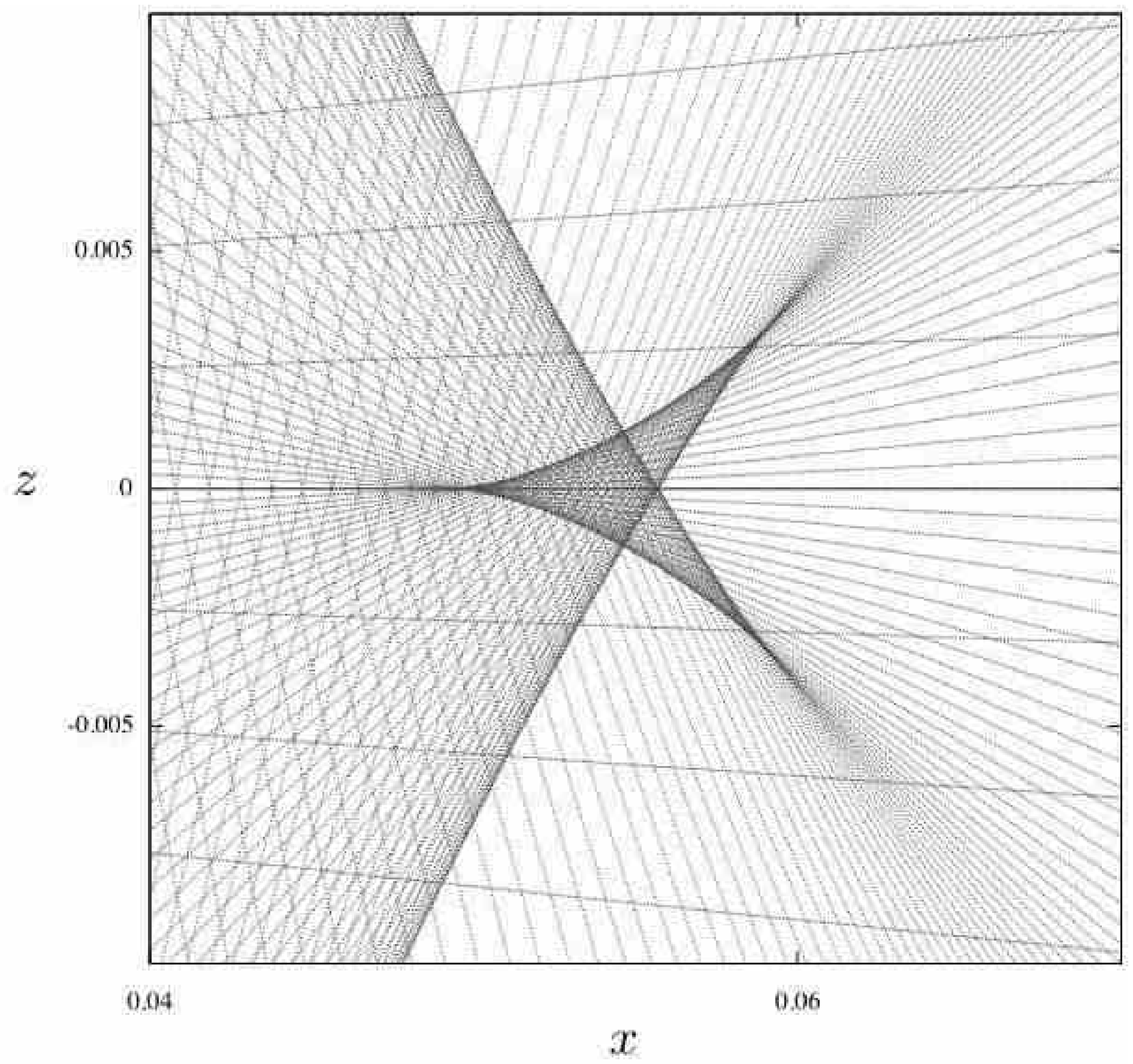}}}
\caption{Cross section of the inner caustic produced 
by the irrotational axially symmetric velocity field
$\vec{v}= - 0.1 \sin\,2\theta\;\hat\theta$.\label{caust2}}
\end{figure}

The inner caustic is a surface of revolution whose cross section is 
shown in Fig.~\ref{caust1-a} for the case $(c_3, g_1) = (0, - 0.0333)$
and in Fig.~\ref{caust2-a} for $(c_3, g_1) = (0, - 0.0667)$.  For the 
sake of brevity, we call this structure a ``tent caustic".  On the
$z$ axis, there is a caustic line which we call the ``tent pole".  The 
remainder of the caustic is called the ``tent roof".  As was mentioned 
in Section II, caustic lines are not generic.  The tent pole is a line 
in Figs.~\ref{caust1-a} and \ref{caust2-a} only because the initial 
velocity field is axially symmetric and irrotational.  We will see 
below that when axial symmetry is broken or when a rotational component 
is added, the tent pole becomes a caustic tube of a specific sort. 

For $|g_1| < g_{1*}$ there is a cusp in the tent roof where it 
meets the $z = 0$ plane and two cusps where the tent roof meets 
the tent pole, one at the top and one at the bottom.   For 
$g_1 < - g_{1*}$ the cusp in the $z = 0$ plane is replaced 
by a \emph{butterfly} catastrophe \cite{cat}.  The butterfly has 
three cusps and three points of self-intersection.  The cusp in the 
$z = 0$ plane transforms into a butterfly by increasing $|g_1|$.  
The latter parameter may therefore be called the ``butterfly 
factor" \cite{Saunders}. 

If $g_1$ is chosen positive instead of negative, the behavior at 
the poles and the equator is reversed [see Eq.~(\ref{axsym})] and 
we have cusps on the $x$ axis and either cusp or butterfly 
catastrophes on the $z$ axis, depending on the magnitude of $g_1$.

\begin{figure}
\resizebox{4in}{4in}{\includegraphics{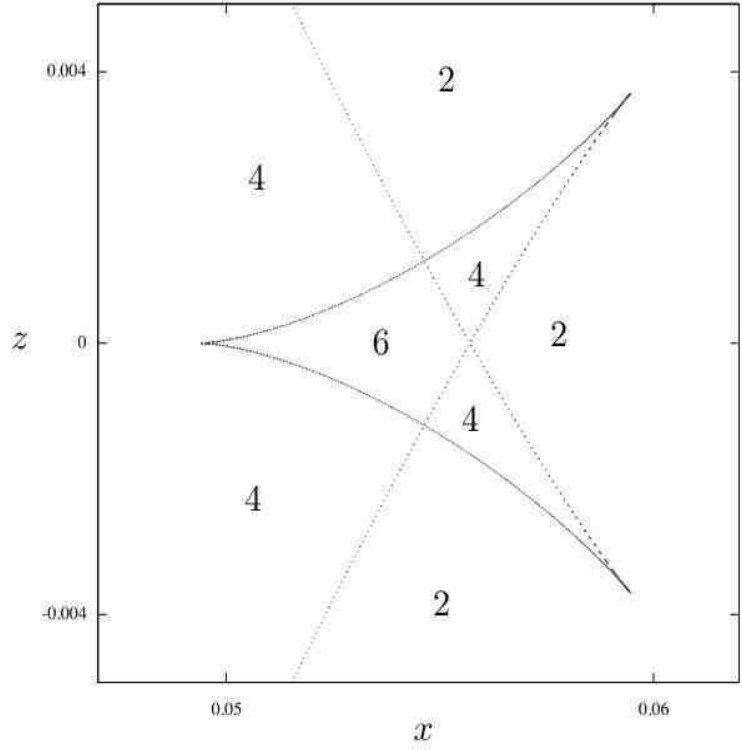}}
\caption{The butterfly catastrophe has three cusps, and three 
points where the caustic intersects itself. The number of flows 
in each region is indicated.\label{bfly}}
\end{figure}

Fig.~\ref{caust1-b} shows the dark matter flows in the vicinity 
of the caustic for $|g_1| < g_{1*}$. There are four flows everywhere 
inside the caustic tent and two flows everywhere outside.  
Fig.~\ref{caust2-b} shows the dark matter flows in the vicinity 
of the butterfly caustic, for $g_1 < - g_{1*}$.  Fig.~\ref{bfly} 
shows the number of flows in each region of the butterfly caustic. 

\subsubsection{Adding a rotational component}

Here we show, in the axially symmetric case, the effect of 
adding a rotational component to the initial velocity field.
On the basis of the discussion in Section IV, we expect 
the tent caustic to transform into a tricusp ring.  
Fig.~\ref{tr_axial} shows the transformation. We start with 
an irrotational velocity field $(c_3 = 0)$ in \ref{tr_axial-a} 
and increase $c_3$ until the rotational component dominates 
the velocity field, in \ref{tr_axial-d}. The tent pole on 
the $z$ axis, which is a line caustic in the irrotational case, 
changes to a tube of circular cross section. The radius of this 
tube increases until the tent caustic becomes indistinguishable 
from a tricusp ring. 

\subsection{Non axially symmetric case}

We start off by discussing the flows and caustics resulting 
from irrotational initial velocity fields.  With $\vec{c} = 0$, 
Eqs.~(\ref{vin}) become
\begin{equation}
\vec{v} = \xi~\sin\theta~\sin(2\phi)~\hat{\phi} + 
\sin(2\theta)(\xi~\sin^2\phi + g)~\hat{\theta}~~~,
\label{irrot}
\end{equation} 
where $\xi \equiv {1 \over 2}(g_2 - g_1)$ and 
$g \equiv g_1 + {1 \over 2}g_2$.  $\xi$ is a measure of $\hat{z}$ 
axial symmetry breaking in the irrotational case.  We first 
let $\xi \ll g$. Next, we explore all of $(g_1, g_2)$ parameter 
space. Finally, we add a rotational component by letting 
$c_3 \neq 0$.

\subsubsection{Irrotational, non axially symmetric perturbations}

We saw in subsection V-A that there is a caustic line on the $z$ axis 
(the tent pole) when the initial velocity field is irrotational and 
axially symmetric.  Fig.~\ref{zp-a} shows the trajectories of the 
particles in the $z = 0$ plane for such a case.  The orbits are radial.  
Indeed all particles have zero angular momentum with respect to the 
$z$ axis when the initial velocity field is irrotational and axially 
symmetric.  Because all trajectories intersect the $z$ axis, there is 
a pile up of particles on that axis and hence a caustic line.

Fig.~\ref{zp-b} shows the trajectories of the particles in the 
$z = 0$ plane for the initial velocity field of Eq.~(\ref{irrot})
with $\xi = 0.01$ and $g = -0.05$.  The particles do have angular 
momentum with respect to the $z$ axis now.  The caustic line on 
the $z$ axis spreads onto a tube whose cross section is the diamond 
shaped envelope of particle trajectories shown in Fig.~\ref{zp-b}.
That envelope has four cusps. The flows and caustic have reflection
symmetry about the $x y$, $x z$ and $y z$ planes because the initial
velocity field has those symmetries.  Fig.~\ref{zp-b} shows four 
flows inside the diamond-shaped caustic and two flows outside. The 
infall of dark matter particles from regions above and below the 
$z=0$ plane will add two more flows at each point, which are not 
shown in Fig.~\ref{zp} for clarity.

Fig.~\ref{ttt} shows the inner caustic in 3 dimensions for the 
initial velocity field of Eq.~(\ref{irrot}) with $\xi = 0.005$ 
and $g = - 0.05$.  The tent pole has spread onto a tube with 
diamond shaped cross section, as in Fig.~\ref{zp-b}.  Fig.~\ref{ttt-b} 
shows a succession of constant $z$ sections.  The (topological) circles, 
which are sections of the tent roof, and the diamond structure can be 
seen clearly.  Near the $z = 0$ plane, there are six flows inside the 
diamond, four flows in the other regions inside the caustic tent, and 
two outside the tent.  Fig.~\ref{tttc} shows $y = 0$ and $x = 0$ 
sections of the tent caustic.

\subsubsection{A hyperbolic umbilic catastrophe}

Let us look more closely at the two regions (top and bottom) in 
Fig.~\ref{tttc} where the tent pole reaches and traverses the tent 
roof. Figs.~\ref{hyp_umb-a} - ~\ref{hyp_umb-d} show $z$ = constant 
sections of the inner caustic in such a region. As $|z|$ is increased, 
the two cusps of the pole which are in the $x = 0$ plane simply pierce 
through the roof, whereas the parts of the pole near the $y = 0$ plane 
traverse the roof by forming with the latter two {\it hyperbolic umbilic} 
($D_{+4}$) catastrophes \cite{cat}, one on the positive $x$ side and one 
on the negative $x$ side.  The sequence through which this happens is 
shown in greater detail in Figs.~\ref{hyp_umb-e} - ~\ref{hyp_umb-h} for 
the hyperbolic umbilic on the positive $x$ side.  The arc (section of the 
tent roof) and the cusp (section of the tent pole) approach each other 
until they overlap [\ref{hyp_umb-g}], forming a corner.  After they 
have crossed, the roof section is cuspy whereas the pole section is 
smooth.  The cusp is transferred from the tent pole to the tent roof 
as the two surfaces pass through one another.  This behaviour is 
characteristic of the hyperbolic umbilic catastrophe.  There are 
four hyperbolic umbilics embedded in the caustic tent structure, 
two ($x > 0$ and $x < 0$) at the top ($z > 0$) and two at the bottom 
($z < 0$).  The hyperbolic umbilic at $z > 0$ and $x < 0$ is shown in 
three dimensions in Fig.~\ref{hyp_umb-i}.  

Let's mention that the particles forming the pole and roof where they 
intersect in the $x = 0$ plane originate from different patches of the 
initial turnarround sphere whereas the particles forming the pole and 
roof near a hyperbolic umbilic originate from the same patch of the 
initial turnaround sphere.

\subsubsection{The $(g_1, g_2)$ landscape}

Here we describe the inner caustic in the irrotational case 
for $g_1$ and $g_2$ far from those values where the flow is 
axially symmetric.  Recall that the flow is symmetric about 
the $z$ axis when $g_1 = g_2$ ($g_3 = - 2 g_1$), about the 
$y$ axis when $g_2 = - 2 g_1$ ($g_3 = g_1$), and about the 
$x$ axis when $g_2 = - {1 \over 2} g_1$ ($g_3 = g_2$).
In terms of $\xi$ and $g$, these conditions for axial symmetry 
are $\xi = 0$, $g = 0$, and $\xi = - g$, respectively.

The first, second and third column of Fig.~\ref{g1g2} show 
respecively the $z = 0$, $y = 0$ and $x = 0$ sections of the 
inner caustic produced by the initial velocity field of Eq.~(\ref{irrot}) 
for various values of $(g_1, g_2)$.  The ratio $g_2/g_1$ decreases 
uniformly from 1 (top row) to  -1/2 (bottom row). Note that the 
third column describes a sequence which is that of the first 
column in reverse, and that the first half of the sequence in the 
second column is the reverse of the sequence in its second half, 
with $x$ and $z$ axes interchanged.

In the first row, the caustic is axially symmetric about the $z$ 
axis.  It is as described earlier in Fig.~\ref{caust1}.  In the 
second row, the axial symmetry is broken and the tent pole has 
acquired a diamond shaped cross section.  The caustic is now as 
described in Figs.~\ref{ttt}~-~\ref{hyp_umb}.  In the third row, 
the cusps on the tent pole that lie in the $x = 0$ plane have 
pierced the tent roof all the way from top to bottom, and the 
hyperbolic umbilics in the $y = 0$ plane have moved towards the $z = 0$ 
plane.  In the fourth row, the hyperbolic umbilics have almost reached
the $z = 0$ plane.  What was the tent roof in the first row is now 
stretched along the $x$ axis, and becomes the tent pole in the fifth 
row when the inner caustic is axially symmetric about the $x$ axis.  
In the process described by Fig.~\ref{g1g2}, in which the inner 
caustic metamorphoses from a tent symmetric about the $z$ axis 
to a tent symmetric about the $x$ axis, the tent roof smoothy 
deforms into the tent pole and vice-versa.

The plots of Fig.~\ref{g1g2} are reminiscent of caustics seen in
gravitational lensing theory~\cite{BN,Petters,Petters2}, and ship
stability analysis~\cite{Poston,Zeeman}.

\subsubsection{Adding a rotational component}

Here we add a rotational component ($c_3 \neq 0$) to the initial 
velocity field of Eq.~(\ref{irrot}) to see the metamorphosis of the 
inner caustic from tent to tricusp ring.  Fig.~\ref{tr_non} shows 
the $z = 0$ cross sections of the inner caustic during such a 
transition. In Fig.~\ref{tr_non-a}, the initial velocity field is 
irrotational and we see a circle and diamond, as before. In 
Fig.~\ref{tr_non-b}, the diamond is skewed because of the rotation 
in the $z=0$ plane introduced by $c_3 \neq 0$. Fig.~\ref{tr_non-c} 
shows the case $c_3=\xi$. As $c_3$ is increased further, the diamond 
transforms into two \emph{swallowtail} $(A_4)$ catastrophes \cite{cat} 
joined back to back [\ref{tr_non-d}]. There are two flows in the central 
region, six flows in the cusped region of each swallowtail, four in the 
other regions inside the circle and two outside the circle. Finally, 
the swallowtails pinch off to form the inner circle of the $z=0$ 
section of the tricusp ring.  Fig.~\ref{tr_non4} shows the transition 
in three dimensions. 

\section{Conclusions\label{six}}

In Section II, we gave a mathematical proof of the statement that 
a cold flow of collisionless particles from all directions in and 
out of a region necessarily produces a caustic.  We call this caustic 
the ``inner caustic" of the in and out flow.  The main purpose of 
our paper was to determine the catastrophe structure of the inner 
caustics formed by cold dark matter particles falling in and out of 
a galactic gravitational potential well.

The structure of the inner caustic depends for the most part on the 
angular momentum distribution of the particles falling in.  It had 
been shown previously by analytical methods \cite{sing} that the 
inner caustic is a tricusp ring when the velocity distribution 
of the infalling particles is dominated by net overall rotation.
However we show in Section III that the leading theory for the 
angular momentum of galaxies, namely tidal torque theory, predicts 
that the velocity field is irrotational 
$(\vec{\nabla} \times \vec{v} = 0)$ to all orders of perturbation 
theory, i.e. to all orders in an expansion in powers of the size 
of density perturbations.  So, tidal torque theory states not only 
that the initial velocity field of the infalling particles is 
not dominated by net overall rotation, it states that the initial 
velocity field is exactly irrotational.

So the question was: what is the structure of the inner caustic when 
the initial velocity field is irrotational?  Or more generally, what 
is that structure when the initial velocity field is not dominated by 
net overall rotation?  We addressed this issue by simulating the flow 
of cold collisionless particles falling in and out of a fixed 
gravitational potential.  In principle we can do this for any 
initial velocity field.  However, we restricted ourselves 
to initial velocity fields of the form $\vec{v} = M \vec{x}$ where 
$\vec{x}$ is initial position and $M$ is a $3 \times 3$ real traceless 
matrix.  As far as uncovering the catastrophe structure of inner 
caustics is concerned, we expect this to be a sufficiently broad 
set of initial velocity fields.  Adding higher order terms to the 
expansion of $\vec{v}$ in powers of $\vec{x}$ is expected to merely
deform the inner caustics obtained when keeping linear terms only. 
$M$ can be written in the form of Eq.~(\ref{M}) which depends on 
five parameters: $\vec{c} = (c_1, c_2, c_3)$, $g_1$ and $g_2$.  
The first three describe net rotation with angular velocity 
$\vec{c}$. The last two describe irrotational flow.

In Section IV, we simulated flows which are dominated by net rotation 
and confirm that the inner caustic is a tricusp ring in that case.  
We show that the tricusp ring is stable under perturbations both in 
the initial velocity field and in the gravitational potential.  This 
stability is not a surprise since it is known that the structure of 
non-degenerate catastrophes is stable under perturbations \cite{Vakif}.  
When the initial velocity field and/or the gravitational potential are 
not axially symmetric, the dimensions of the tricusp vary along the ring.  
In the neighborhood of a point where the tricusp dimensions have shrunk 
to zero, the catastrophe structure is the full elliptic umbilic.

\begin{figure}

\subfigure[Caustic structure.]{\label{caust1-a}\resizebox{4in}{4in}
{\includegraphics{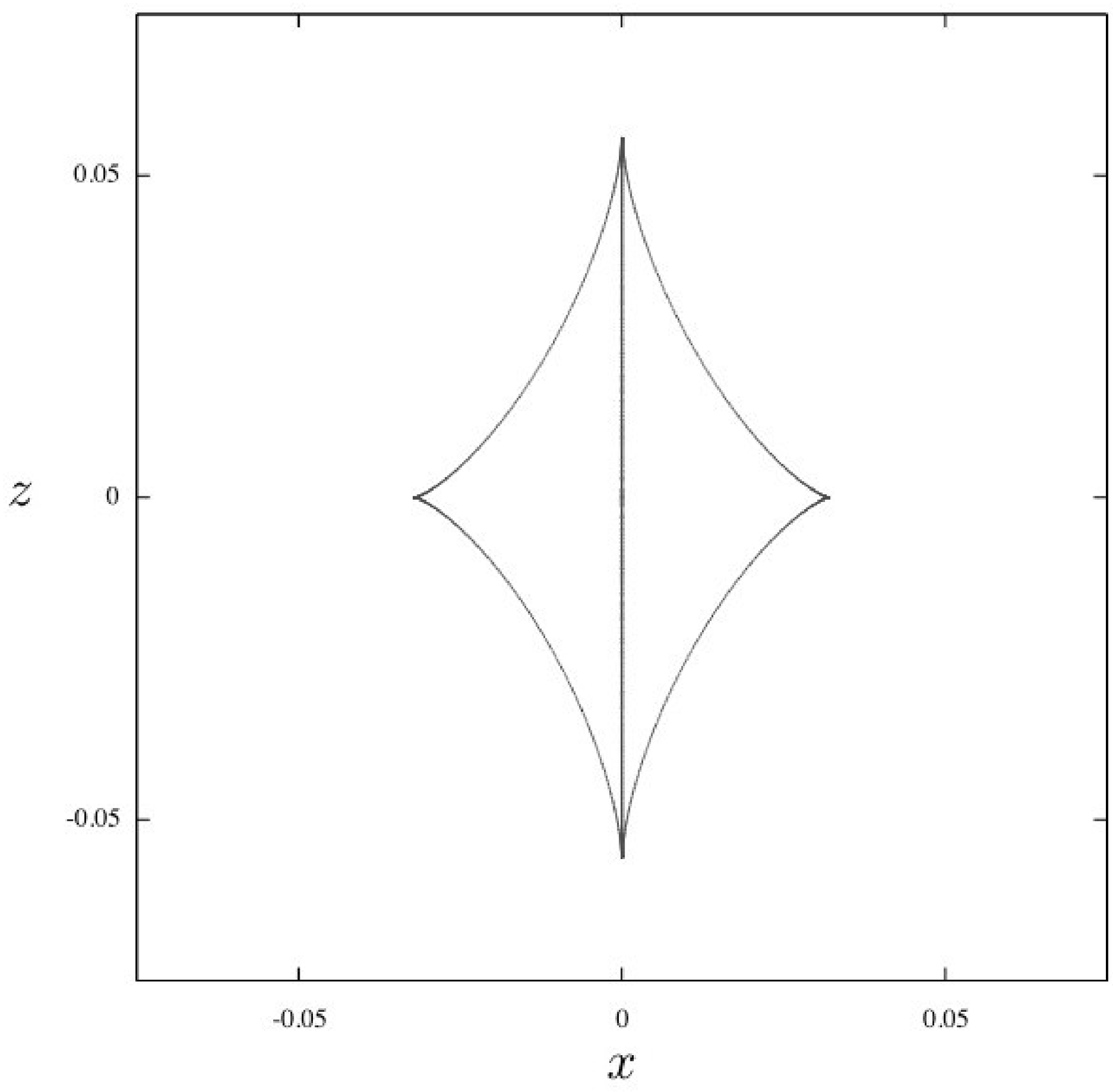}}}
\subfigure[Dark matter flows.]{\label{caust1-b}
\resizebox{4in}{3in}{\includegraphics{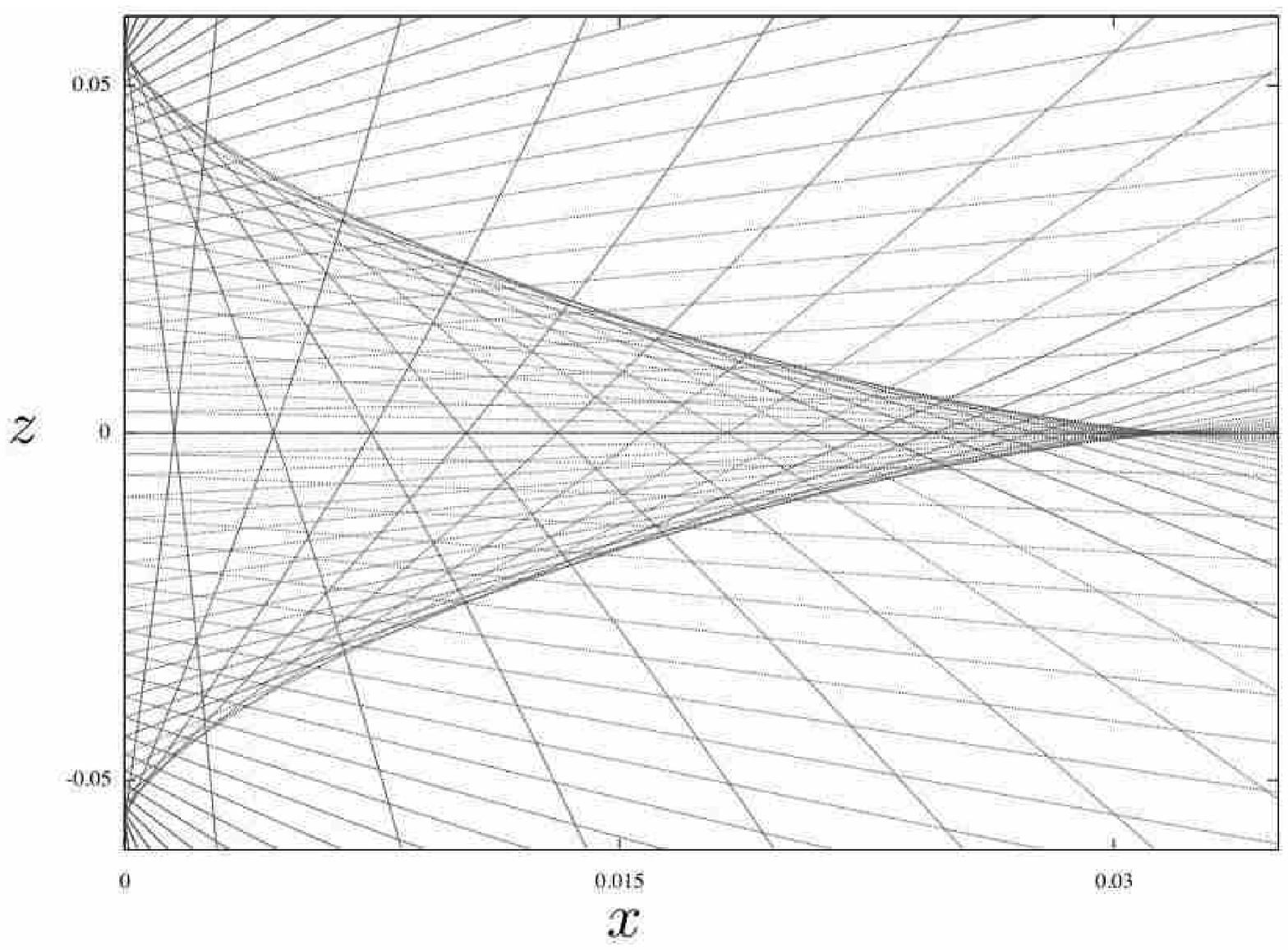}}}
\caption{Cross section of the inner caustic produced 
by the irrotational axially symmetric velocity field  
$\vec{v}= - 0.05 \sin\,2\theta\;\hat\theta$.\label{caust1}}
\end{figure}

In Section V, we simulated flows which are not dominated by a 
rotational component.  We started off simulating flows which 
are both irrotational ($\vec{c} = 0$) and axially symmetric 
($g_1 =g_2$).  For such flows the inner caustic is shown in
Fig.~\ref{caust1} for $|g_1| < g_{1*} \simeq 0.05$.  We call 
this structure a ``tent caustic".  It has a caustic line on the 
axis of symmetry, which we call the ``tent pole", connected to 
a caustic surface which we call the ``tent roof".  When 
$g_1 < - g_{1*}$, the inner caustic is the same as for 
$|g_1| < g_{1*}$ except the cusp in the equatorial plane is 
replaced by a butterfly caustic.  When $g_1 > g_{1*}$, the 
cusps on the symmetry axis are replaced by butterfly caustics.

\begin{figure}

\subfigure{\label{zp-a}\resizebox{3.5in}{3.5in}
{\includegraphics{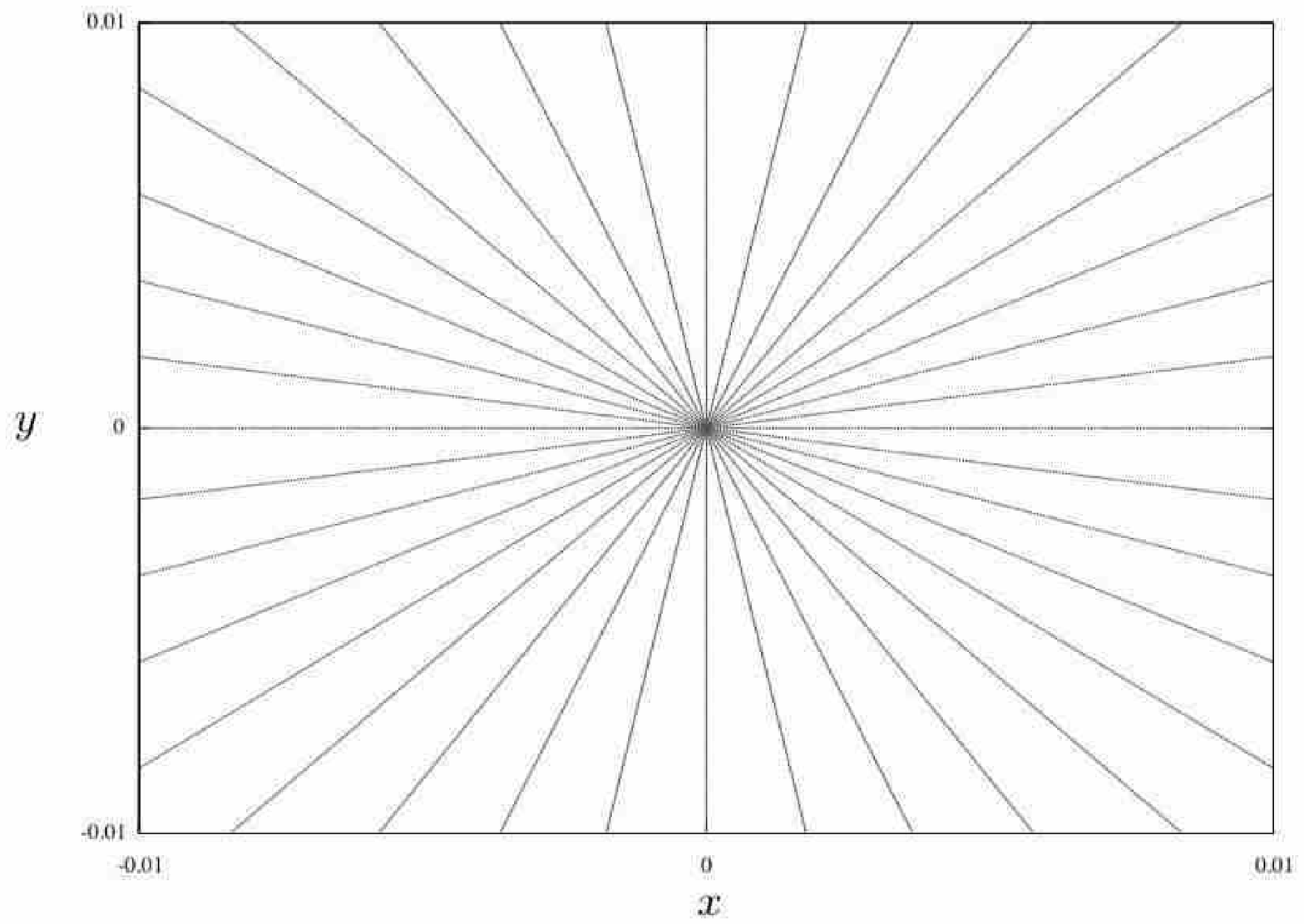}}}
\subfigure{\label{zp-b}\resizebox{3.5in}{3.5in}
{\includegraphics{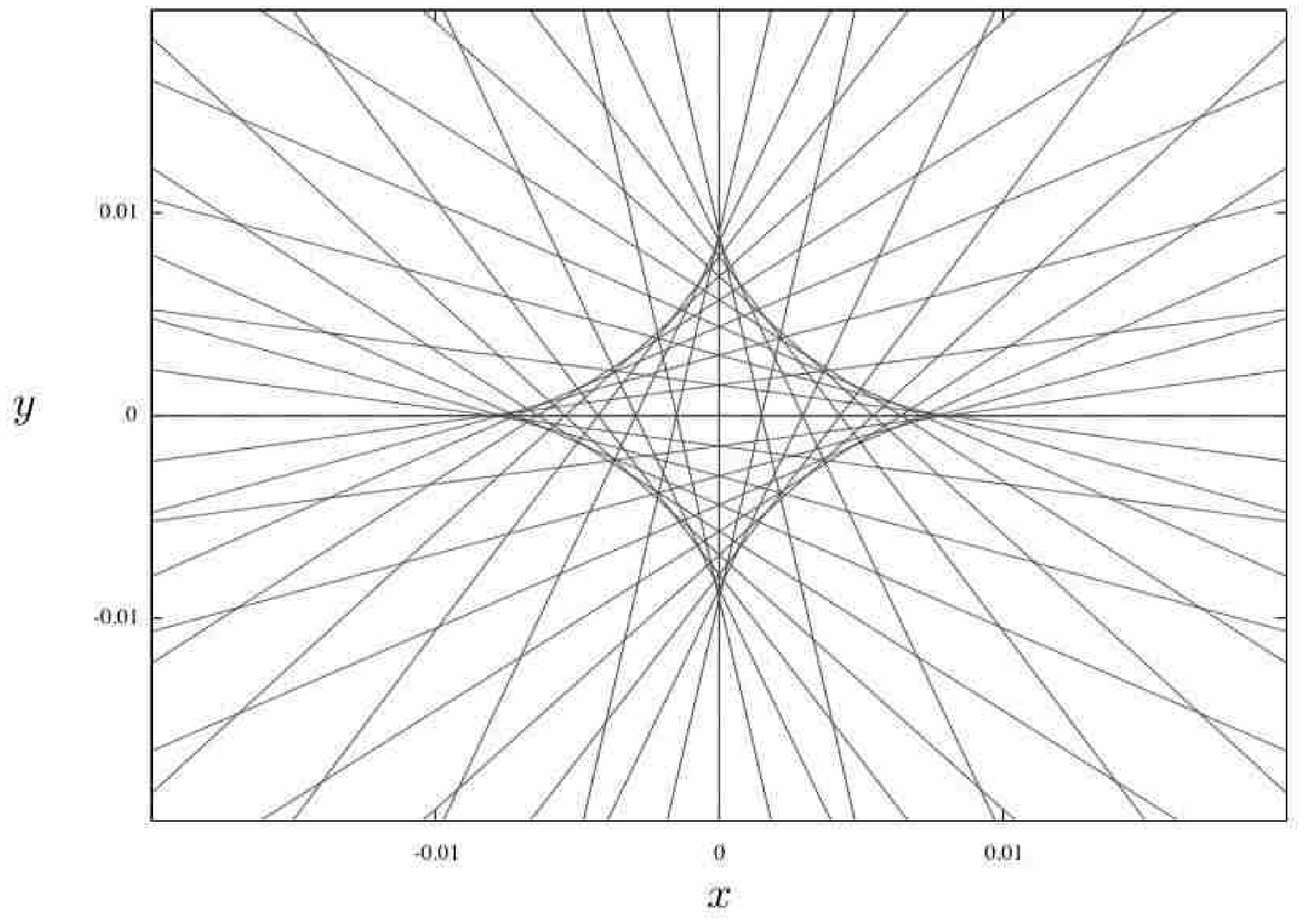}}}

\caption{Dark matter flows in the $z = 0$ plane for the 
initial velocity field of Eq.~(\ref{irrot}) when (a) 
$\xi=0, g=-0.05$, and (b) $\xi=0.01,g=-0.05$.  In (a)
the particles move on radial orbits.  In (b) the particles 
have angular momentum about the $z$ axis.\label{zp}}
\end{figure}

\begin{figure}

\subfigure{\label{ttt-a}
\resizebox{3in}{3in}{\includegraphics{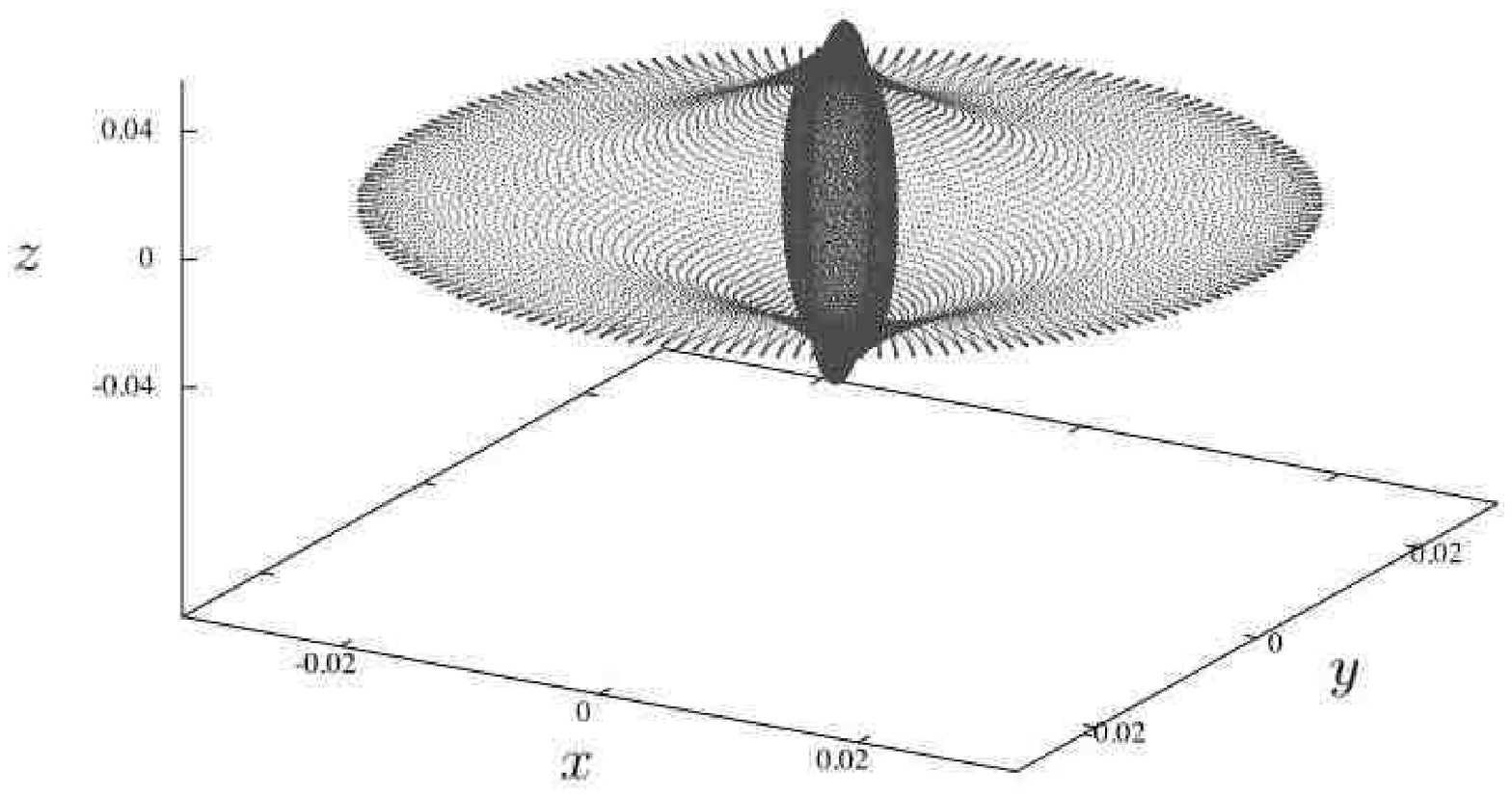}}}
\subfigure{\label{ttt-b}
\resizebox{3in}{3in}{\includegraphics{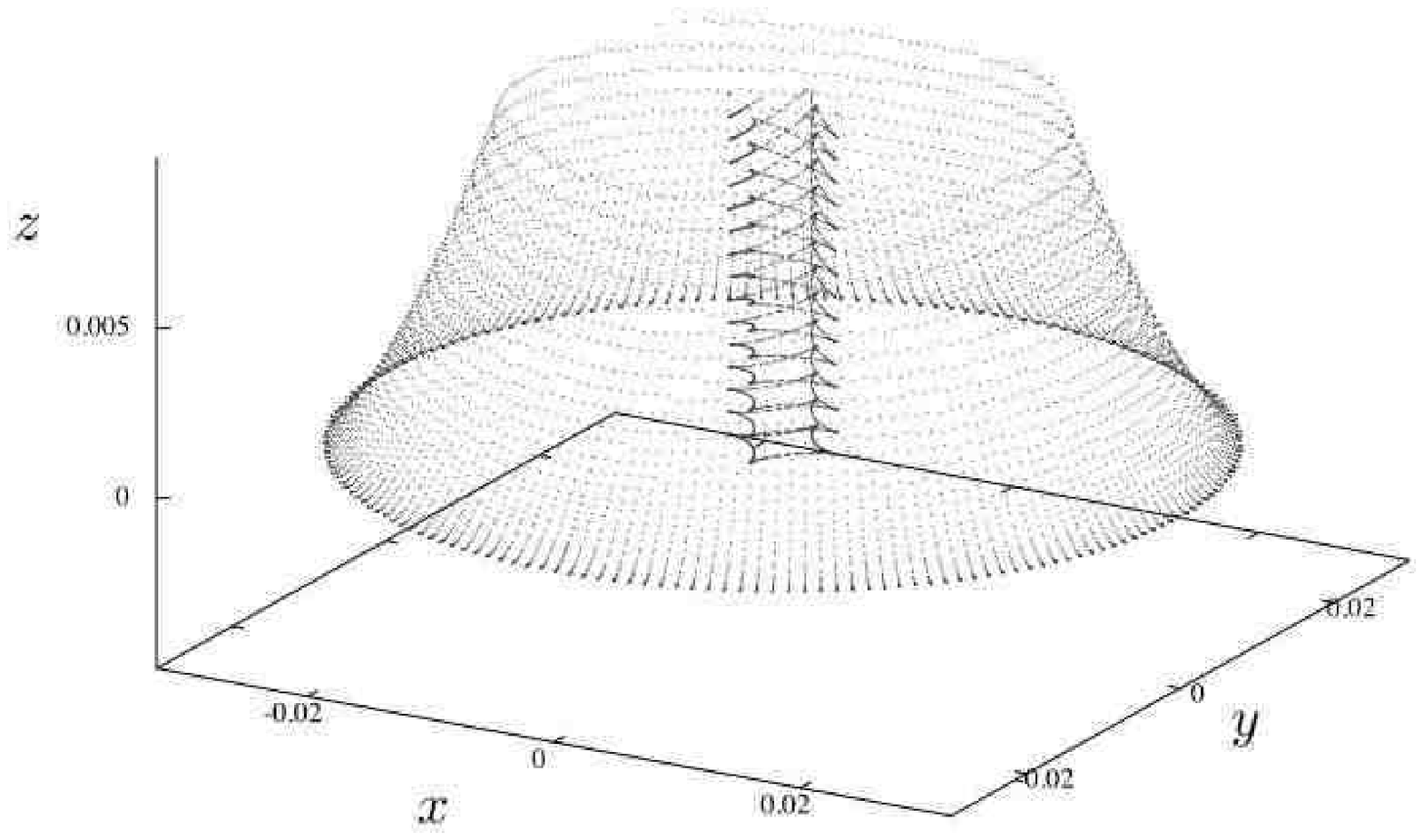}}}

\caption{The inner caustic for the initial velocity 
field of Eq.~(\ref{irrot}) with $\xi = 0.005, g = -0.05$:
(a) the tent caustic in 3 dimensions; (b) a succession of 
constant $z$ sections over the range $0 \leq z \leq 0.01$.\label{ttt}}

\end{figure}

\begin{figure}

\subfigure{\label{tttc-a}
\resizebox{3in}{3in}{\includegraphics{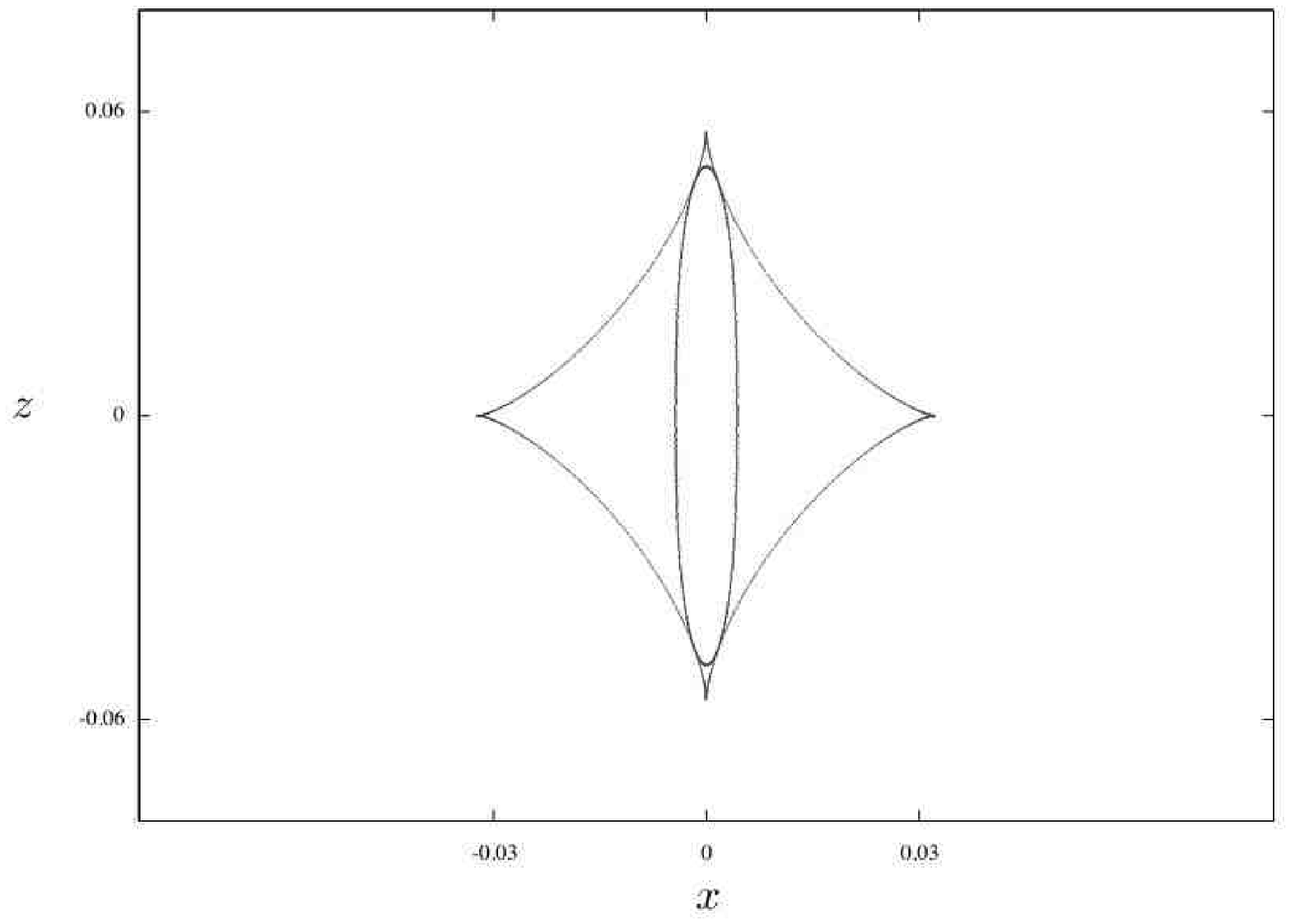}}}
\subfigure{\label{tttc-b}
\resizebox{3in}{3in}{\includegraphics{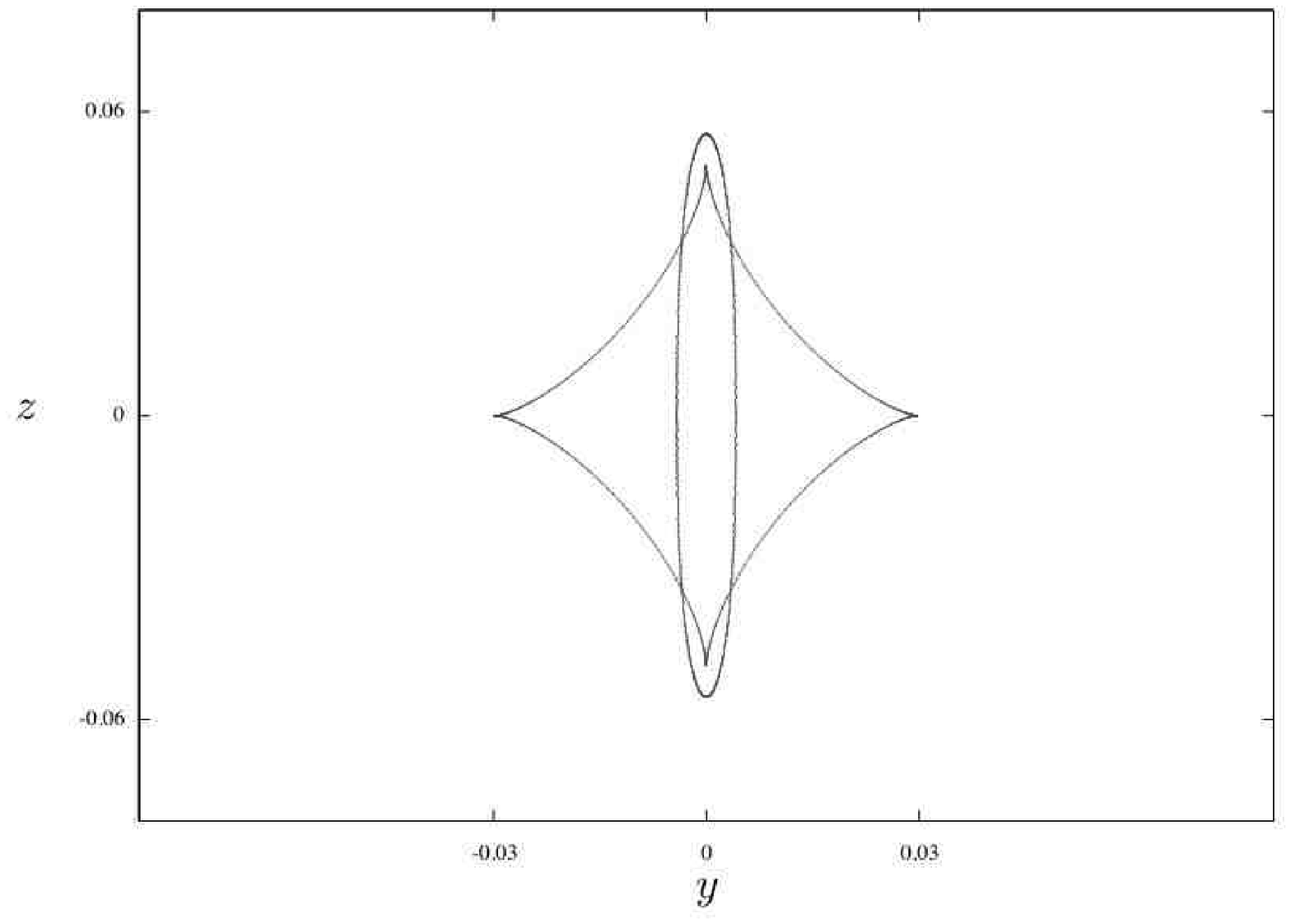}}}

\caption{The tent caustic of Fig.~\ref{ttt} in (a) $y = 0$ 
cross section and (b) $x = 0$ cross section.  The manner in 
which the tent pole is connected to the tent roof is described 
in detail in Fig.~\ref{hyp_umb}.\label{tttc}}
\end{figure}

\begin{figure}

\subfigure[]{\label{hyp_umb-a}\resizebox{1.5in}{1.5in}
{\includegraphics{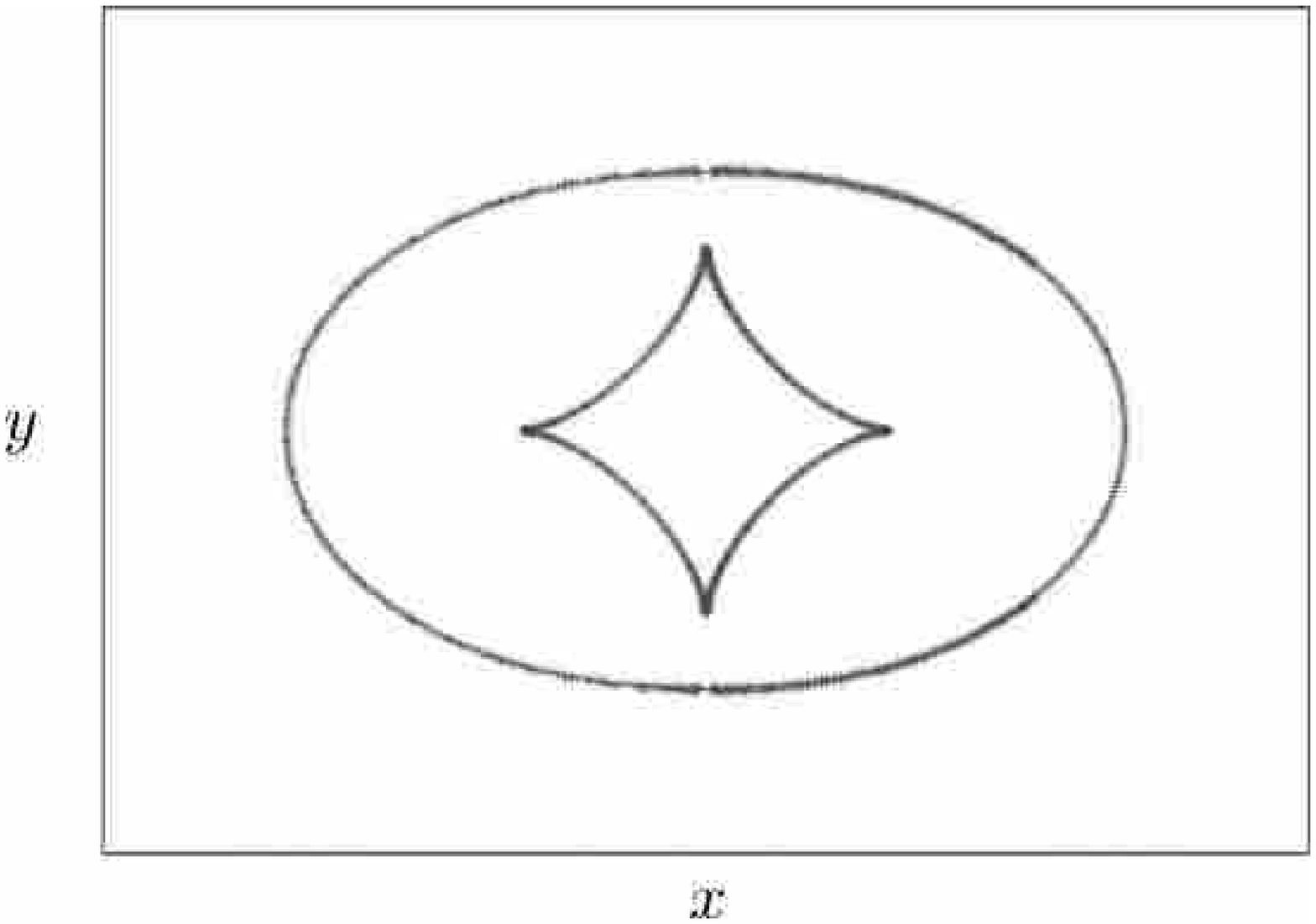}}}
\subfigure[]{\label{hyp_umb-b}\resizebox{1.5in}{1.5in}
{\includegraphics{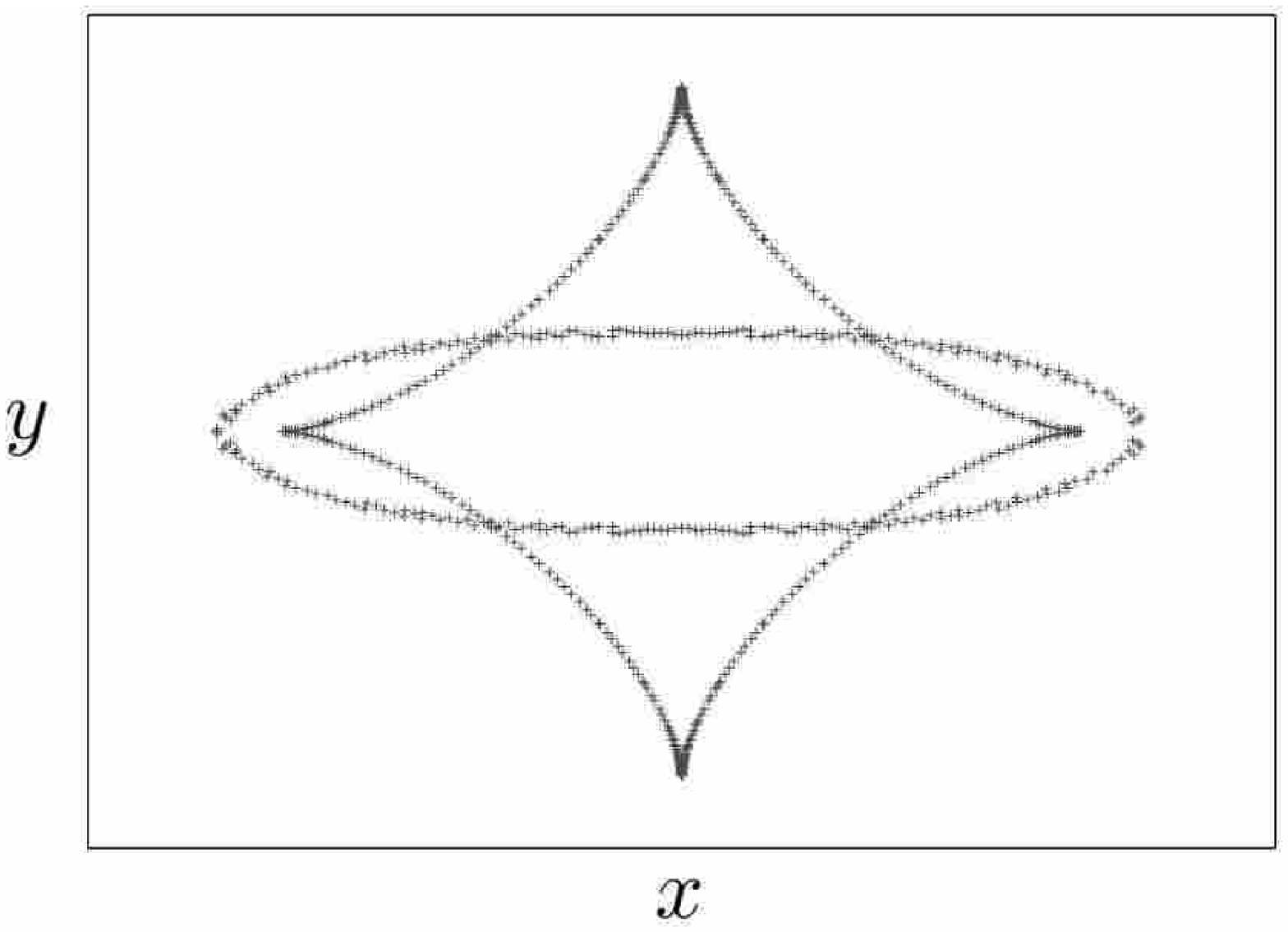}}}
\subfigure[]{\label{hyp_umb-c}\resizebox{1.5in}{1.5in}
{\includegraphics{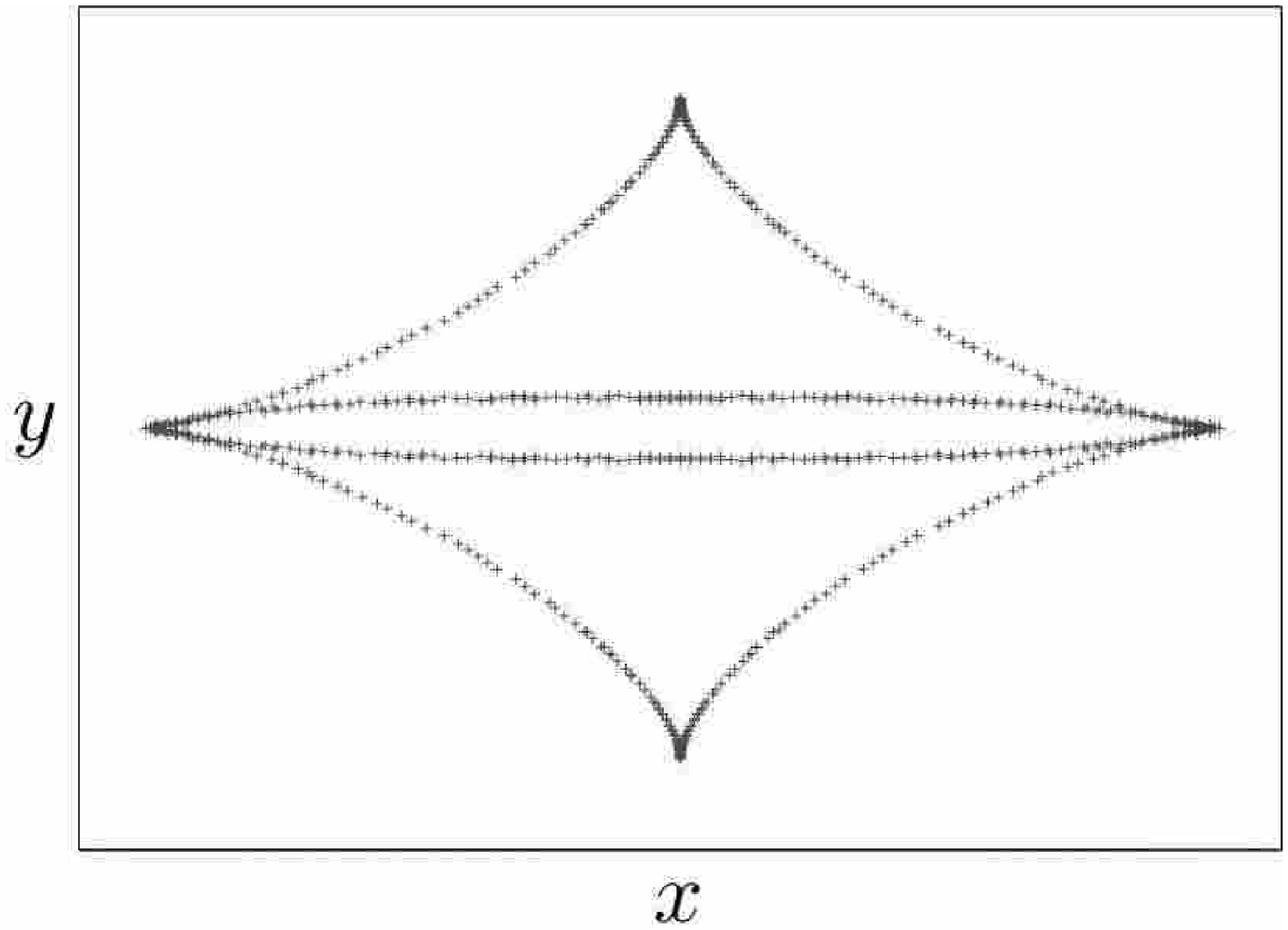}}}
\subfigure[]{\label{hyp_umb-d}\resizebox{1.5in}{1.5in}
{\includegraphics{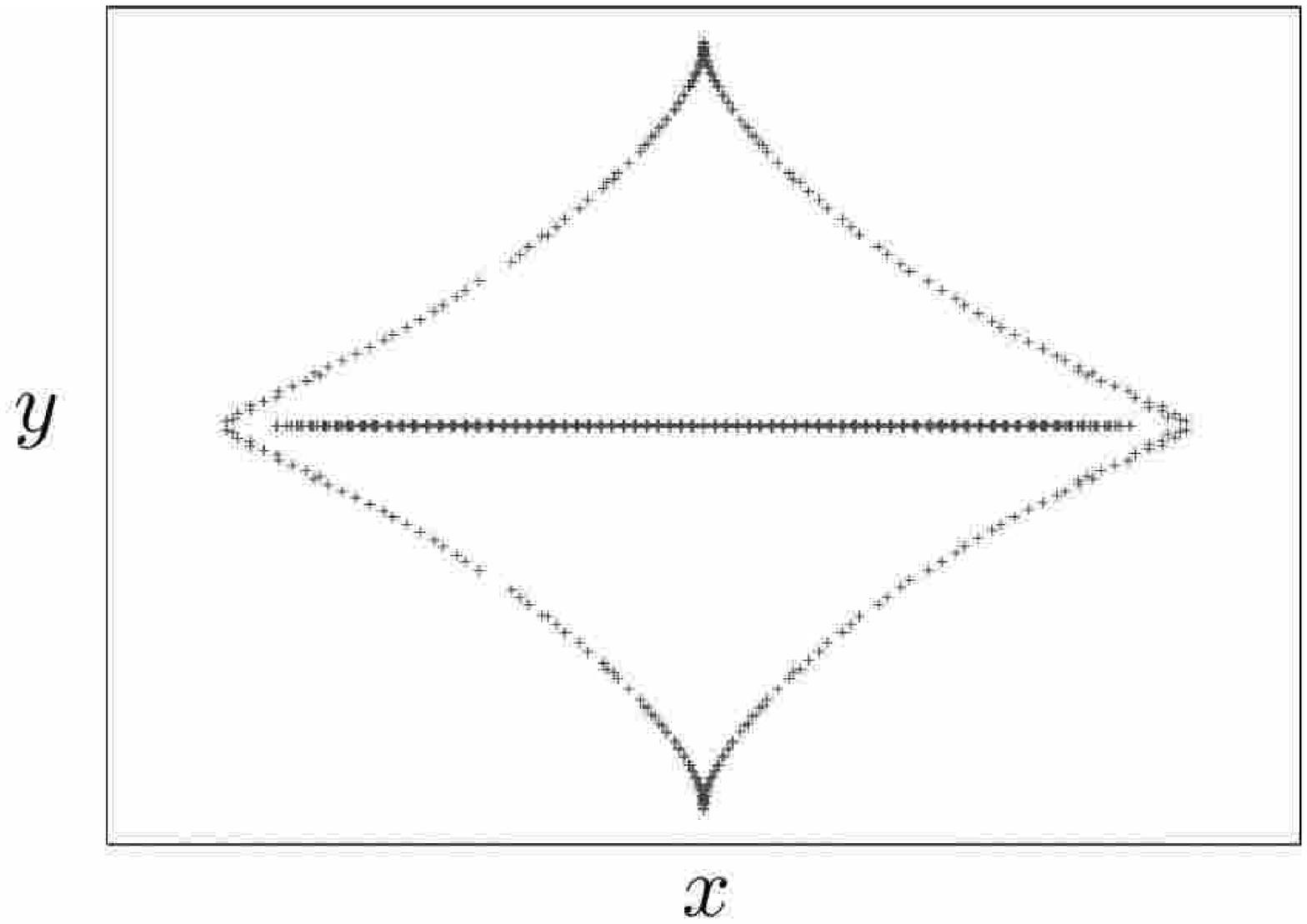}}}\\
\hspace{-.2in}
\subfigure[]{\label{hyp_umb-e}\resizebox{1.55in}{1.55in}
{\includegraphics{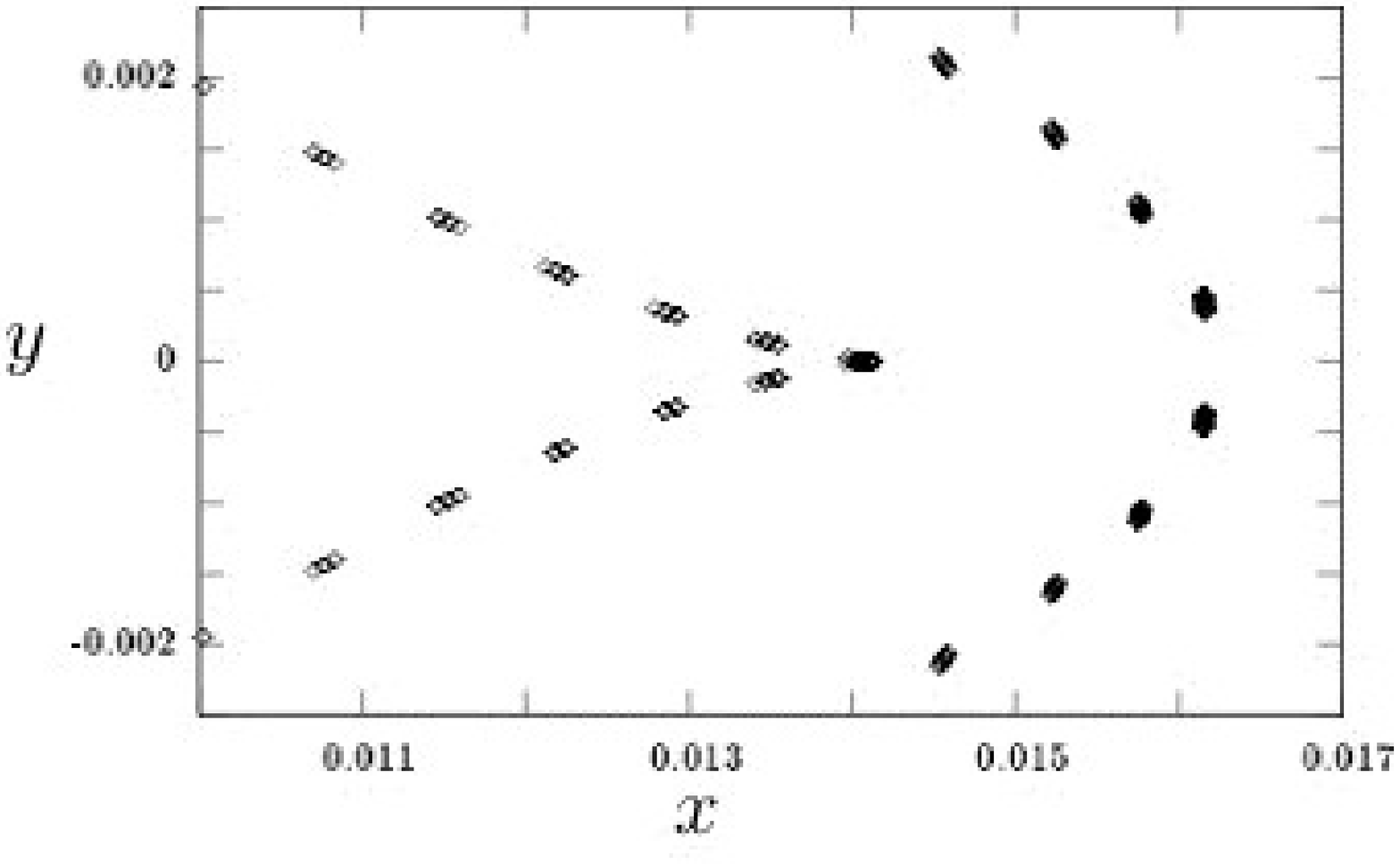}}}
\subfigure[]{\label{hyp_umb-f}\resizebox{1.55in}{1.55in}
{\includegraphics{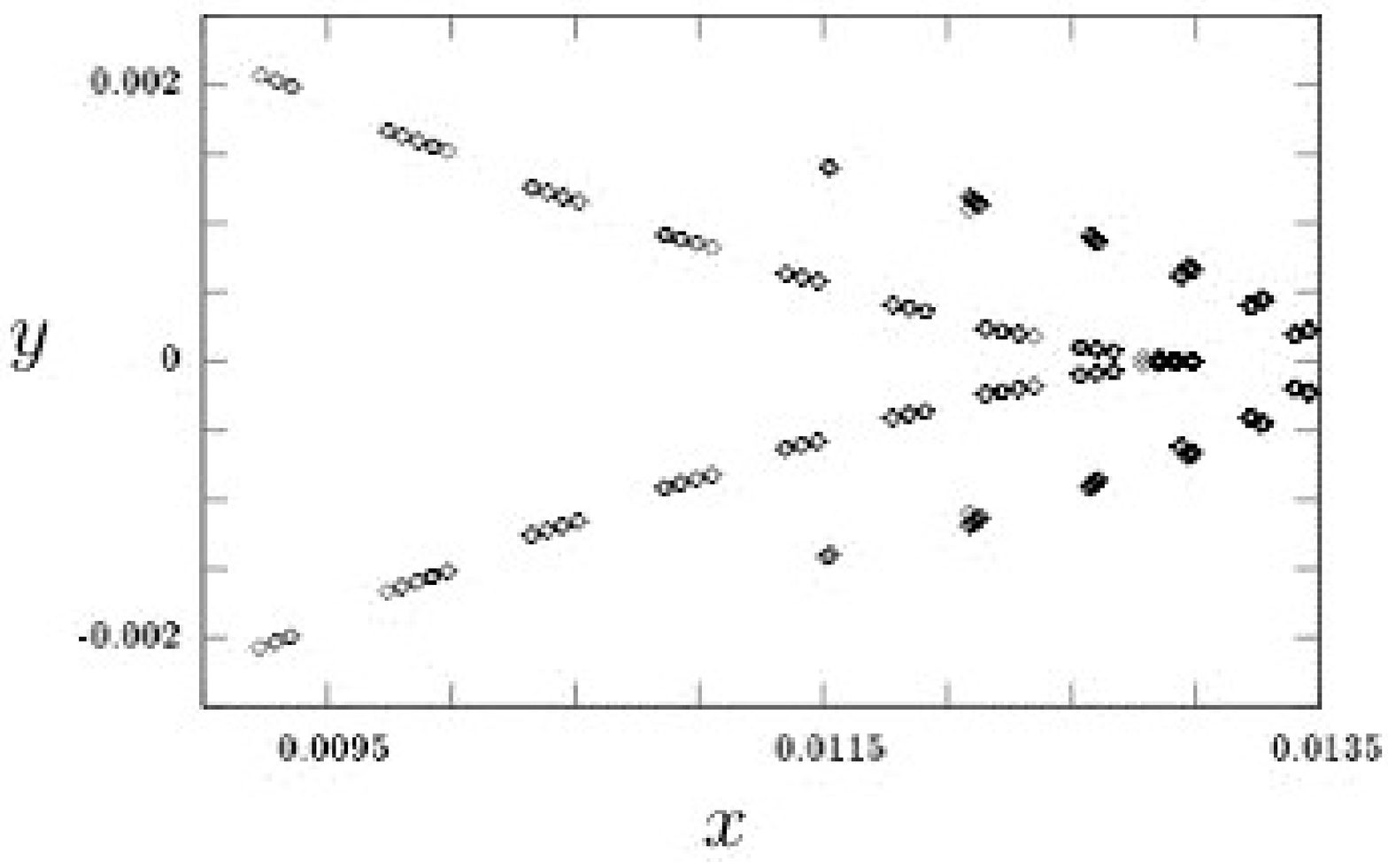}}}
\subfigure[]{\label{hyp_umb-g}\resizebox{1.55in}{1.55in}
{\includegraphics{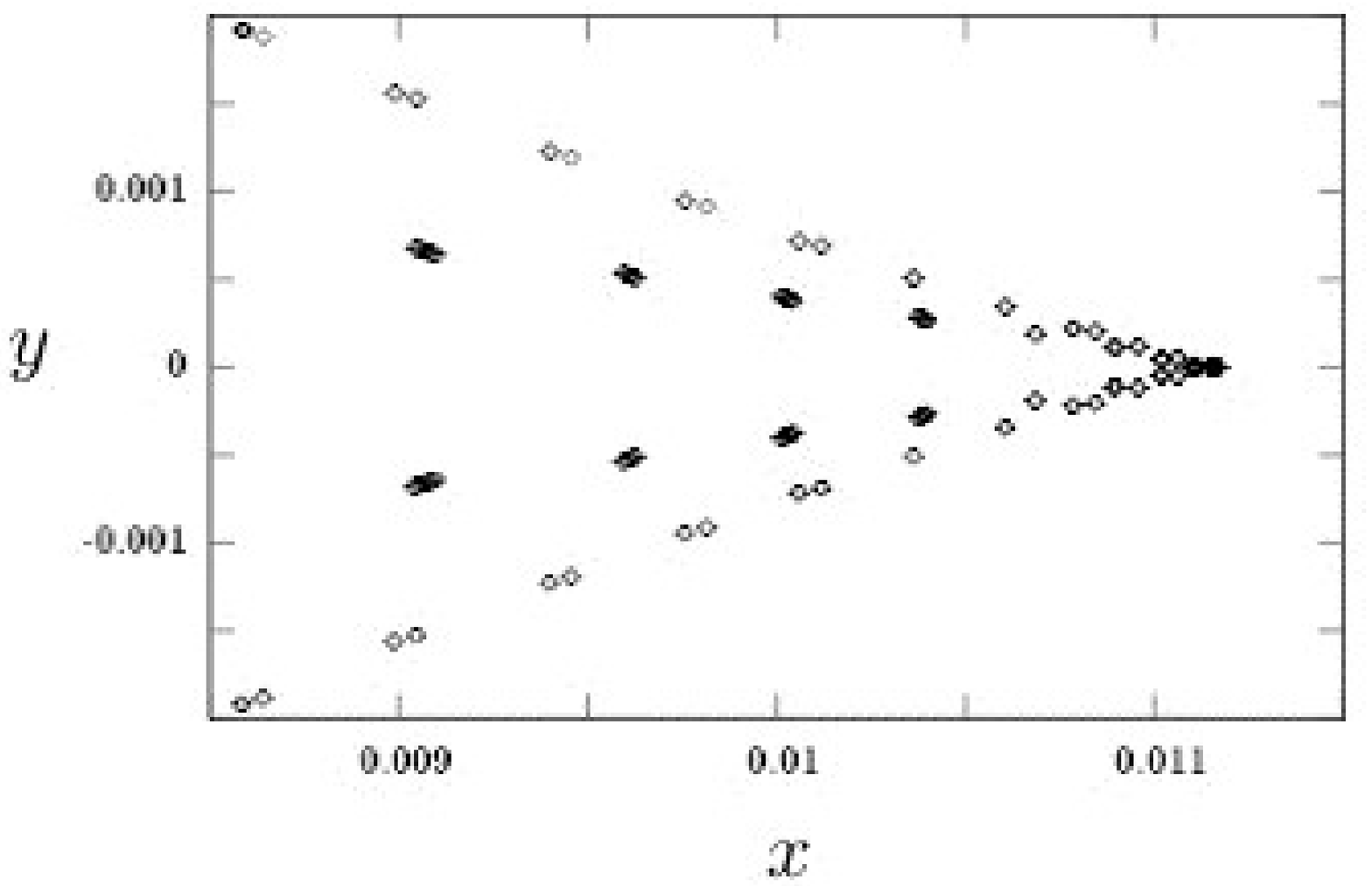}}}
\subfigure[]{\label{hyp_umb-h}\resizebox{1.55in}{1.55in}
{\includegraphics{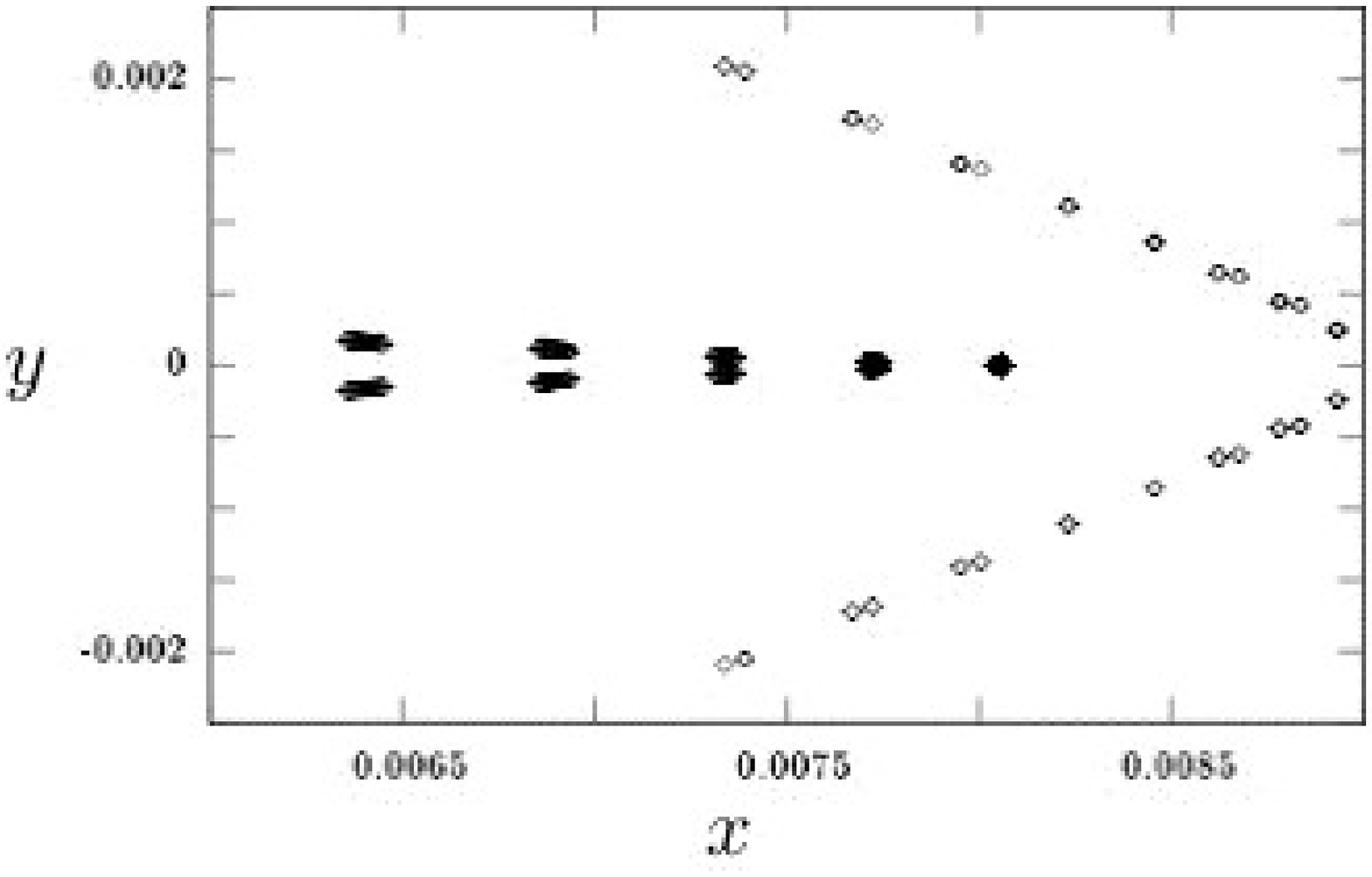}}}\\
\subfigure[]{\label{hyp_umb-i}\resizebox{3in}{3in}
{\includegraphics{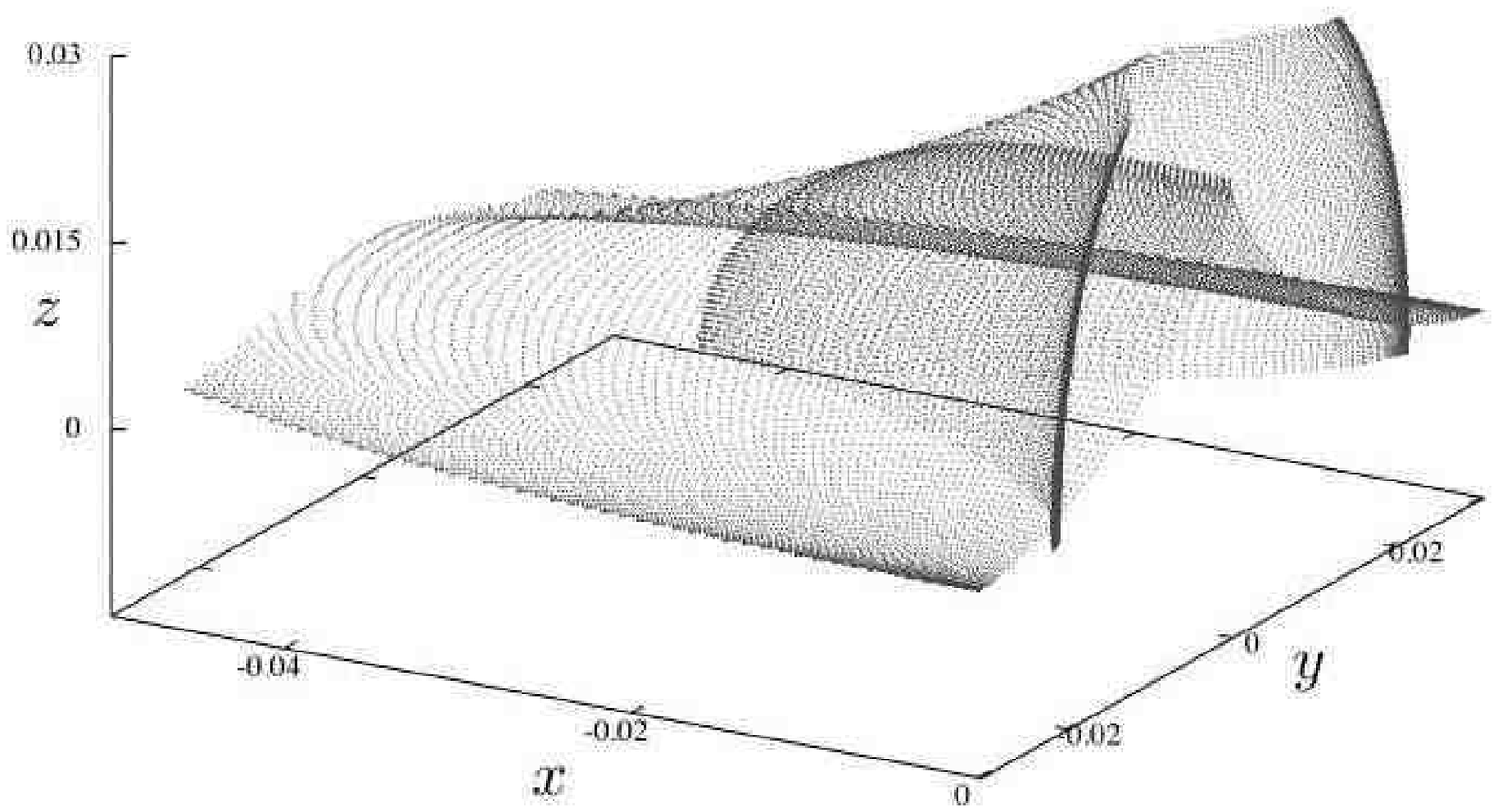}}}

\caption{(a)-(d) Constant $z$ sections of the tent caustic 
of Fig.~\ref{ttt} in the region where the tent pole connects
with the tent roof.  In (a) the pole is entirely inside the 
roof. In (b) the cusps of the pole which are in the $x = 0$
plane have pierced the roof.  From (b) to (d) the parts of 
the pole and the roof near the $y = 0$ plane traverse each 
other by forming hyperbolic umbilic catastrophes, one on 
the $x > 0$ side and one on the $x < 0$ side.  (e)-(h) Constant 
$z$ sections of the tent caustic near the hyperbolic umbilic on 
the $x > 0$ side, for greater detail.  (i) the hyperbolic umbilic
at $z > 0, x < 0$ in three dimensions.\label{hyp_umb}}
\end{figure}

\begin{figure}

\subfigure{\label{g1g2-a}\resizebox{1.4in}{1.4in} 
{\includegraphics{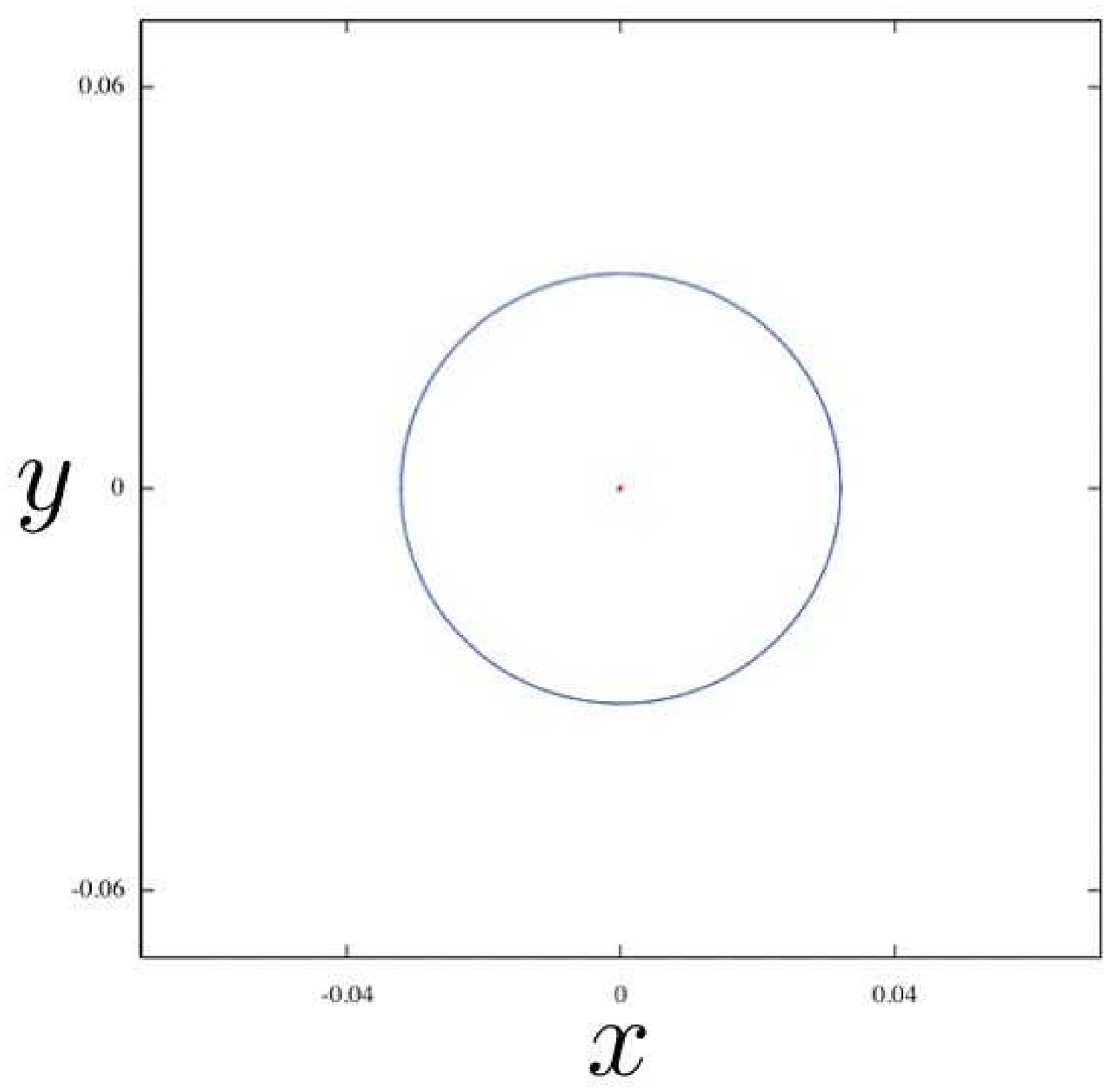}}}
\hspace{.3in} 
\subfigure{\label{g1g2-f}\resizebox{1.3in}{1.3in}
{\includegraphics{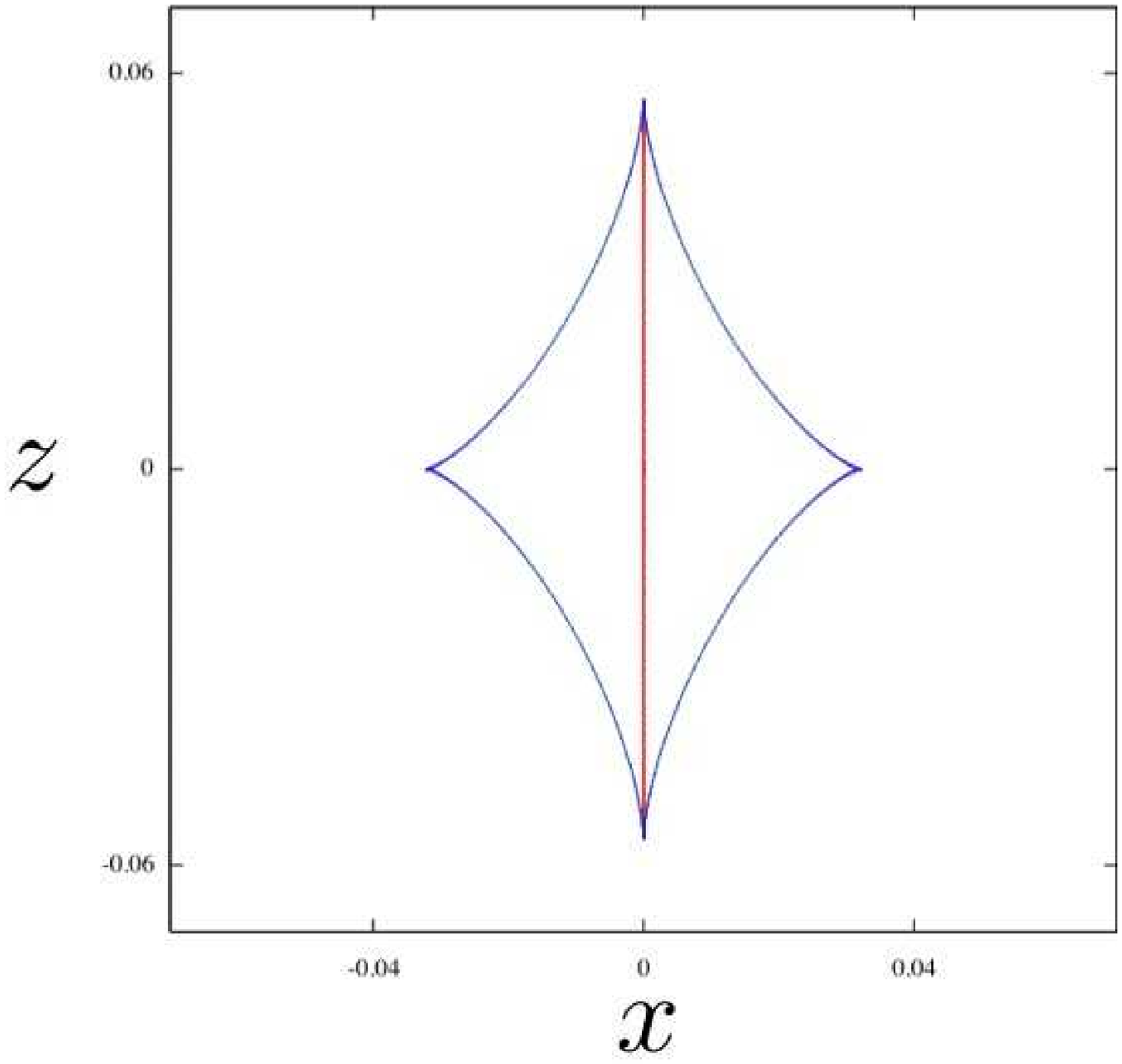}}}
\hspace{.3in}
\subfigure{\label{g1g2-k}\resizebox{1.3in}{1.3in}
{\includegraphics{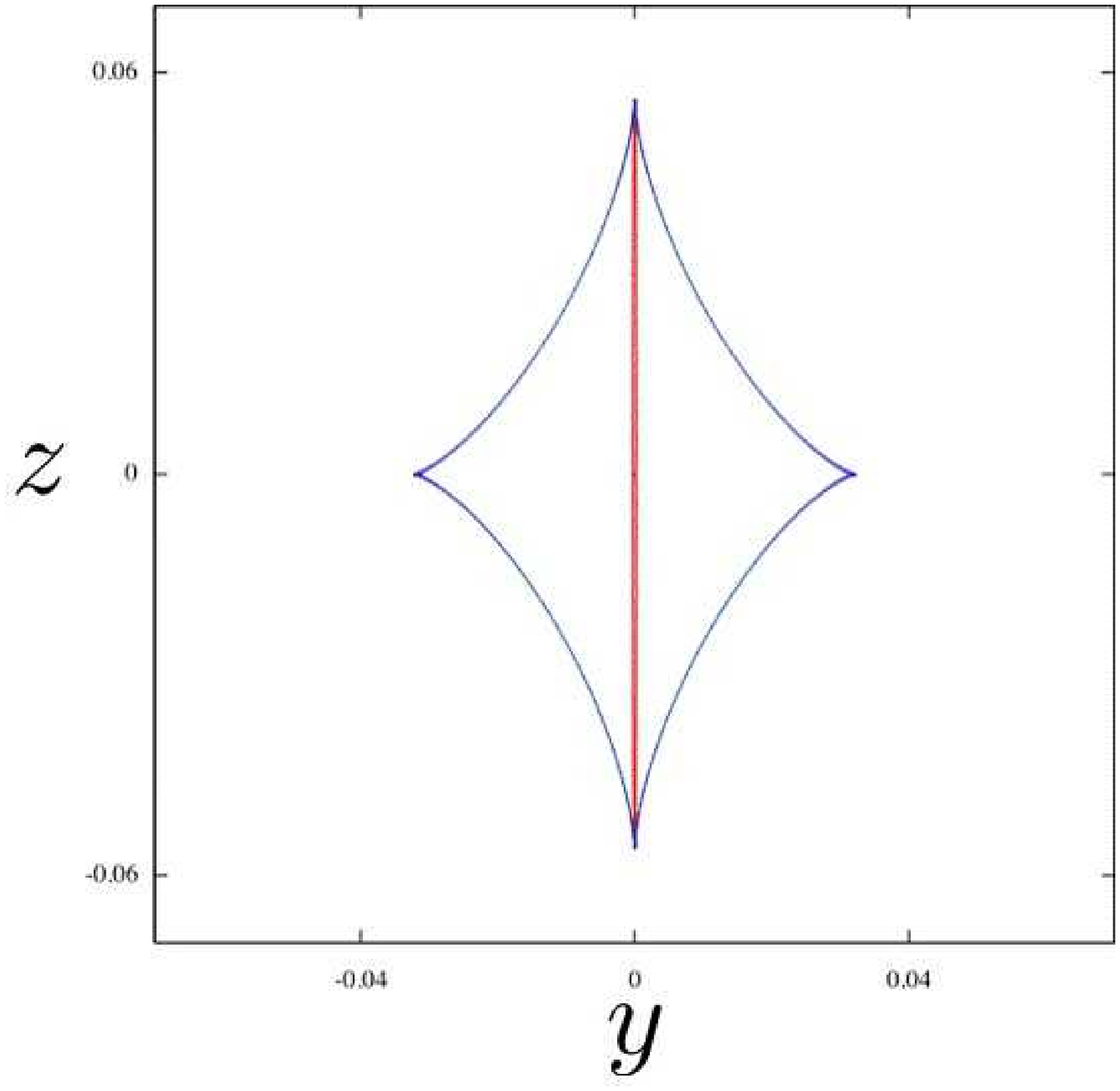}}}\\
\subfigure{\label{g1g2-b}\resizebox{1.3in}{1.3in}
{\includegraphics{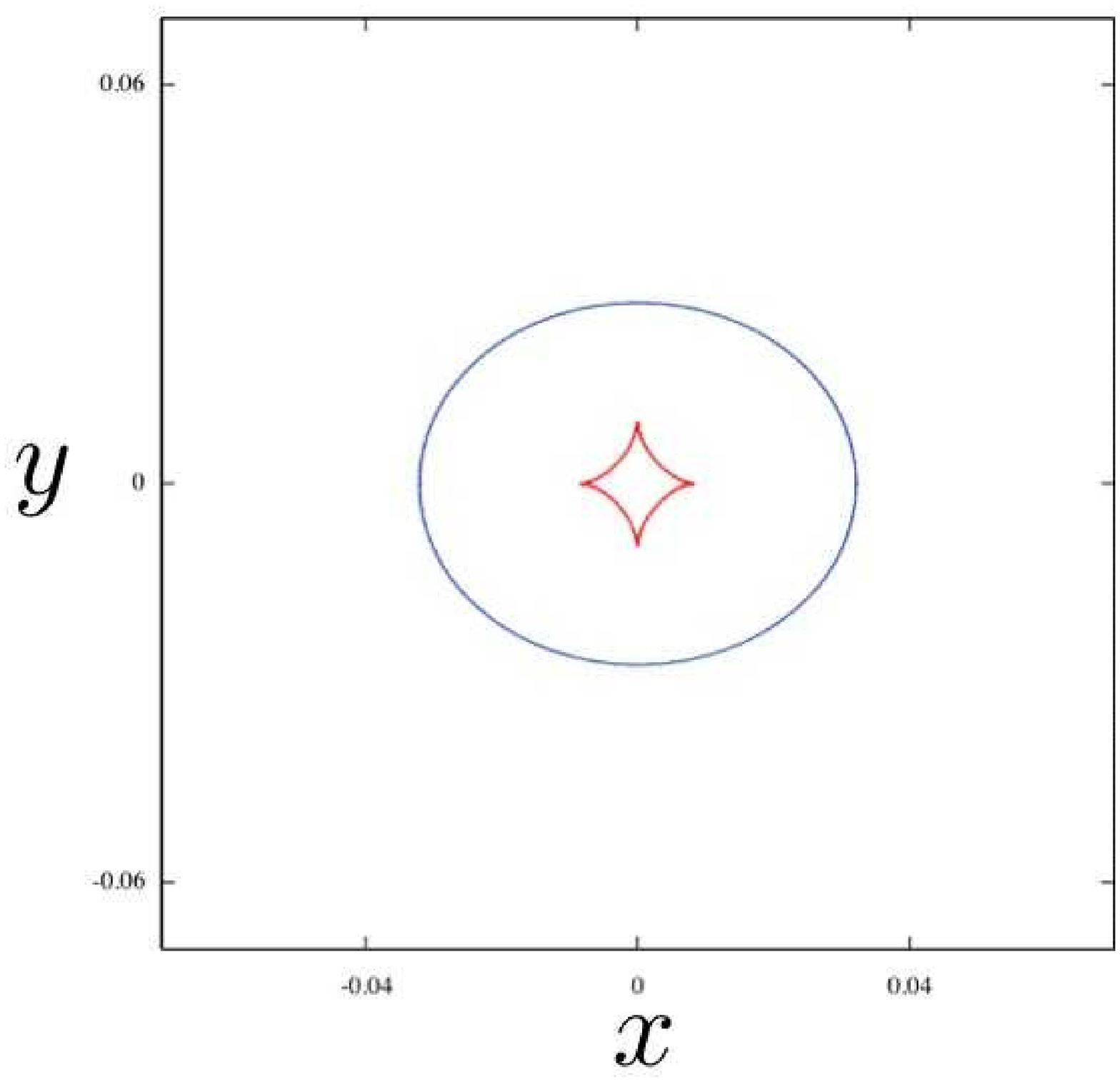}}} 
\hspace{.3in}
\subfigure{\label{g1g2-g}\resizebox{1.3in}{1.3in}
{\includegraphics{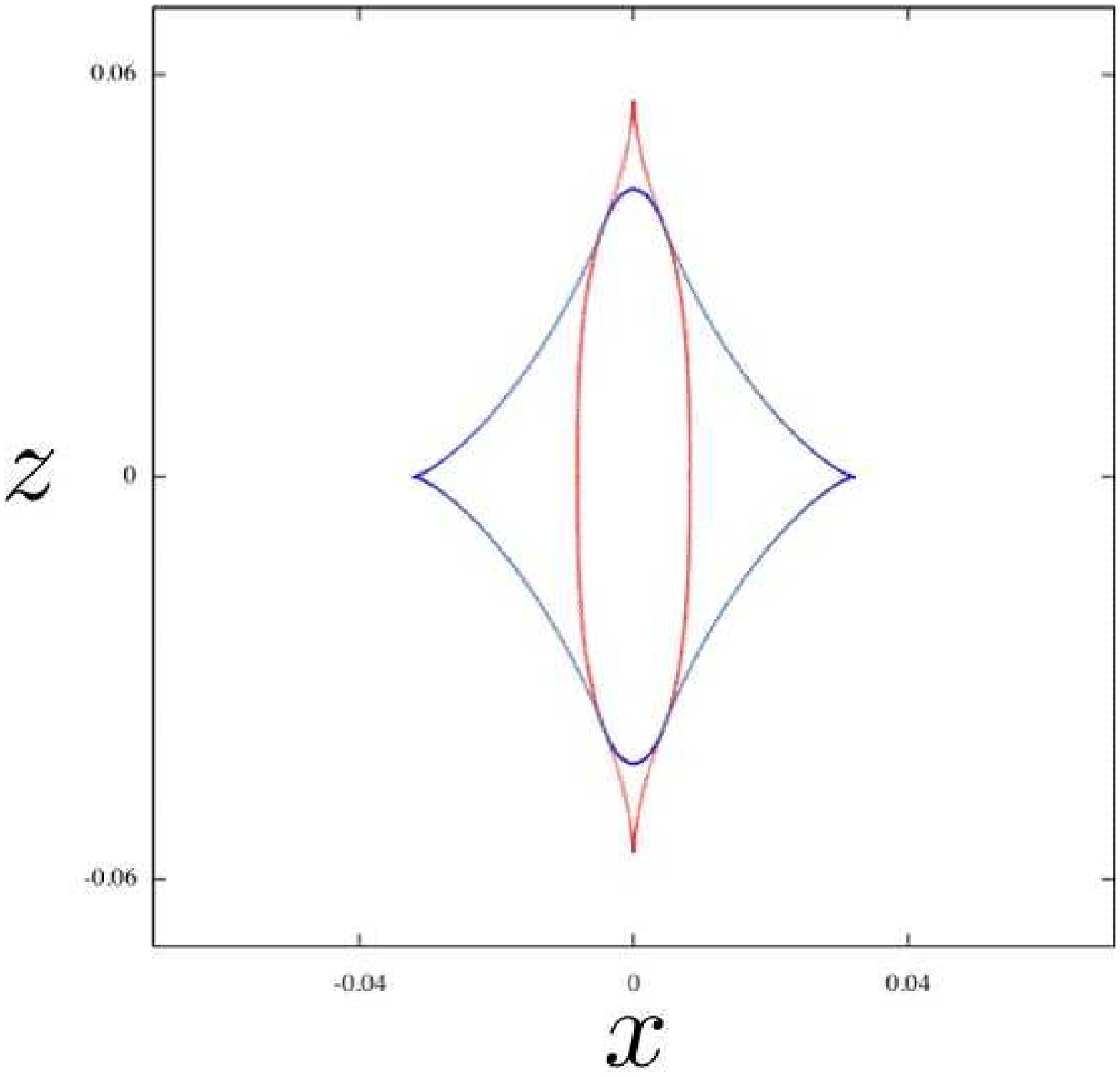}}}
\hspace{.3in}
\subfigure{\label{g1g2-l}\resizebox{1.3in}{1.3in}
{\includegraphics{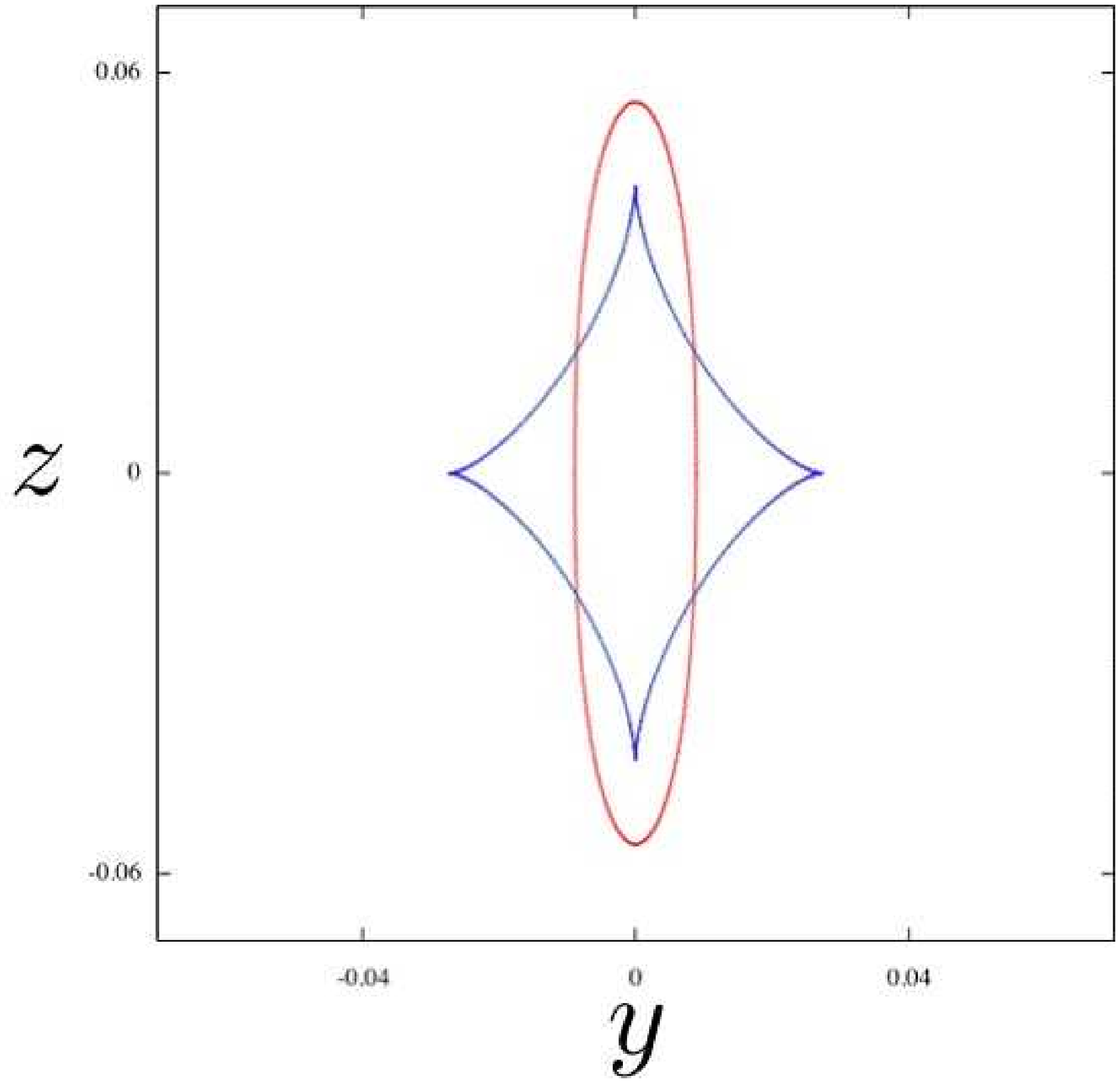}}}\\
\subfigure{\label{g1g2-c}\resizebox{1.3in}{1.3in} 
{\includegraphics{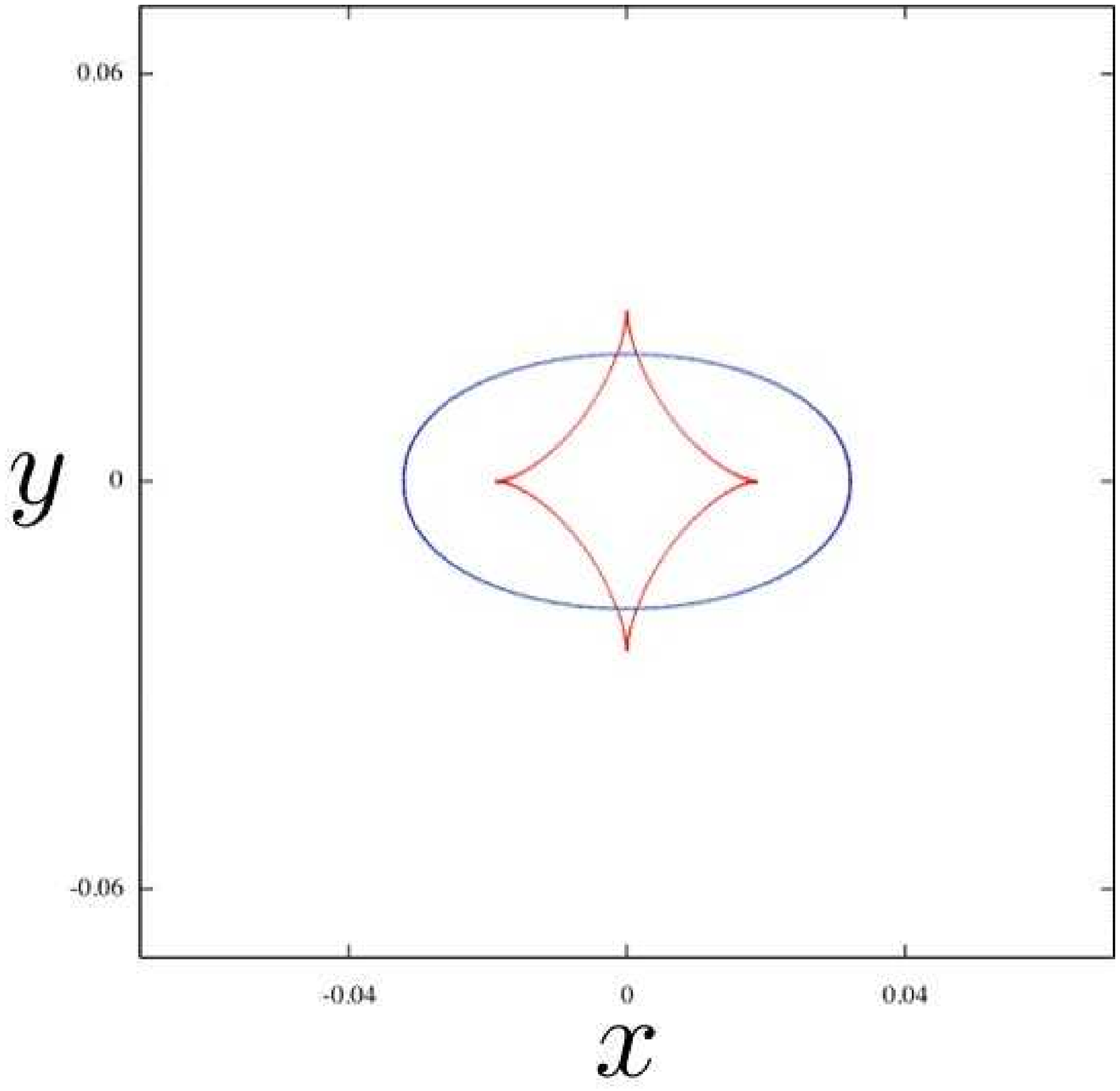}}}
\hspace{.3in}
\subfigure{\label{g1g2-h}\resizebox{1.3in}{1.3in}
{\includegraphics{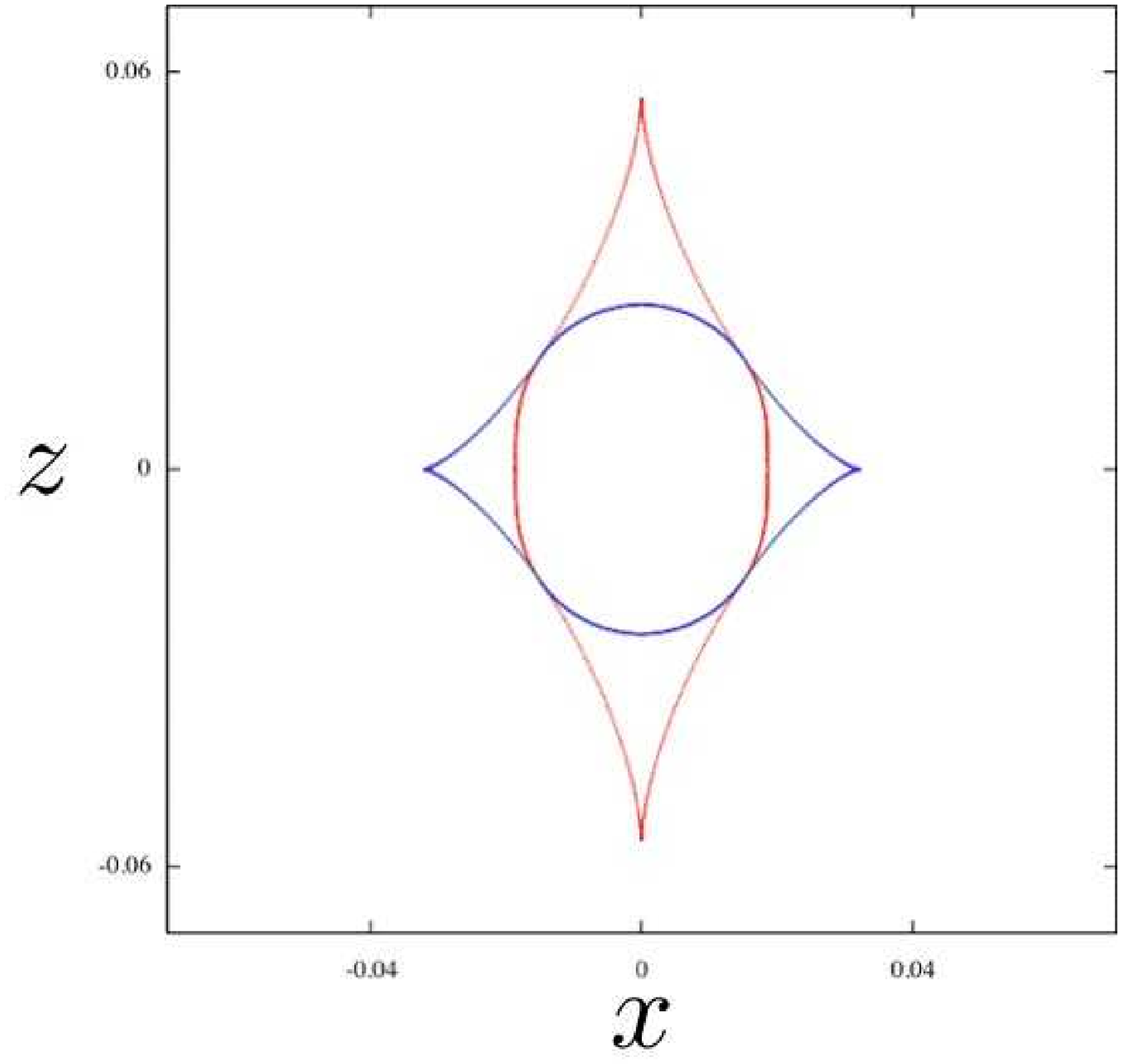}}}
\hspace{.3in}
\subfigure{\label{g1g2-m}\resizebox{1.3in}{1.3in}
{\includegraphics{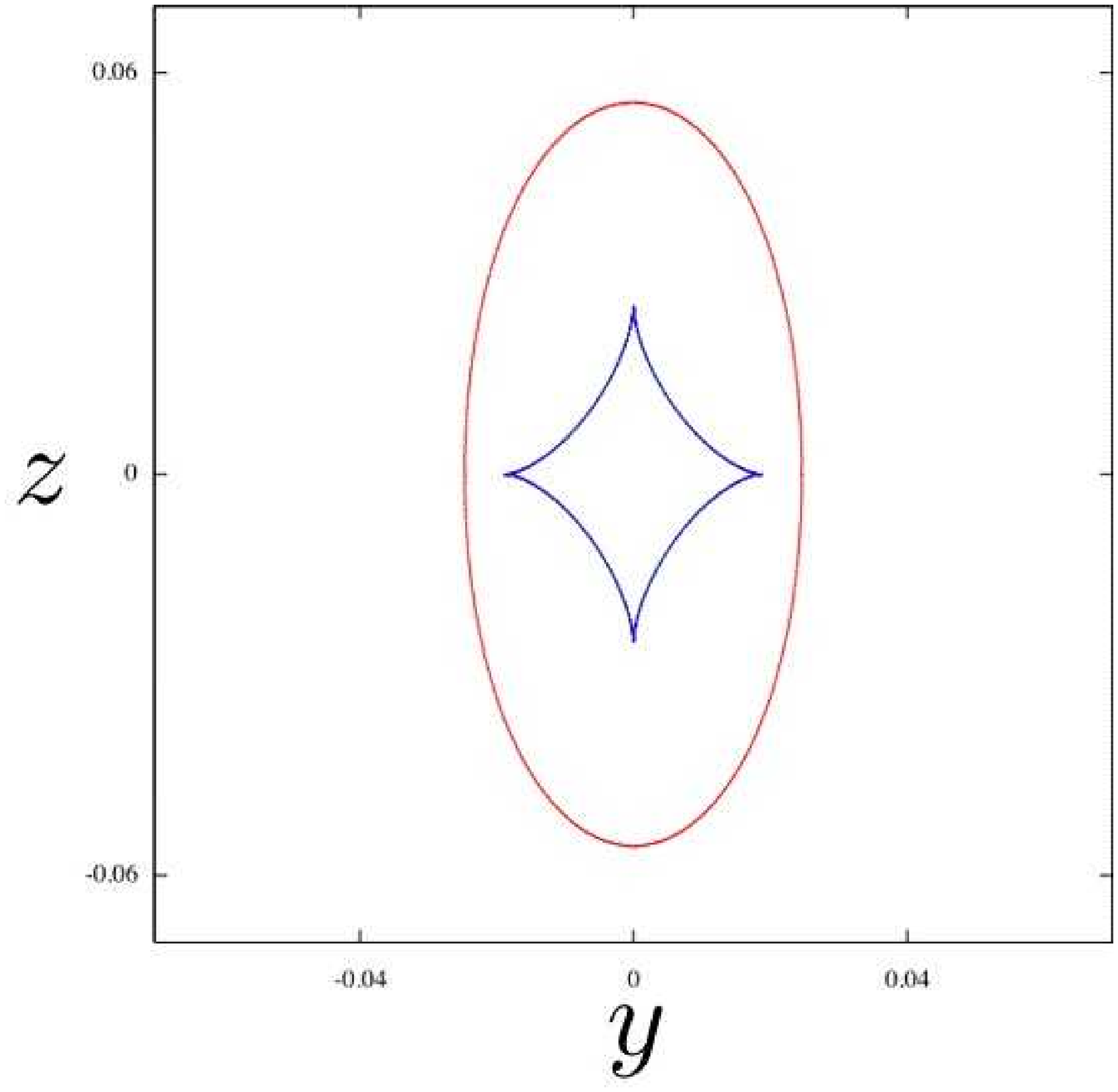}}}\\
\subfigure{\label{g1g2-d}\resizebox{1.3in}{1.3in} 
{\includegraphics{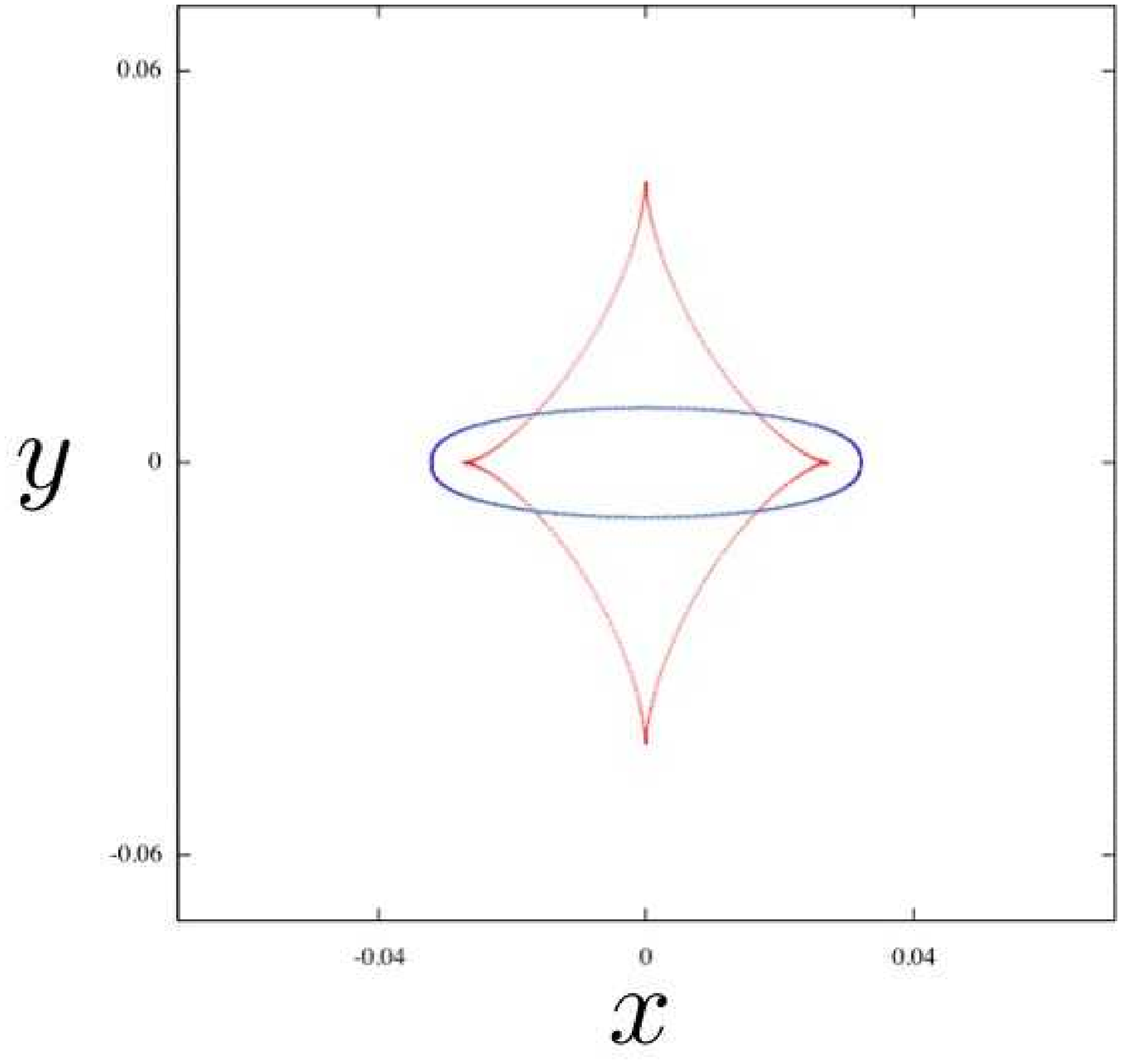}}}
\hspace{.3in}
\subfigure{\label{g1g2-i}\resizebox{1.3in}{1.3in}
{\includegraphics{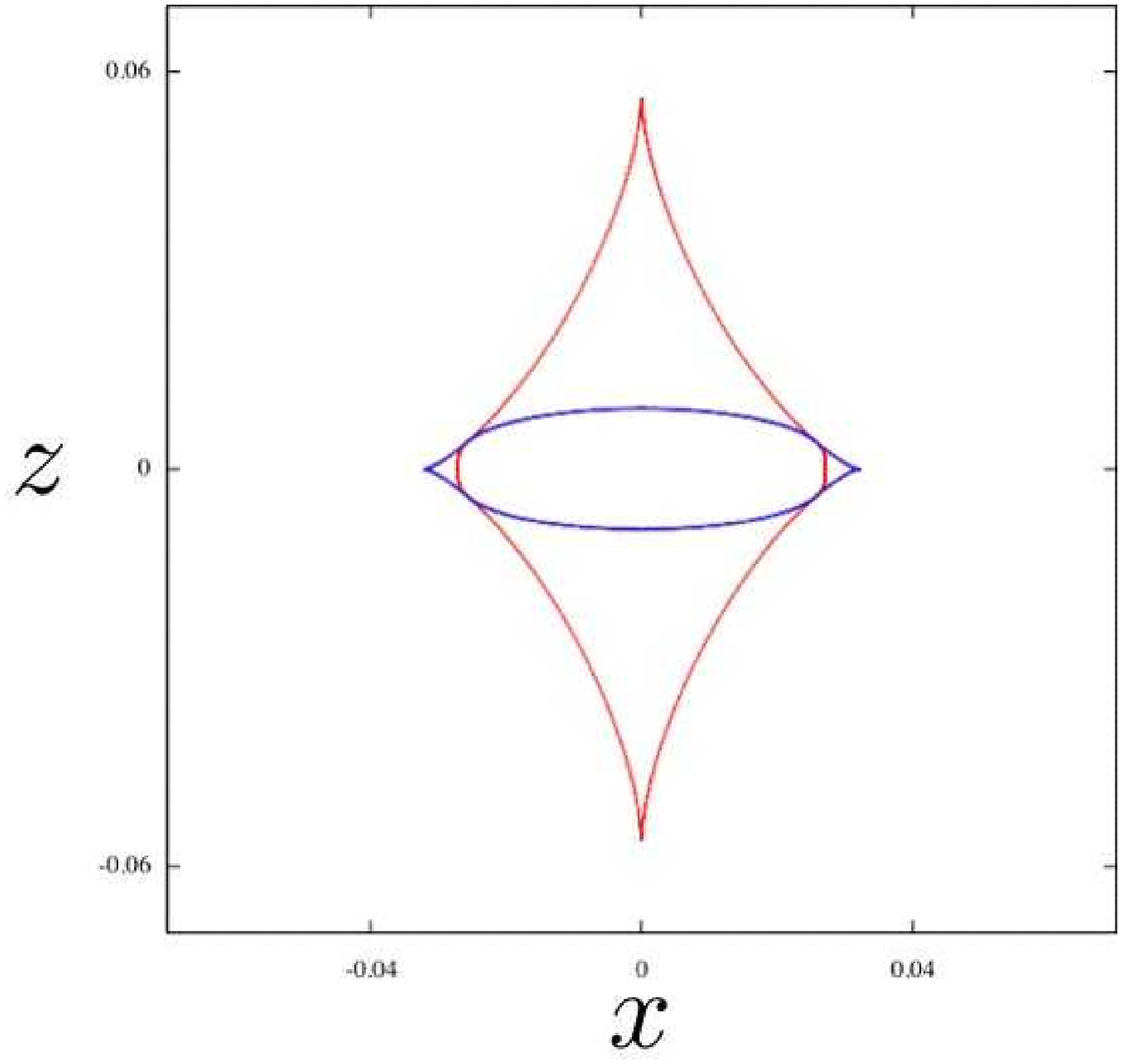}}}
\hspace{.3in}
\subfigure{\label{g1g2-n}\resizebox{1.3in}{1.3in}
{\includegraphics{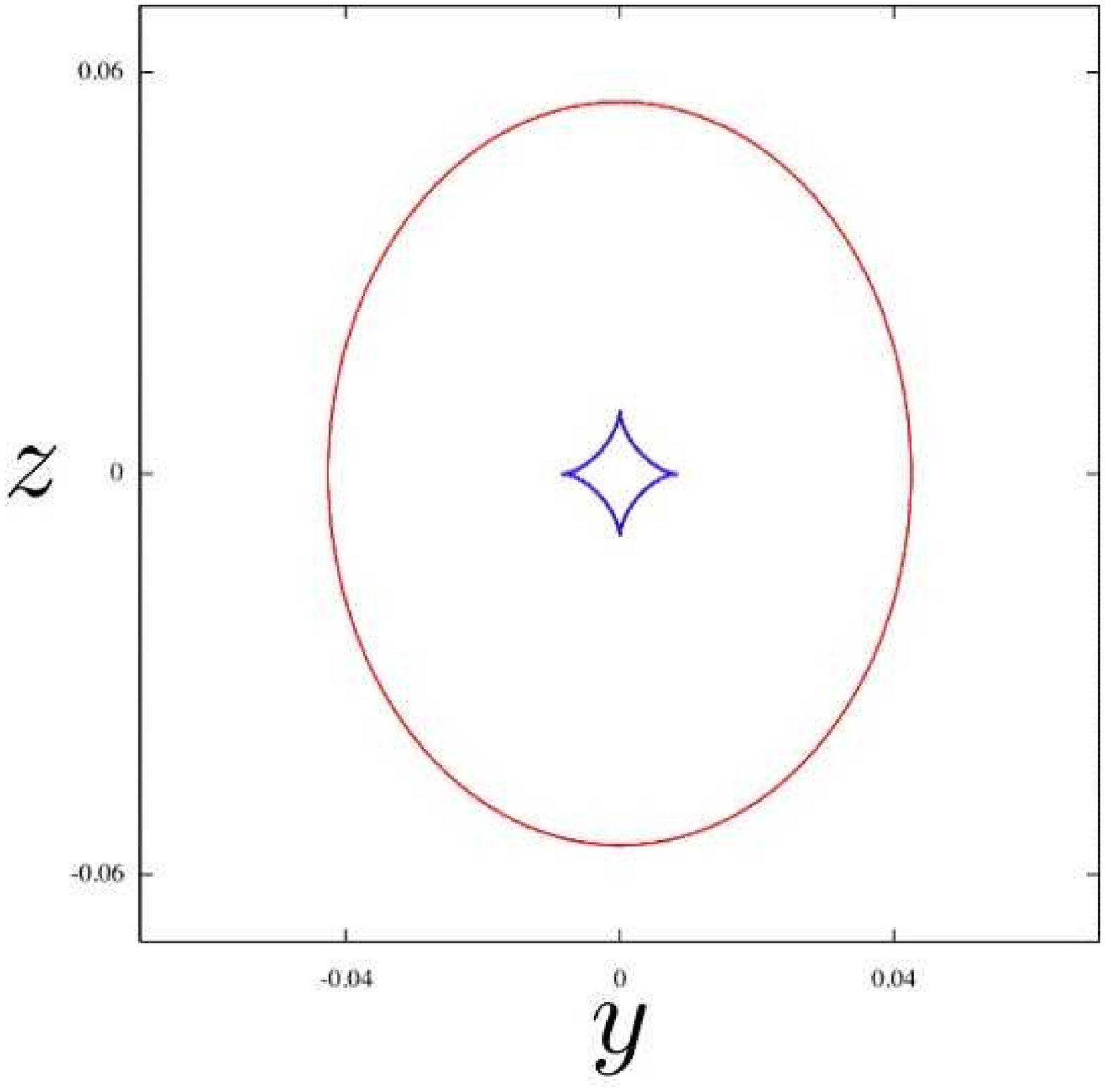}}}\\
\subfigure{\label{g1g2-e}\resizebox{1.3in}{1.3in}
{\includegraphics{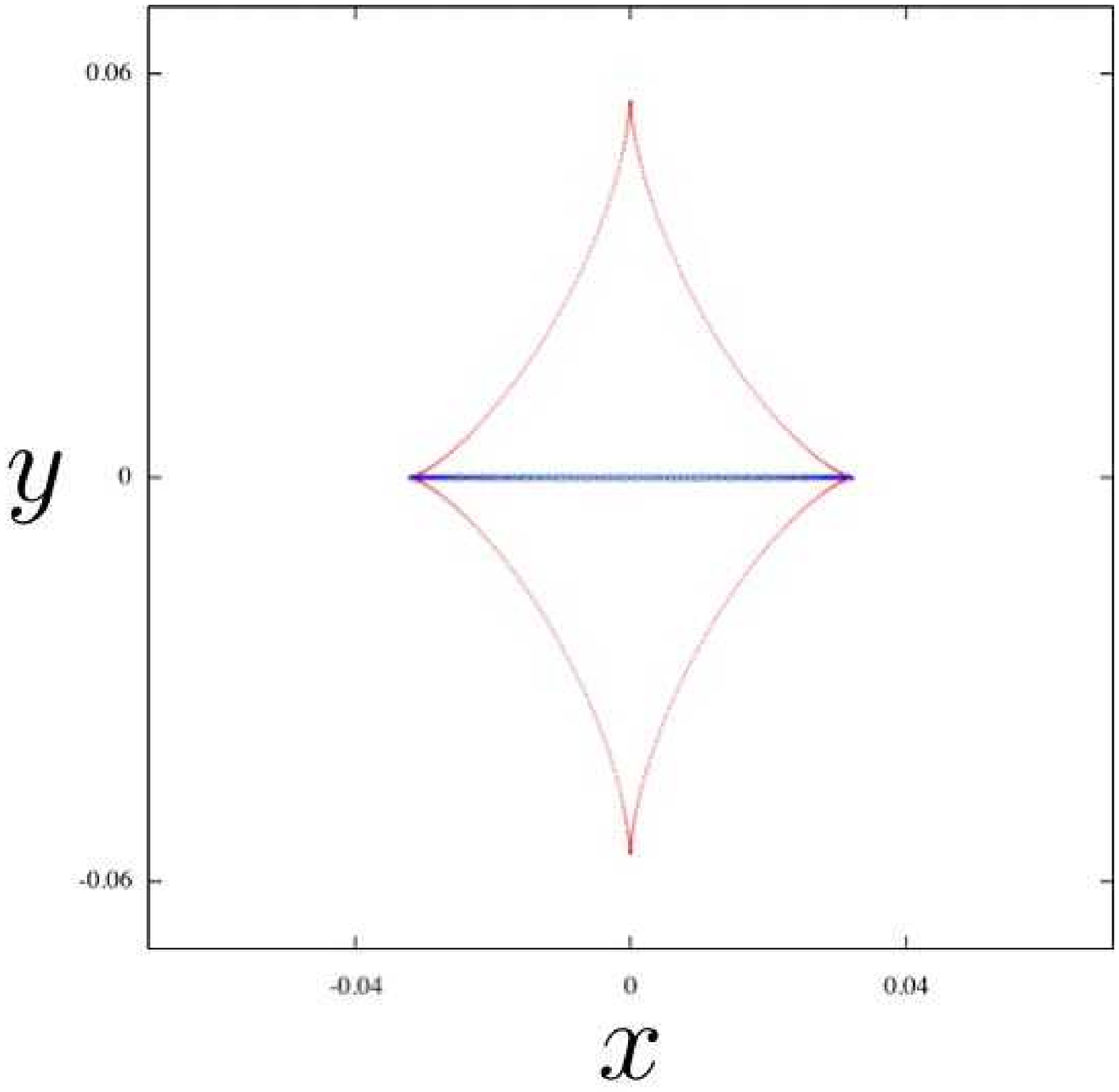}}} 
\hspace{.3in}
\subfigure{\label{g1g2-j}\resizebox{1.3in}{1.3in} 
{\includegraphics{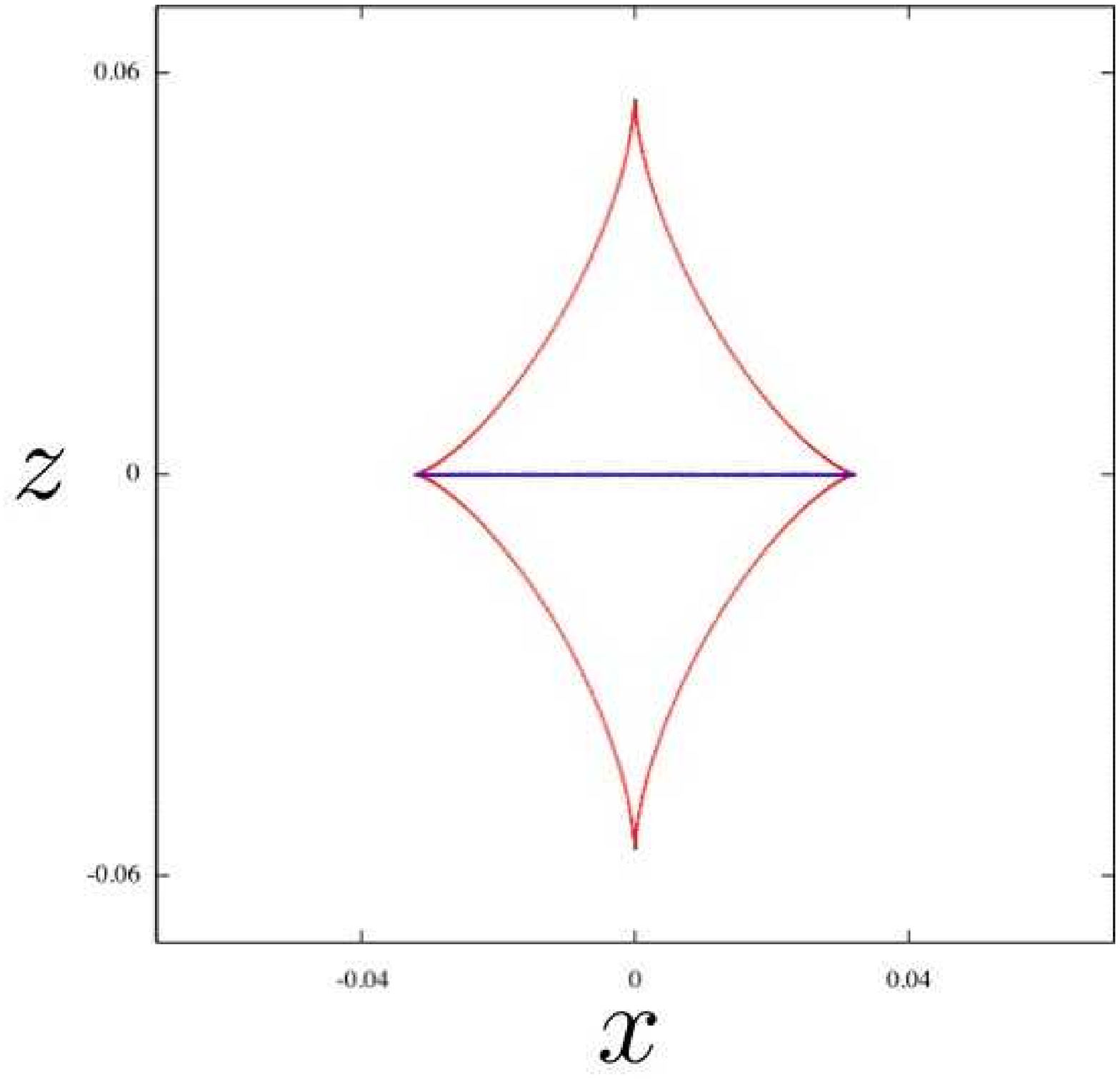}}} 
\hspace{.3in}
\subfigure{\label{g1g2-o}\resizebox{1.3in}{1.3in} 
{\includegraphics{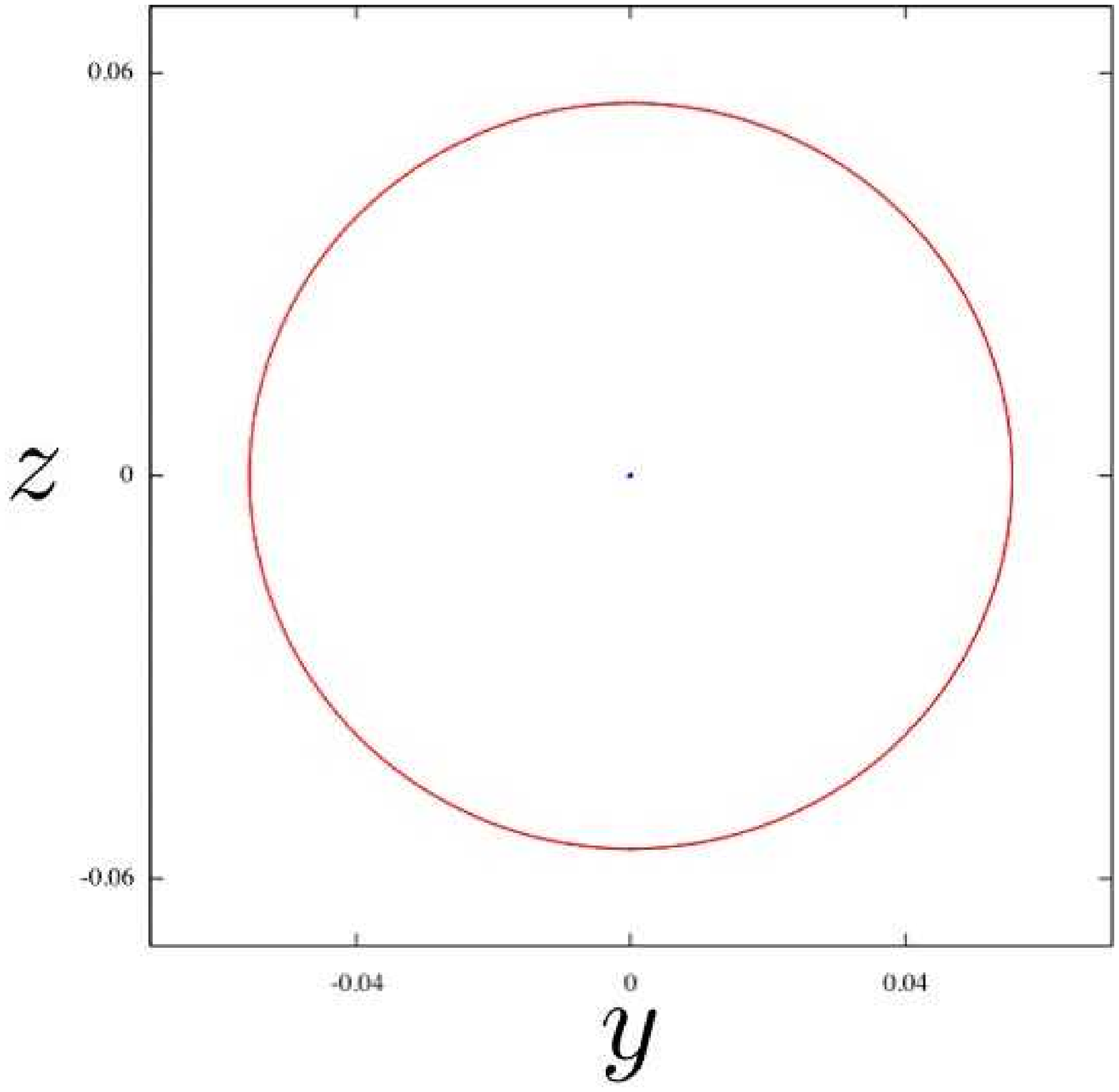}}}

\caption{The first, second and third columns show
respectively the $z = 0$, $y = 0$ and $x = 0$ 
cross sections of the inner caustics produced 
by the irrotational initial velocity field of
Eq.~(\ref{irrot}) for $(g_1, g_2)$ = (-0.033, -0.033),
(-0.04,-0.02), (-0.05, 0), (-0.06, 0.02) and 
(- 0.067, 0.033) in five rows from top to bottom.
During this sequence, the caustic transforms from 
a tent symmetric about the $z$ axis (first row) to 
a tent symmetric about the $x$ axis (last row).  
The roof of the caustic in the first row becomes 
the pole of the caustic in the last row (blue) 
and vice-versa (red).
\label{g1g2}} 
\end{figure}

When $g_2$ is different from but close to $g_1$, axial symmetry 
is slightly broken.  In that case the caustic pole spreads onto a 
tube whose cross section has four cusps forming a diamond shape.
See Figs.~\ref{zp} and \ref{ttt}.  In a region where the tent 
pole connects with the tent roof, two hyperbolic umbilic 
catastrophes appear, as described in Fig.~\ref{hyp_umb}.  
Fig.~\ref{g1g2} shows what happens when $g_2$ is very different 
from $g_1$.  The inner caustic metamorphoses from a tent which is 
symmetric about the $z$ axis when $g_2 = g_1$ to a tent symmetric 
about the $x$ axis when $g_2 = - {1 \over 2} g_1$.  

\begin{figure}

\subfigure[]{\label{tr_axial-a}\resizebox{2.7in}{2.7in}
{\includegraphics{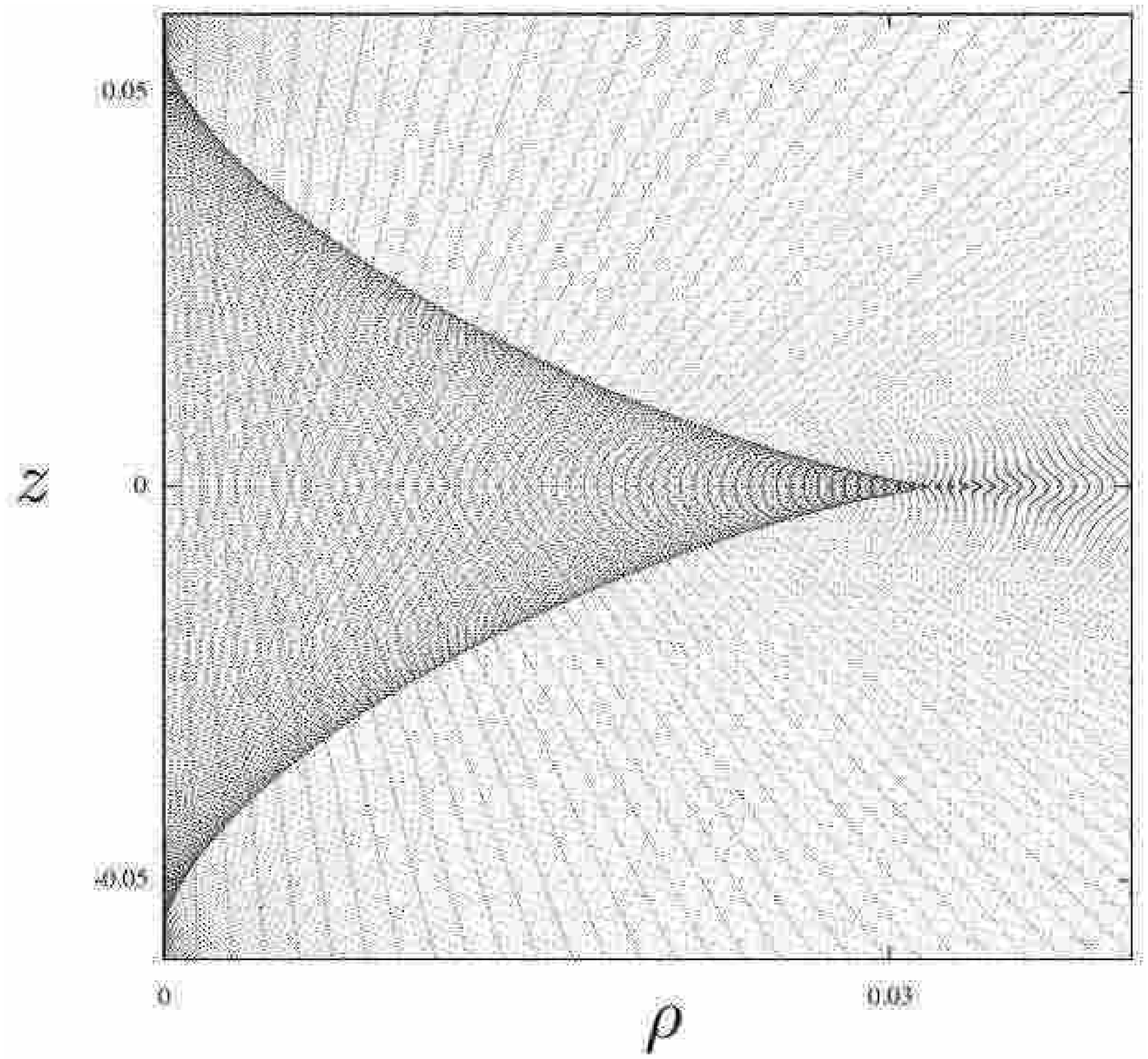}}}
\hspace{.3in}
\subfigure[]{\label{tr_axial-b}\resizebox{2.7in}{2.7in}
{\includegraphics{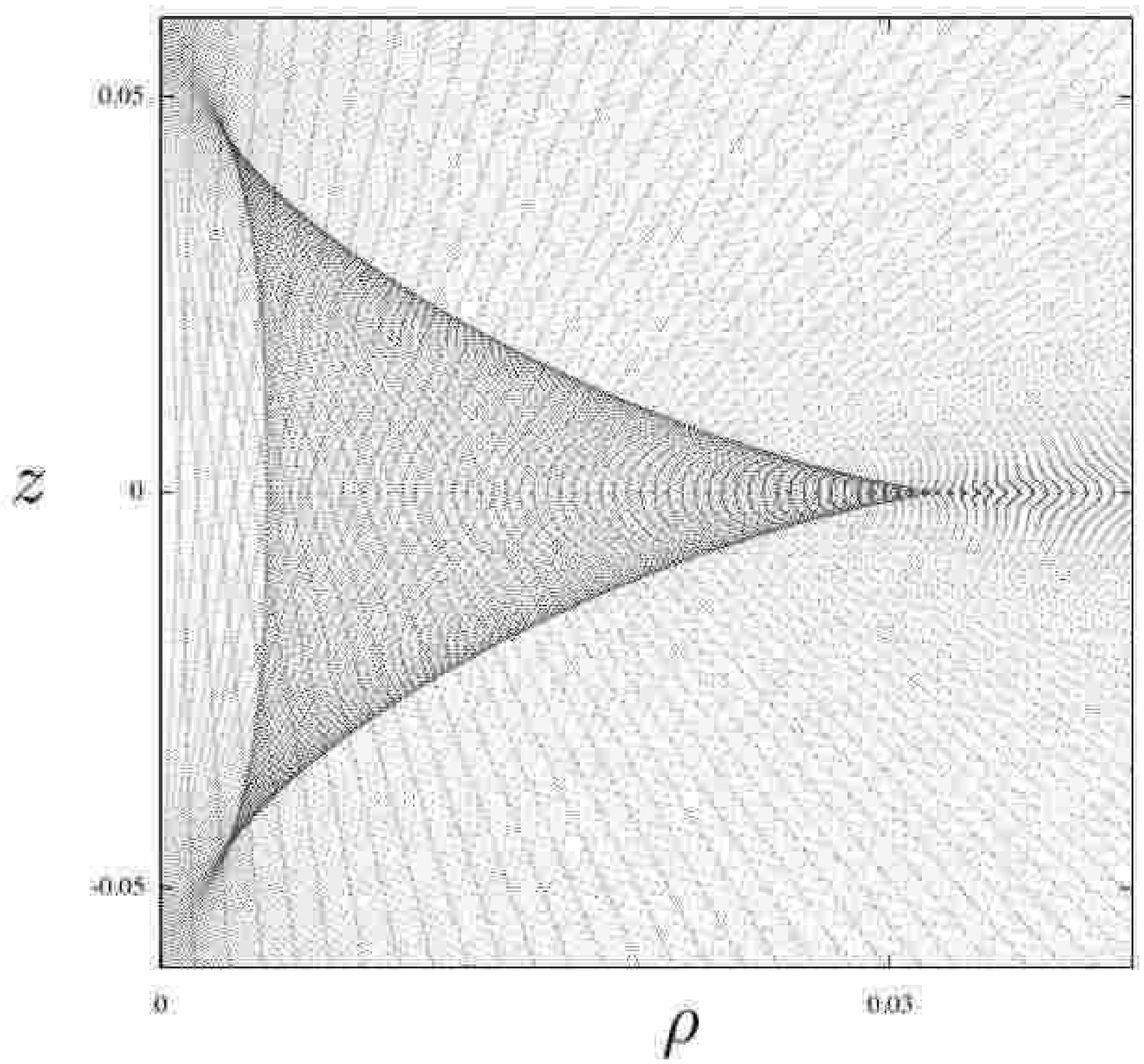}}}\\
\subfigure[]{\label{tr_axial-c}\resizebox{2.7in}{2.7in}
{\includegraphics{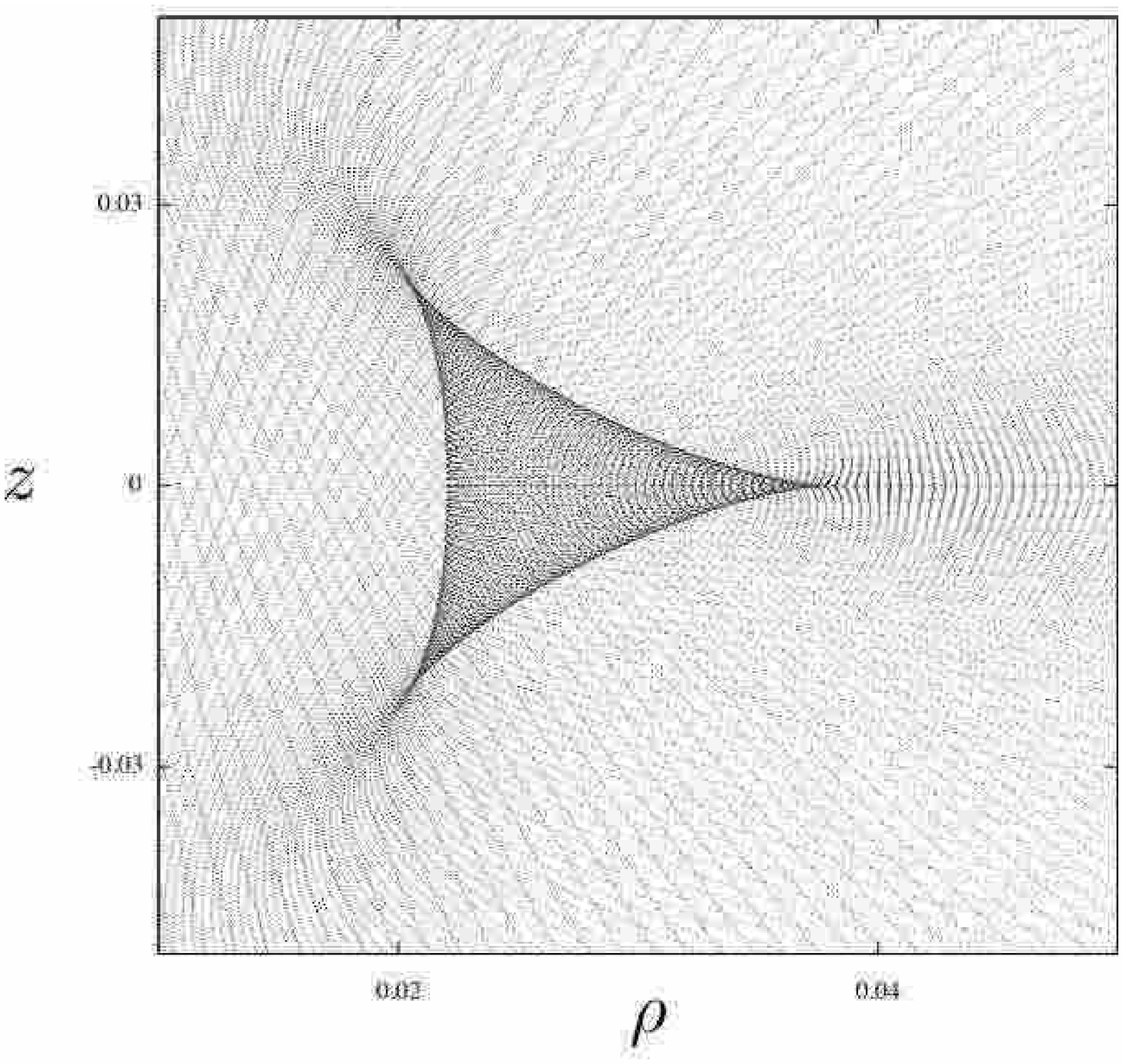}}}
\hspace{.3in}
\subfigure[]{\label{tr_axial-d}\resizebox{2.7in}{2.7in}
{\includegraphics{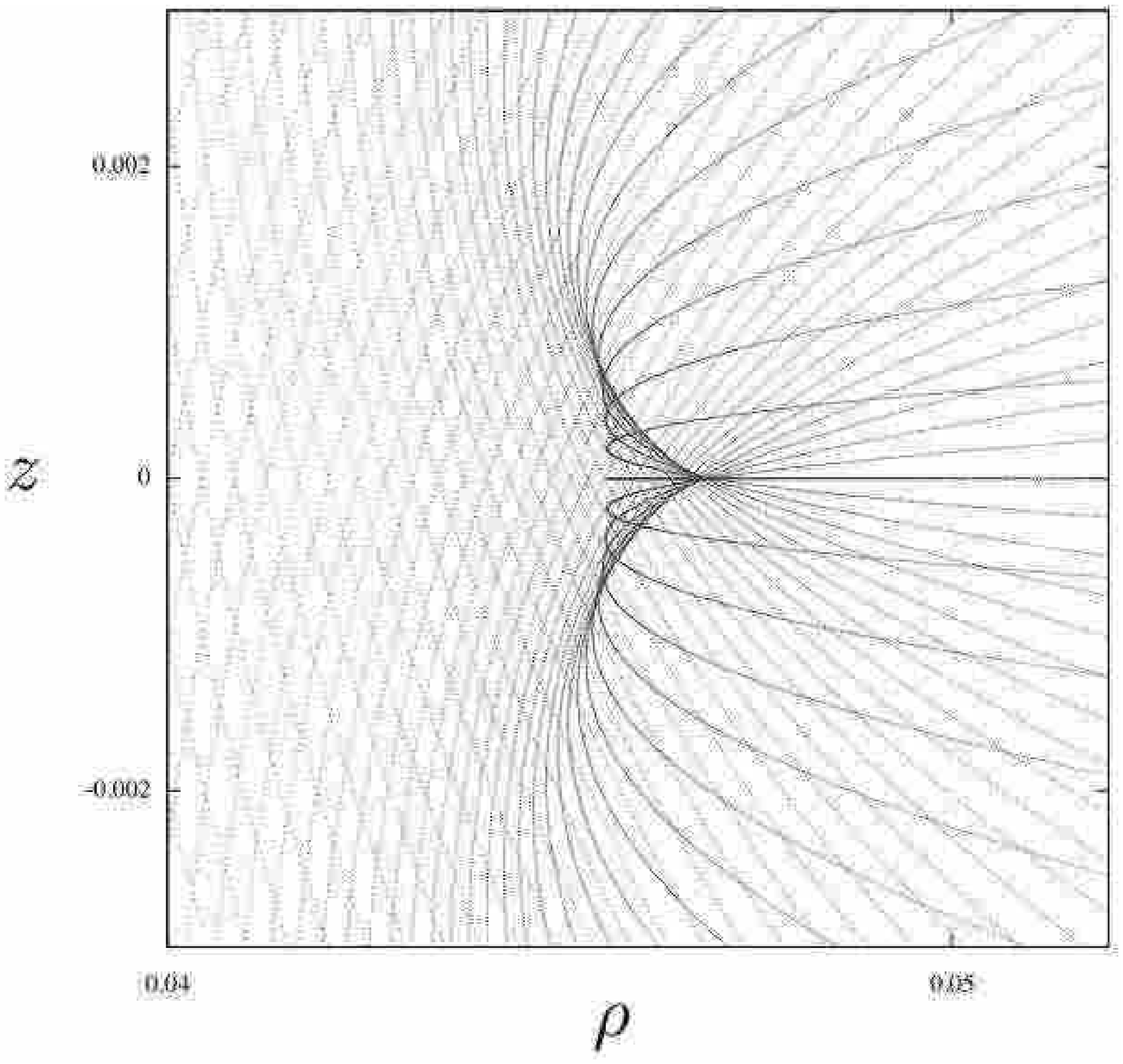}}}

\caption{Cross sections of the inner caustics produced
by the axially symmetric initial velocity field of 
Eq.~(\ref{axsym}) with $g_1 =  - 0.033$, and (a) $c_1 = 0$, 
(b) $c_2 = 0.01$, (c) $c_3 = 0.05$, (d) $c_3 = 0.1$. Increasing 
the rotational component of the initial velocity field causes
the tent caustic (a) to transform into a tricusp ring (d).
\label{tr_axial}}
\end{figure}

\begin{figure}

\subfigure[]{\label{tr_non-a}\resizebox{2.1in}{2.1in}
{\includegraphics{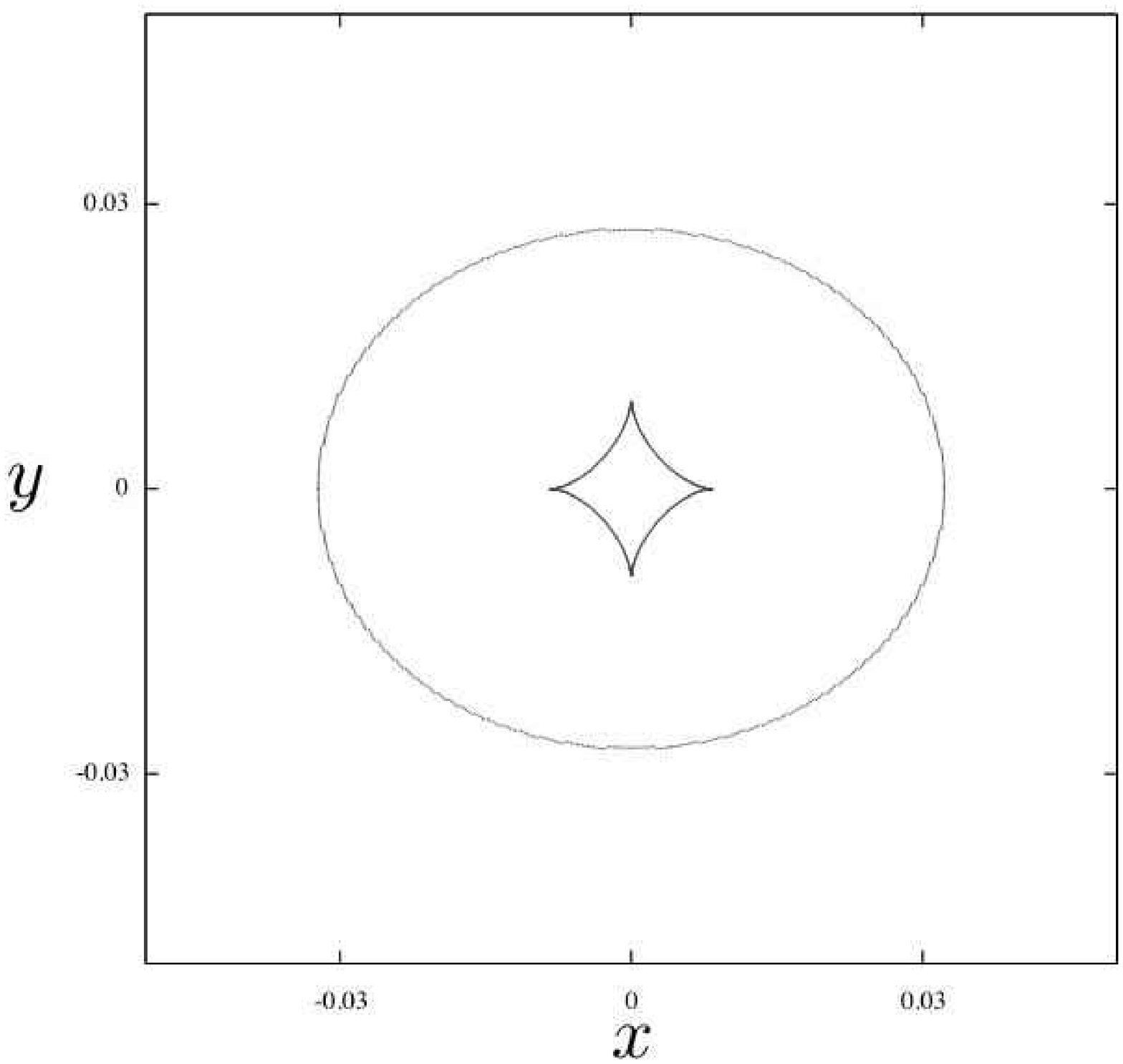}}}
\hspace{.5in}
\subfigure[]{\label{tr_non-b}\resizebox{2.1in}{2.1in}
{\includegraphics{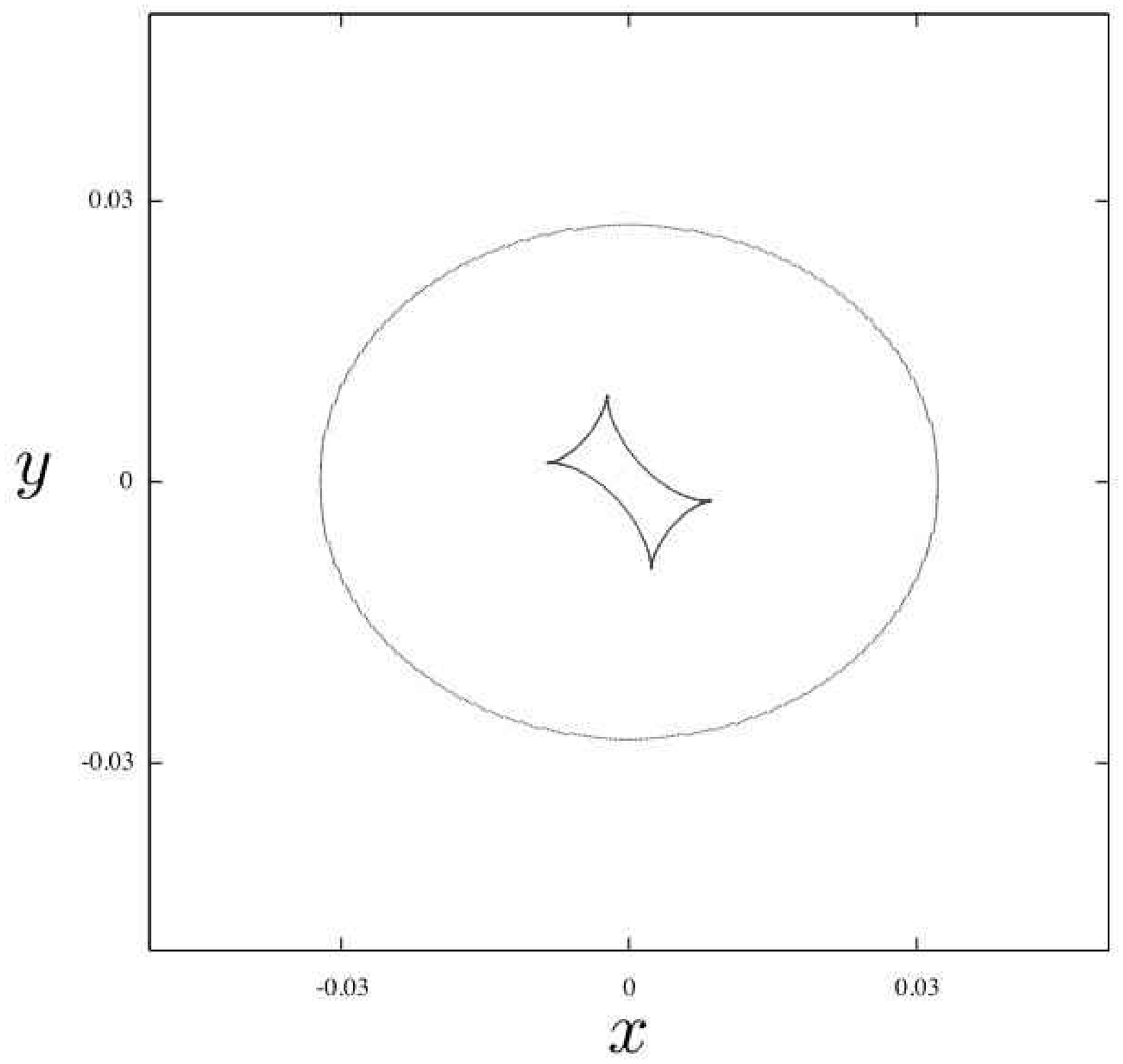}}}\\
\subfigure[]{\label{tr_non-c}\resizebox{2.1in}{2.1in}
{\includegraphics{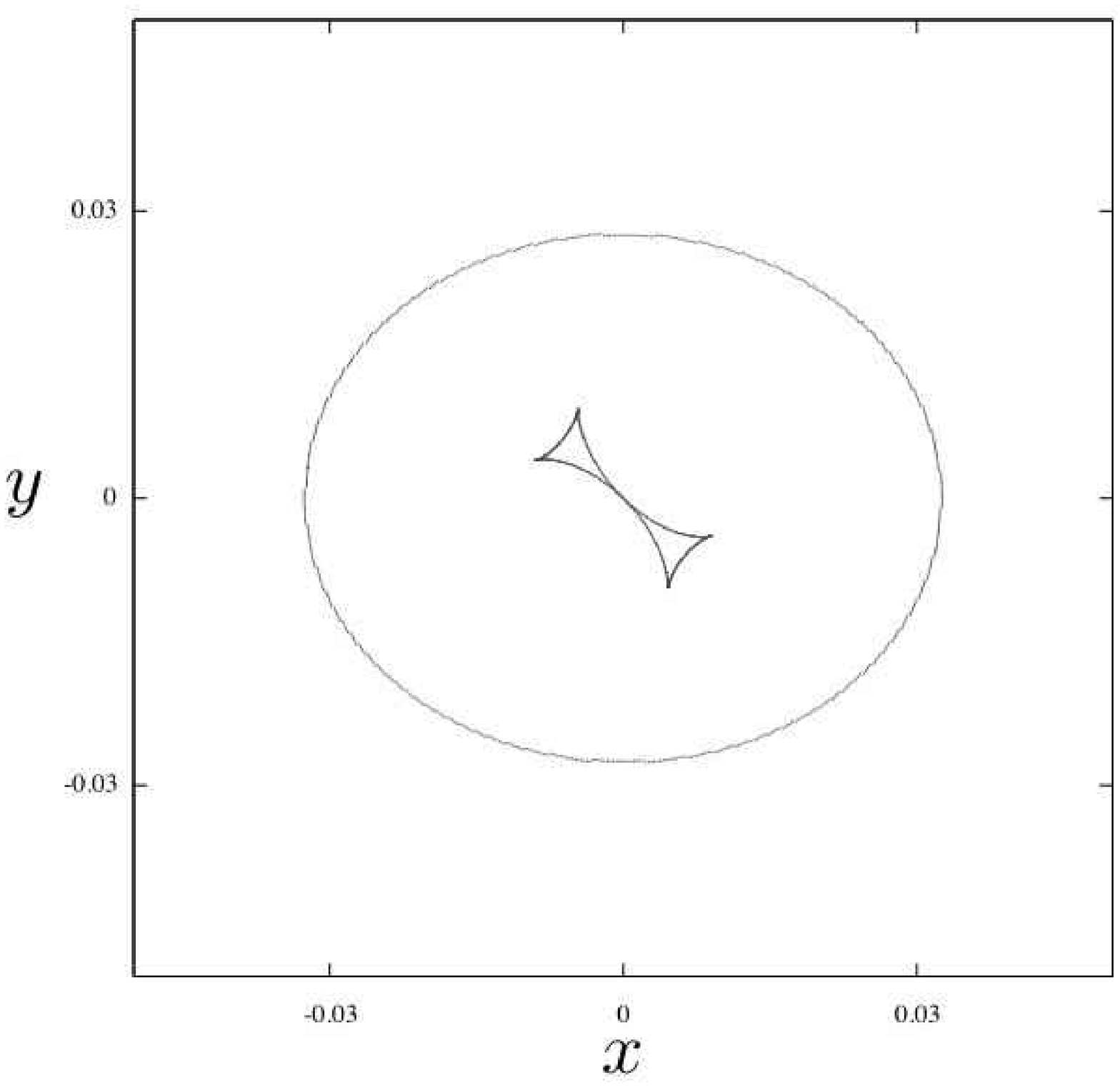}}}
\hspace{.5in}
\subfigure[]{\label{tr_non-d}\resizebox{2.1in}{2.1in}
{\includegraphics{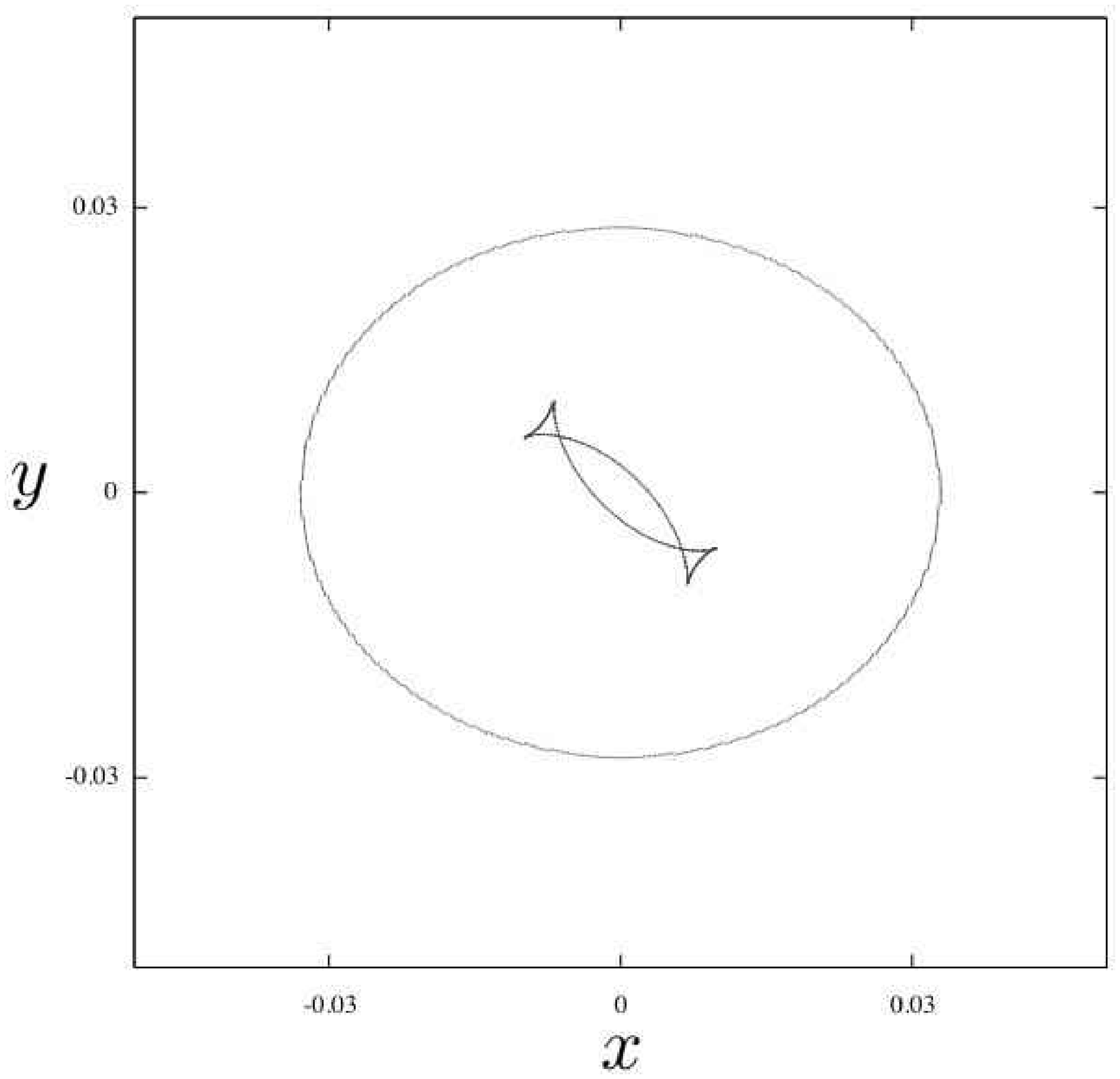}}}\\
\subfigure[]{\label{tr_non-e}\resizebox{2.1in}{2.1in}
{\includegraphics{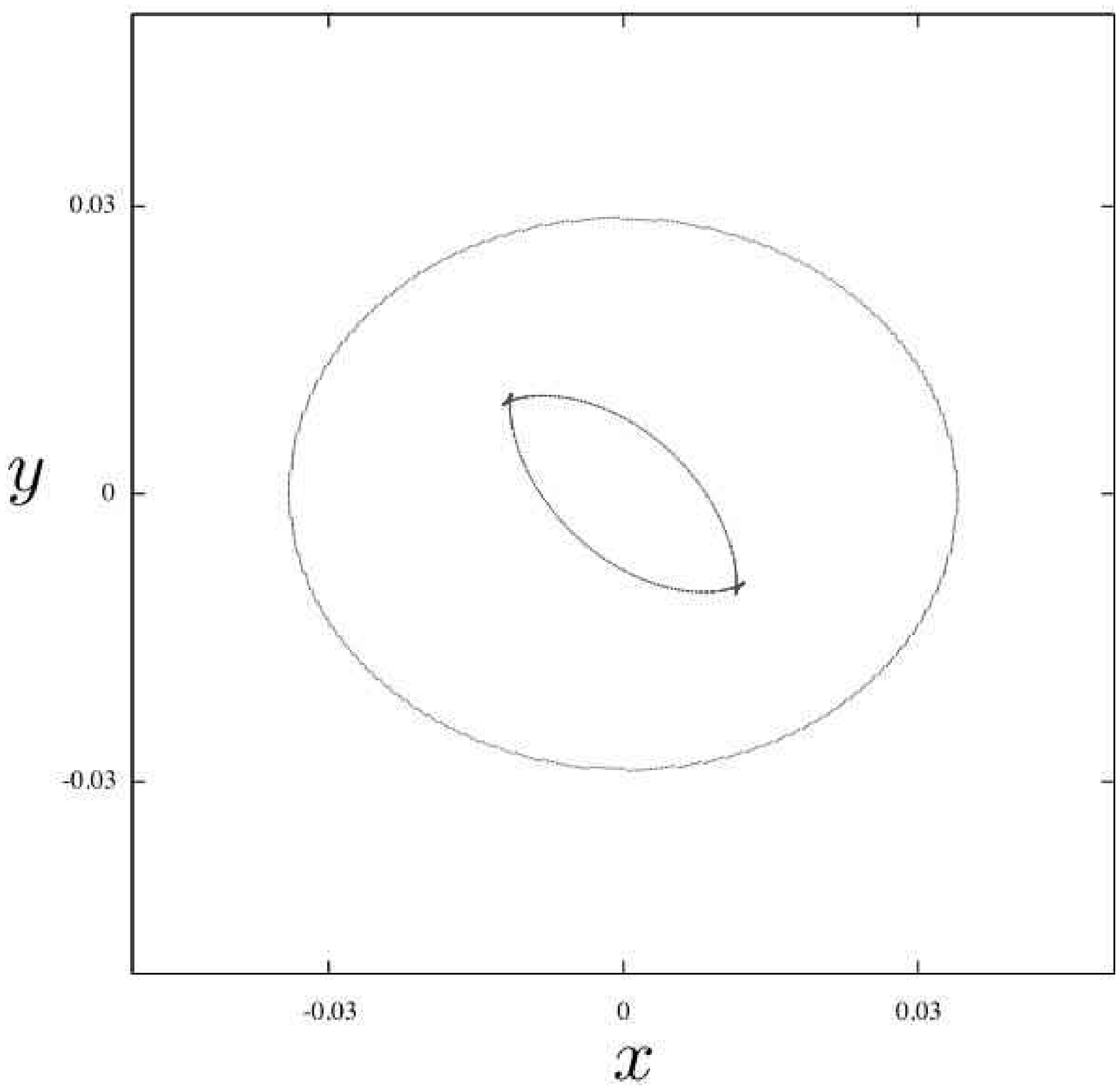}}}
\hspace{.5in}
\subfigure[]{\label{tr_non-f}\resizebox{2.1in}{2.1in}
{\includegraphics{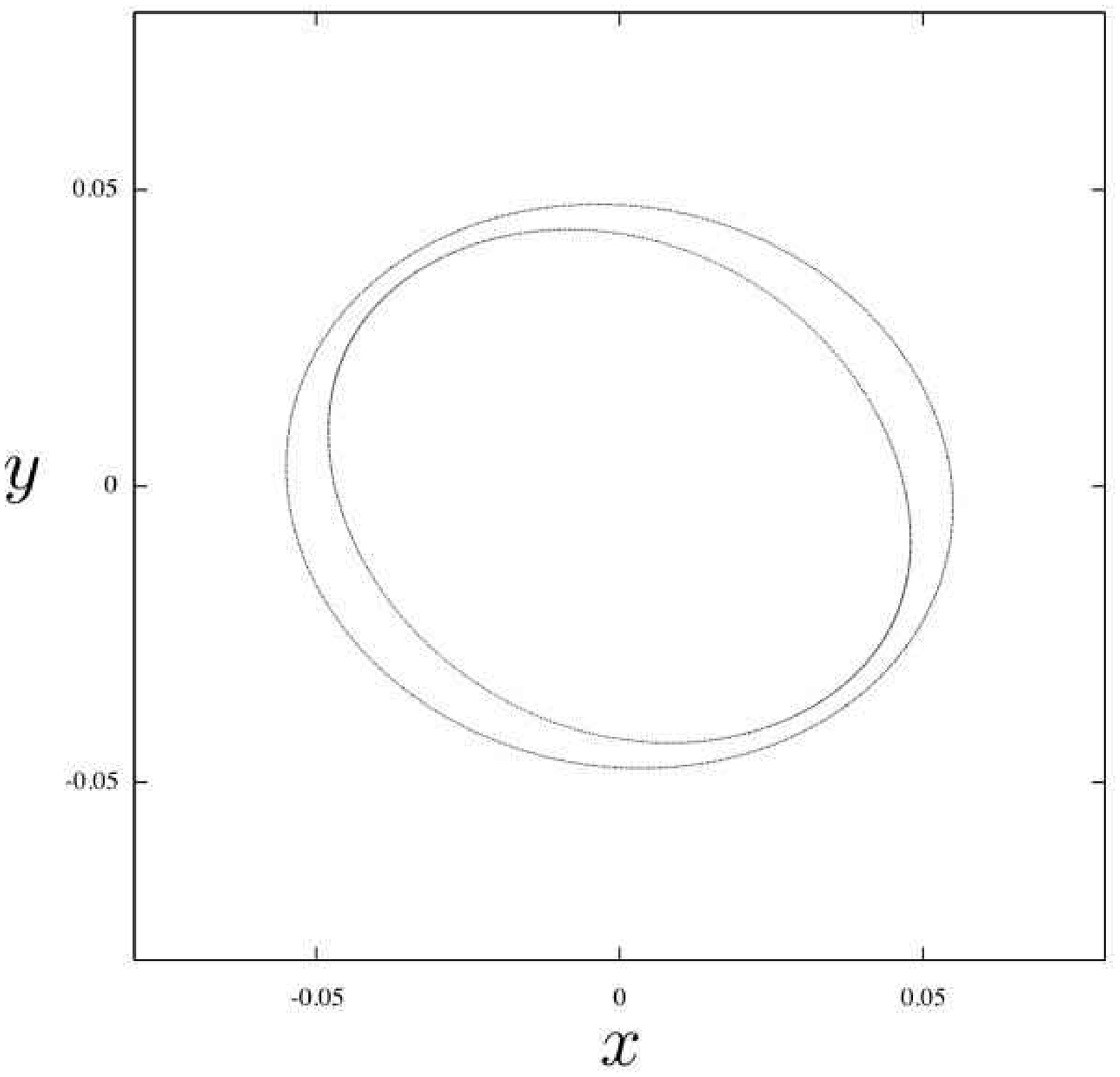}}}

\caption{$z=0$ sections of the inner caustics produced 
by the initial velocity field of Eqs.~(\ref{vin}) for 
$c_1 = c_2 = 0$, $\xi=0.01$, $g = -0.05$, and increasing 
values of $c_3$: (a) $c_3 = 0$, (b) $c_3 = 0.005 $, 
(c) $c_3 = 0.01$, (d) $c_3 = 0.015$, (e) $c_3 = 0.025$, 
(f) $c_3 = 0.1$.\label{tr_non}}
\end{figure}

\begin{figure}

\subfigure[]{\label{tr_non4-a}\resizebox{2.8in}{2.8in}
{\includegraphics{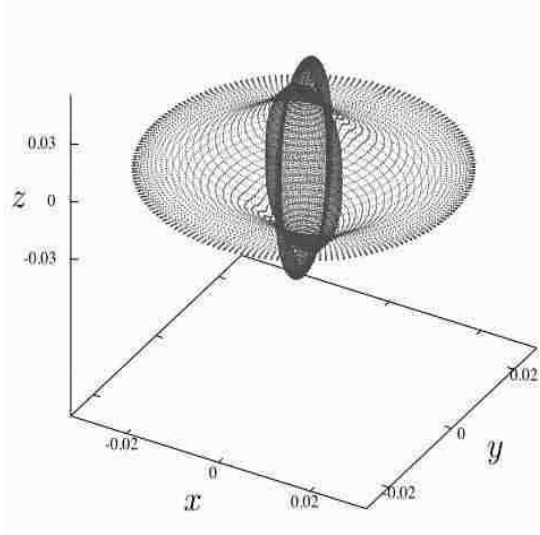}}}
\subfigure[]{\label{tr_non4-b}\resizebox{2.8in}{2.8in}
{\includegraphics{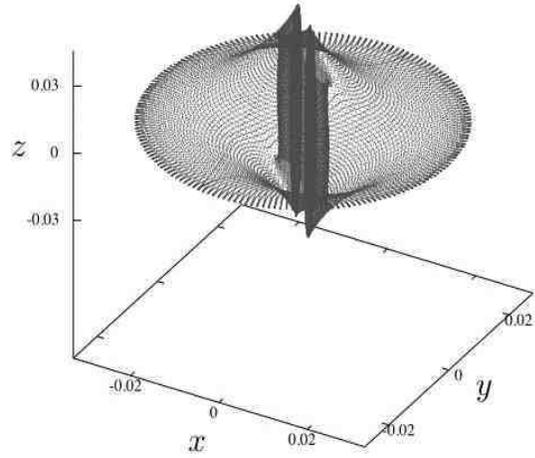}}}
\subfigure[]{\label{tr_non4-c}\resizebox{2.8in}{2.8in}
{\includegraphics{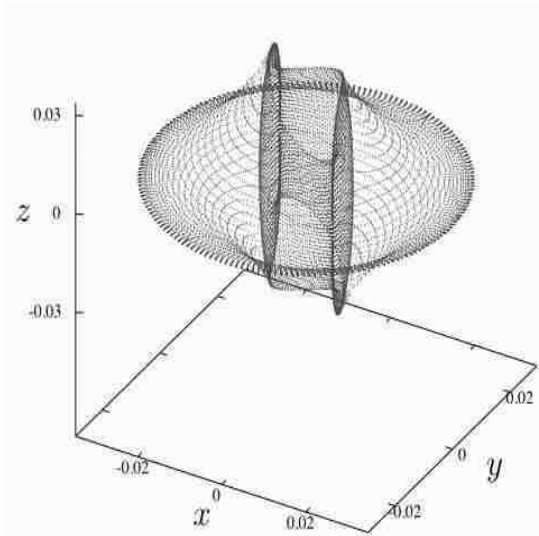}}}
\subfigure[]{\label{tr_non4-d}\resizebox{2.8in}{2.8in}
{\includegraphics{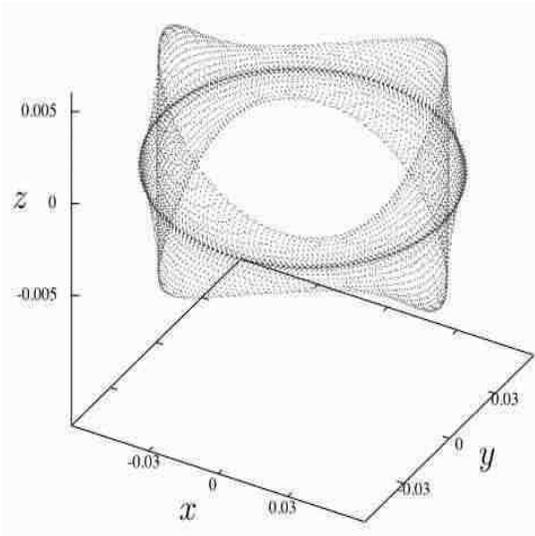}}}

\caption{Inner caustics produced by the initial 
velocity field of Eqs.~(\ref{vin}) for $\xi=0.01$,
$g=-0.05$, $c_1 = c_2 = 0$ and increasing values of 
$c_3$: (a) $c_3=0$, (b) $c_3=0.015$, (c) $c_3=0.03$, 
(d) $c_3=0.12$.
\label{tr_non4}}
\end{figure}

Finally, we investigated the smooth transformation of the tent 
caustic into a tricusp ring when a rotational component is 
added ($\vec{c} \neq 0$).  In the axially symmetric case (see
Fig.~\ref{tr_axial}) the tent pole spreads onto a tube of 
circular cross section.  The radius of this tube becomes the 
inner radius of the tricusp ring.  So, in the axially symmetric 
case, we may think of the tent caustic as a tricusp ring whose 
inner radius has shrunk to zero.  In the non axially symmetric 
case, the metamorphosis of tent caustic to tricusp ring is shown 
in Figs.~\ref{tr_non} and ~\ref{tr_non4}.  The diamond shaped 
cross section of the tent pole transforms to a (topological)
circle by passing through an intermediate stage where it 
consists of two swallowtail catastrophes back to back.

\acknowledgments{This work was supported in part by the U.S. 
Department of Energy under grant DE-FG02-97ER41029, by an IBM 
Einstein Endowed Fellowship at the Institute for Advanced Study, 
and a J.Michael Harris graduate student award at the University 
of Florida.  A.N. thanks Jim Fry and Karthik Shankar for helpful 
discussions.  P.S. gratefully acknowledges the Aspen Center of 
Physics for its hospitality while he was working on this project.}

\vspace{2in}

\begin{figure}

\subfigure[]{\label{inf-a}\includegraphics[width=2.0in,height=2.0in]
{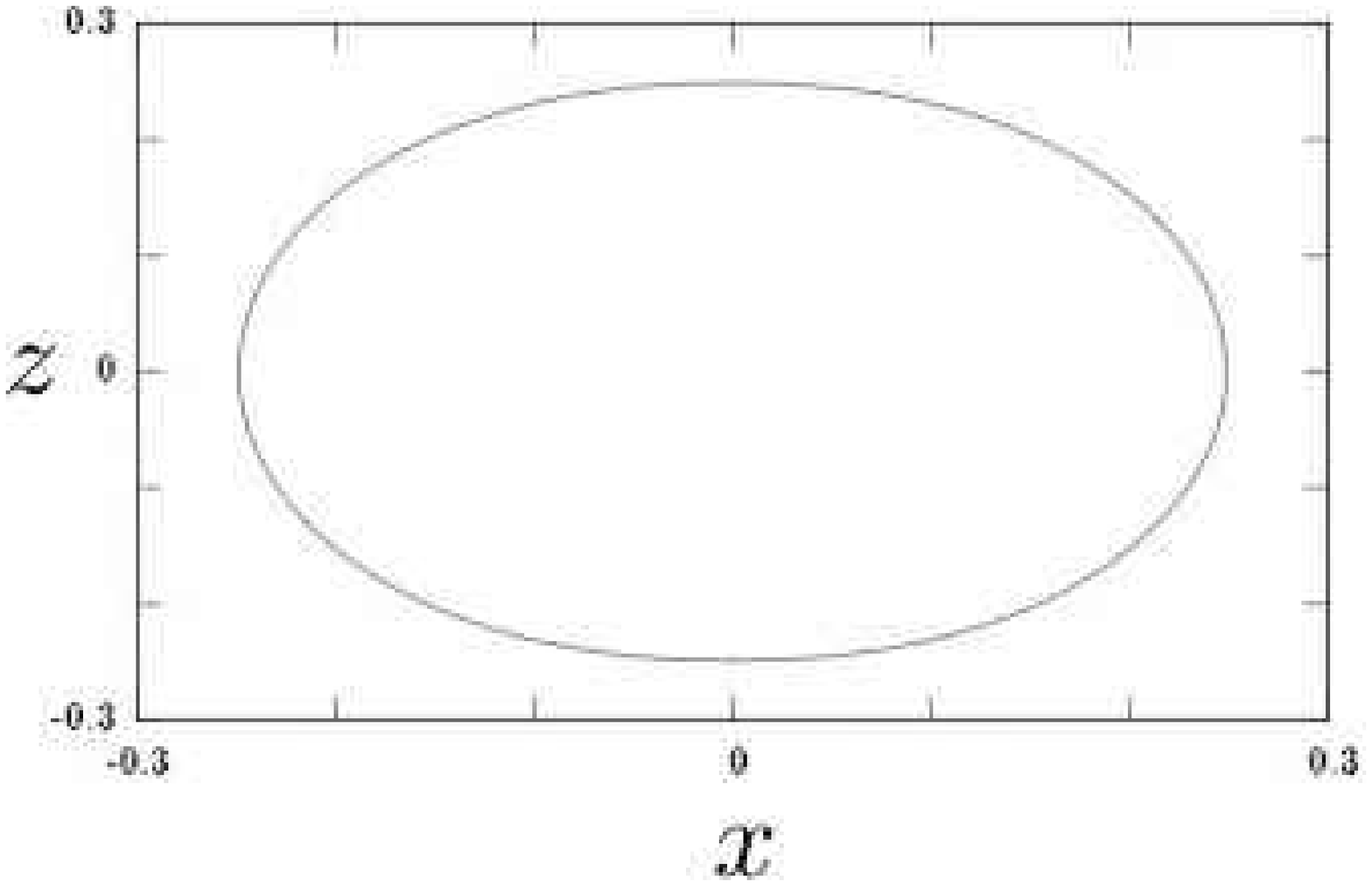}}
\subfigure[]{\label{inf-b}\includegraphics[width=2.0in,height=2.0in]
{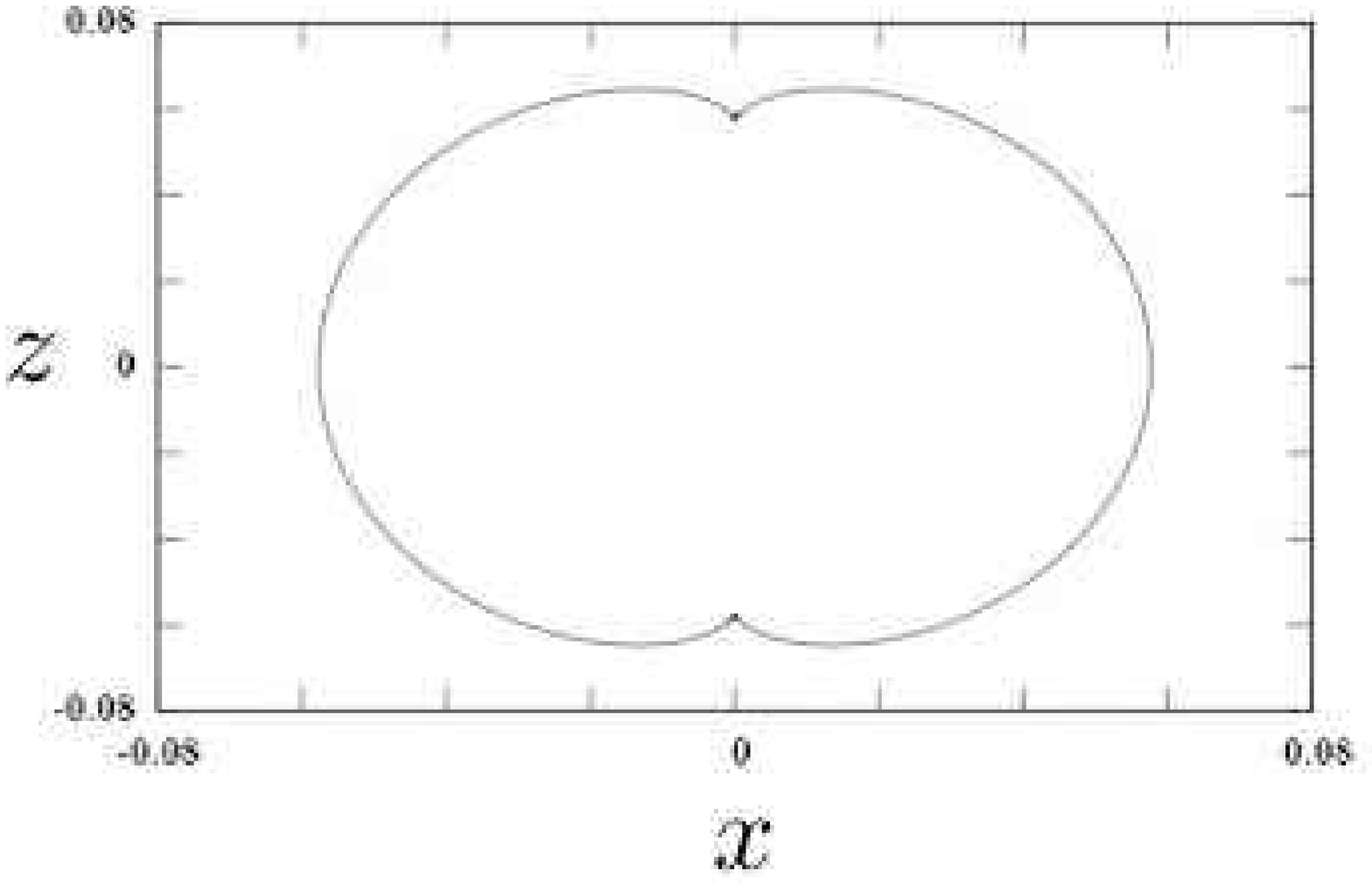}}
\subfigure[]{\label{inf-c}\includegraphics[width=2.0in,height=2.0in]
{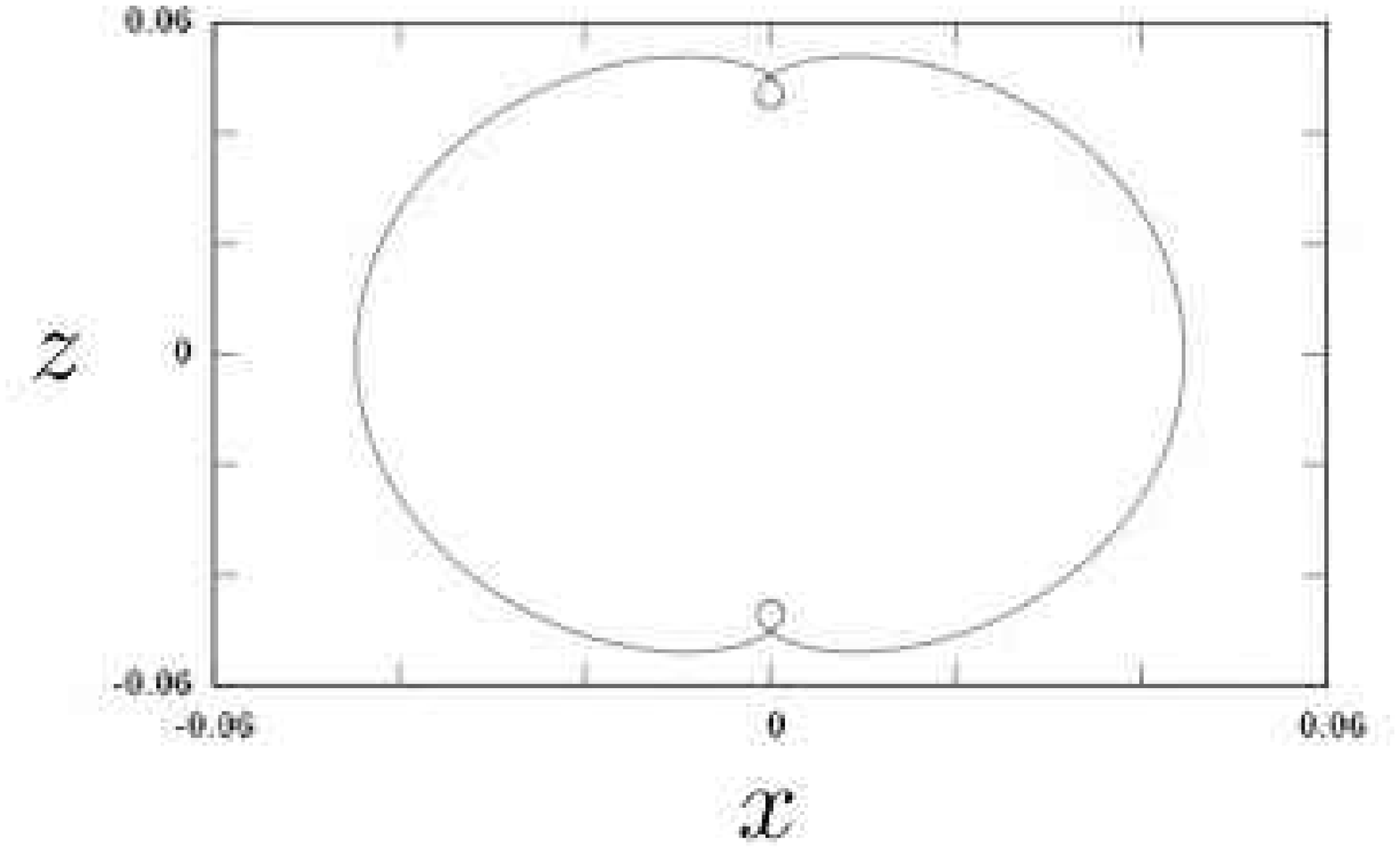}}\\
\subfigure[]{\label{inf-d}\includegraphics[width=2.0in,height=2.0in]
{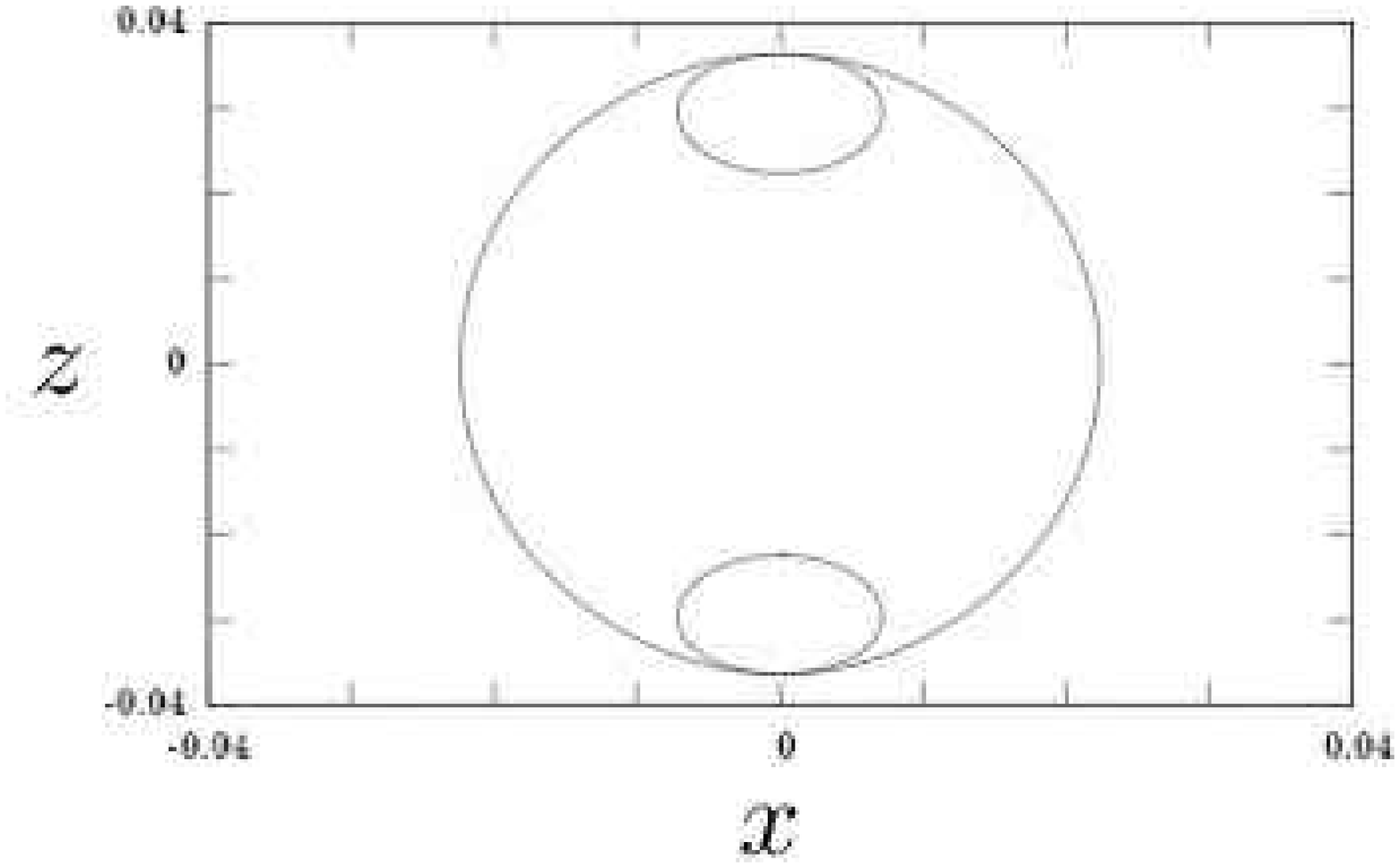}}
\subfigure[]{\label{inf-e}\includegraphics[width=2.0in,height=2.0in]
{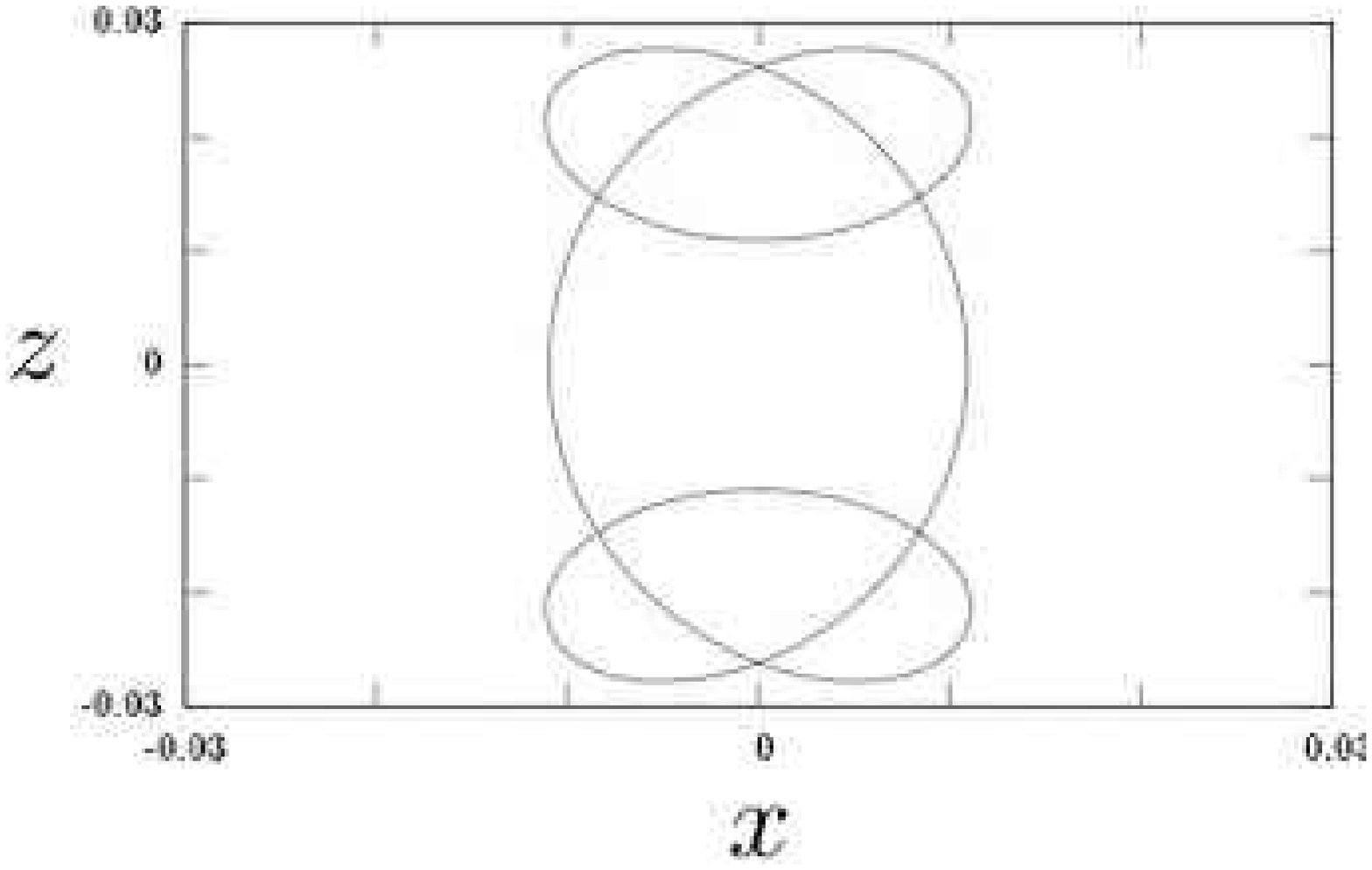}}
\subfigure[]{\label{inf-f}\includegraphics[width=2.0in,height=2.0in]
{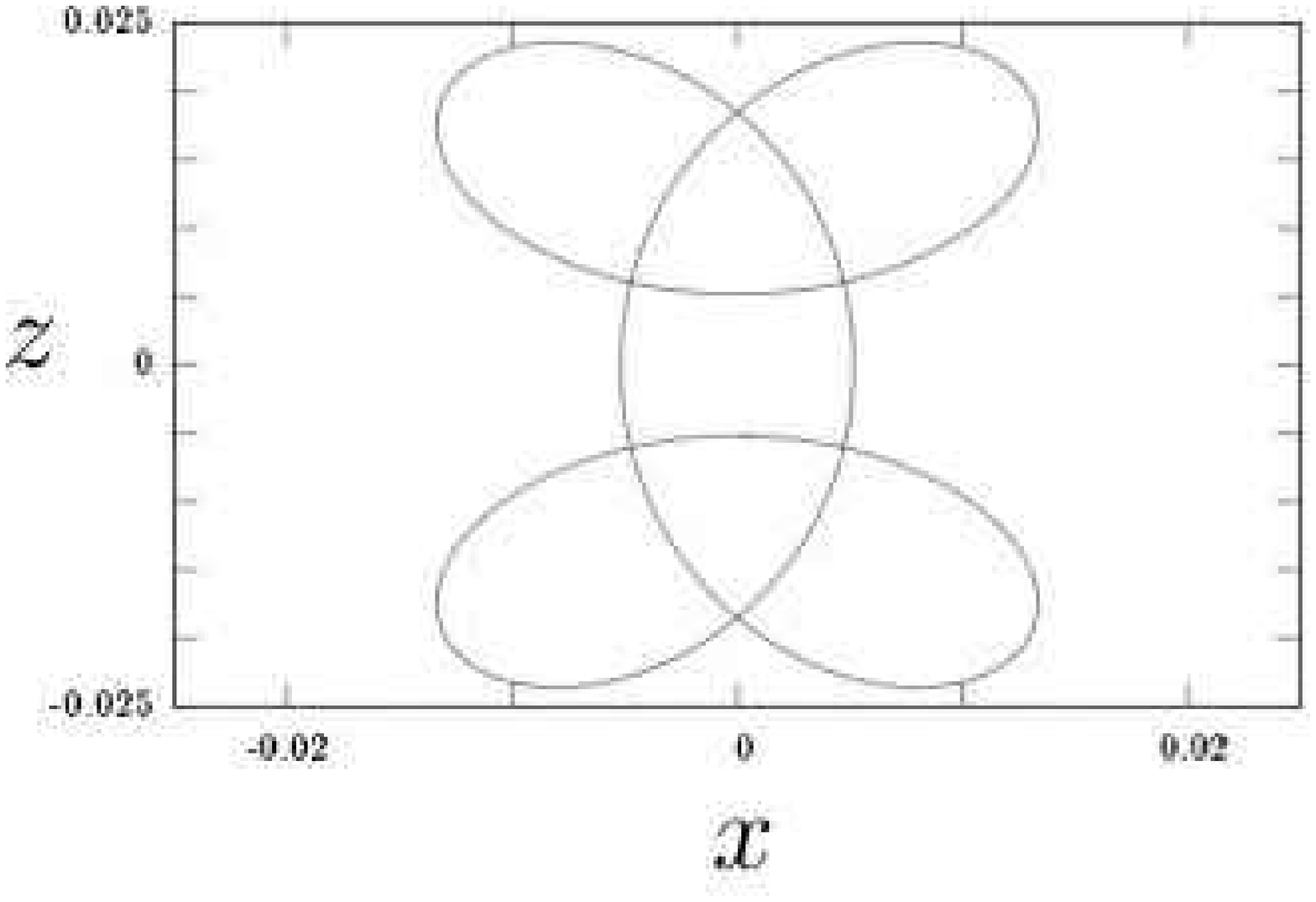}}\\
\subfigure[]{\label{inf-g}\includegraphics[width=2.0in,height=2.0in]
{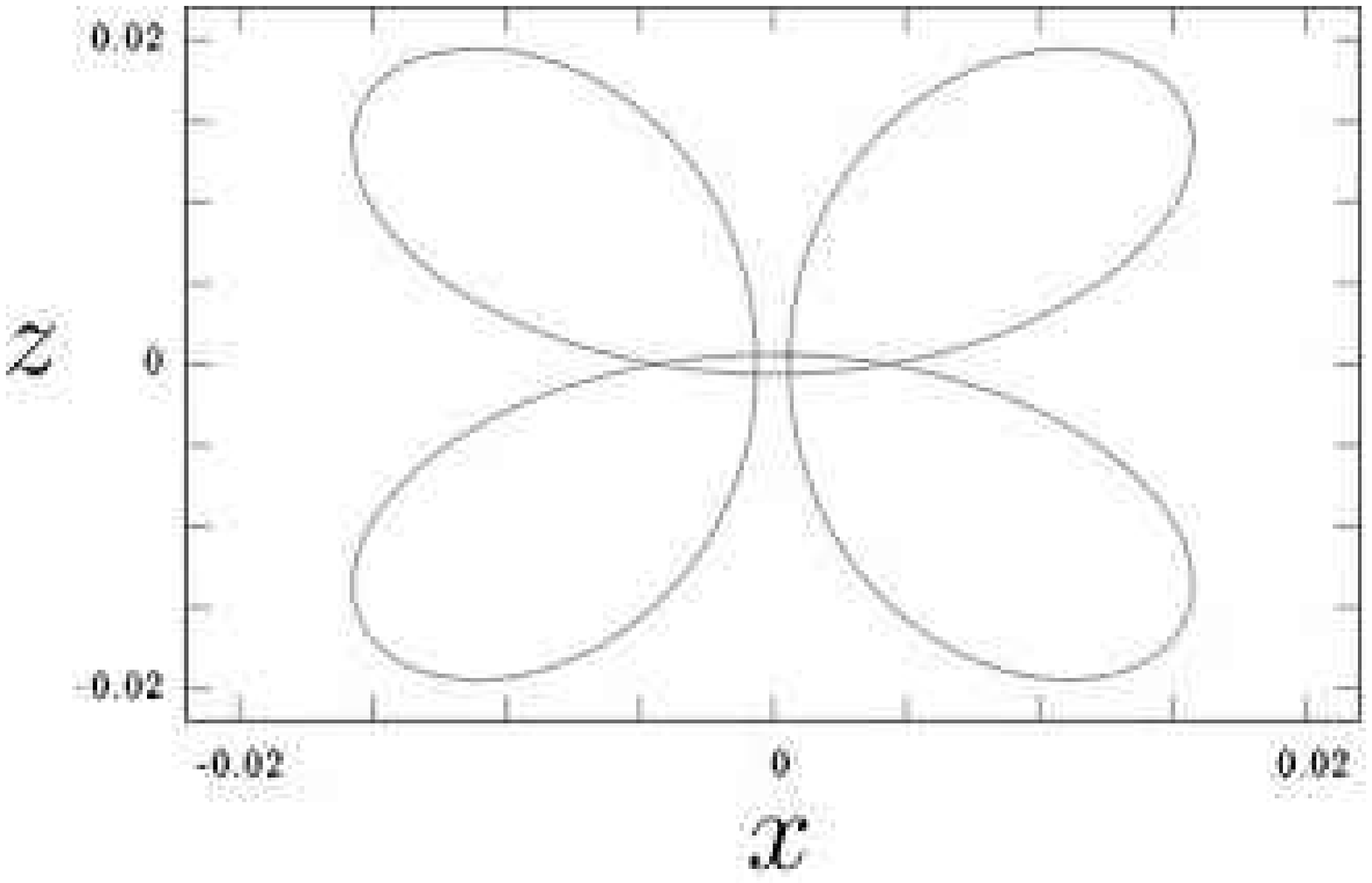}}
\subfigure[]{\label{inf-h}\includegraphics[width=2.0in,height=2.0in]
{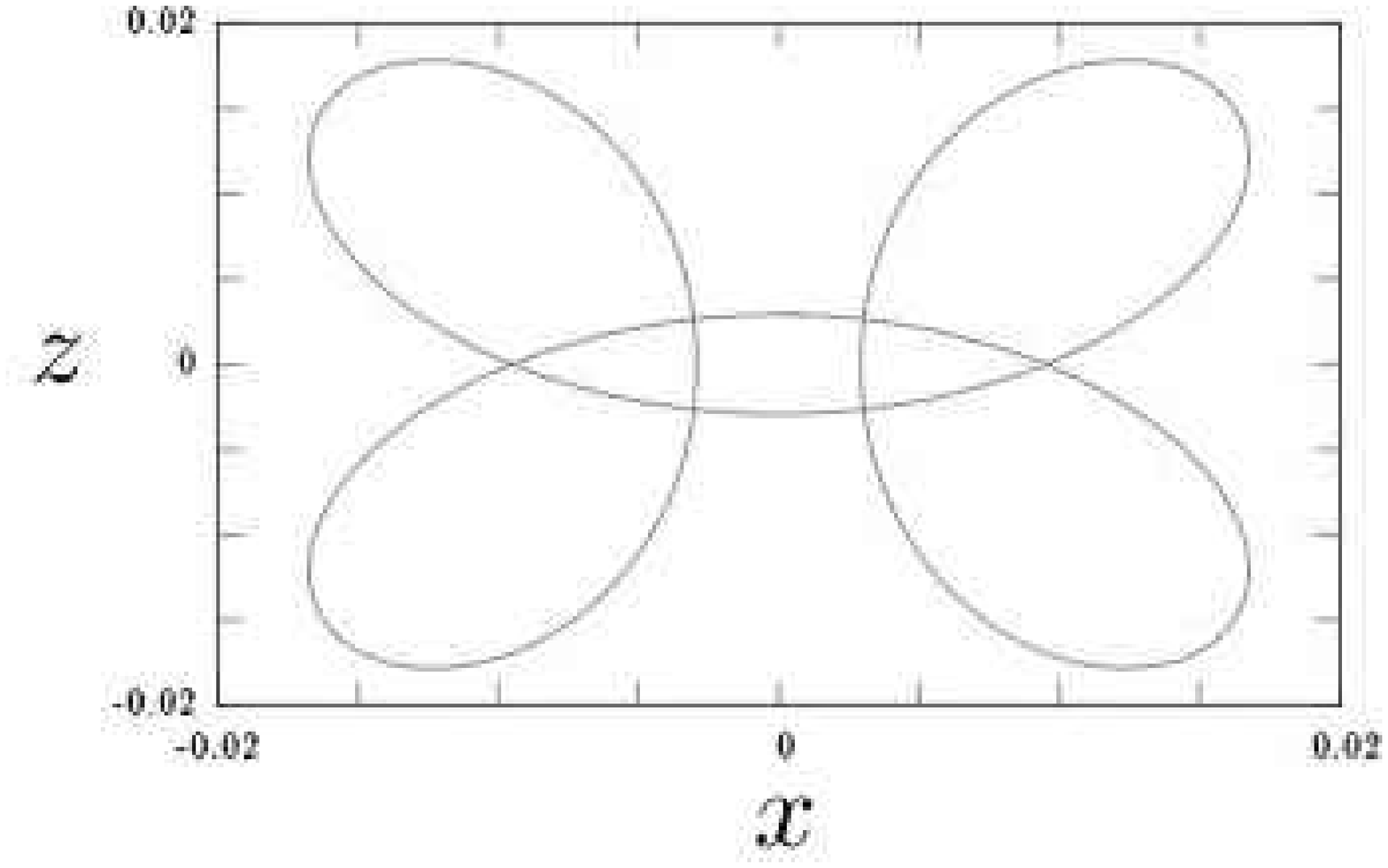}}

\caption{Infall of a cold collisionless shell for the axially symmetric 
initial velocity field $\vec{v} = - 0.05~\sin(2\theta)~\hat{\theta}$.
The shell is shown in $x z$ cross section at times $t^\prime$ 
= 1.132, 1.235, 1.241, 1.251, 1.256, 1.258, 1.260 and 1.261.  
The further time evolution is shown in Fig. \ref{case1}.  The 
continuous infall of many such shells produces the caustic shown 
in Fig. \ref{caust1-a}.\label{inf}}
\end{figure}
\begin{figure}

\subfigure[]{\label{case1-a}\includegraphics[width=2.0in,height=2.0in]
{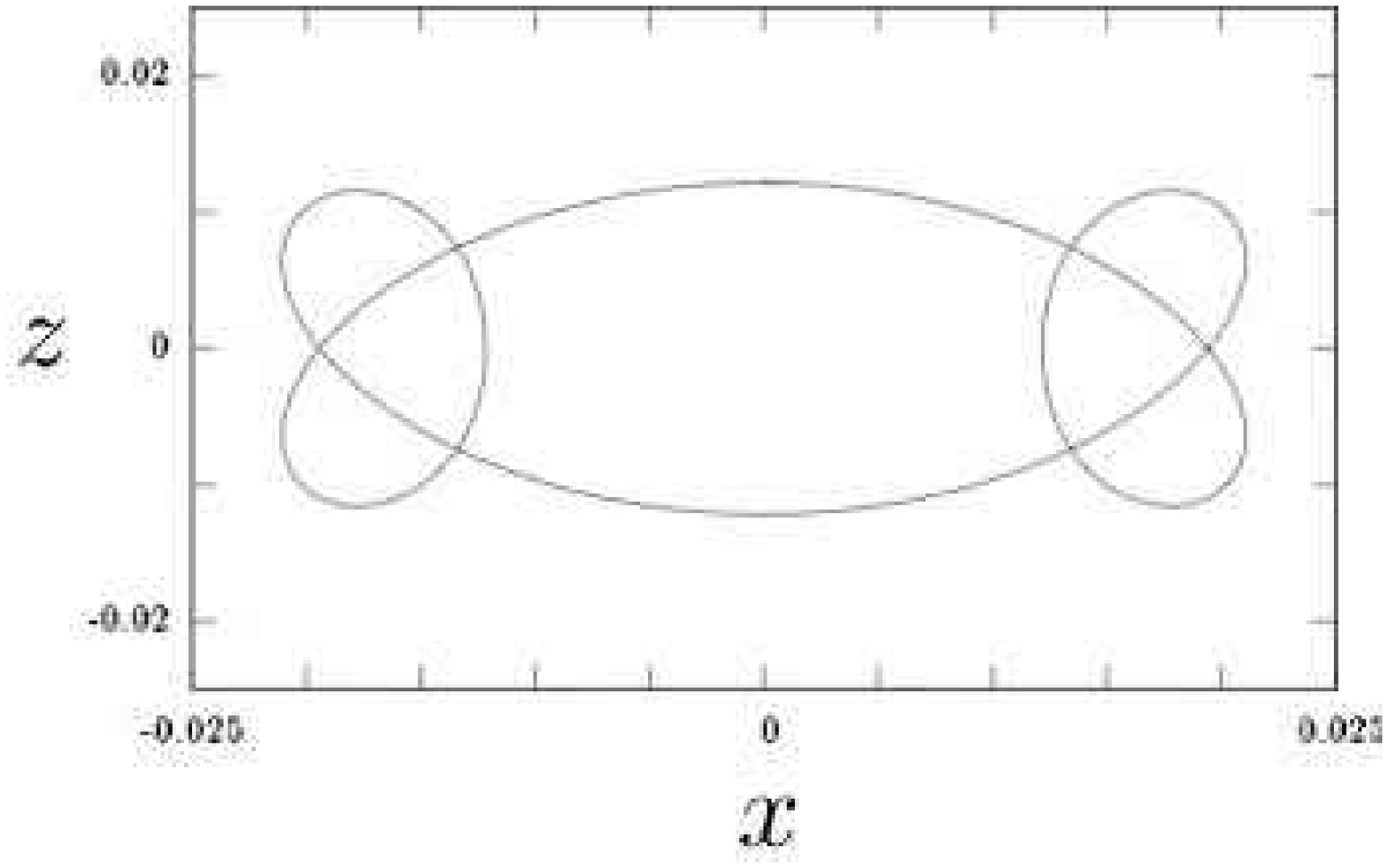}}
\subfigure[]{\label{case1-b}\includegraphics[width=2.0in,height=2.0in]
{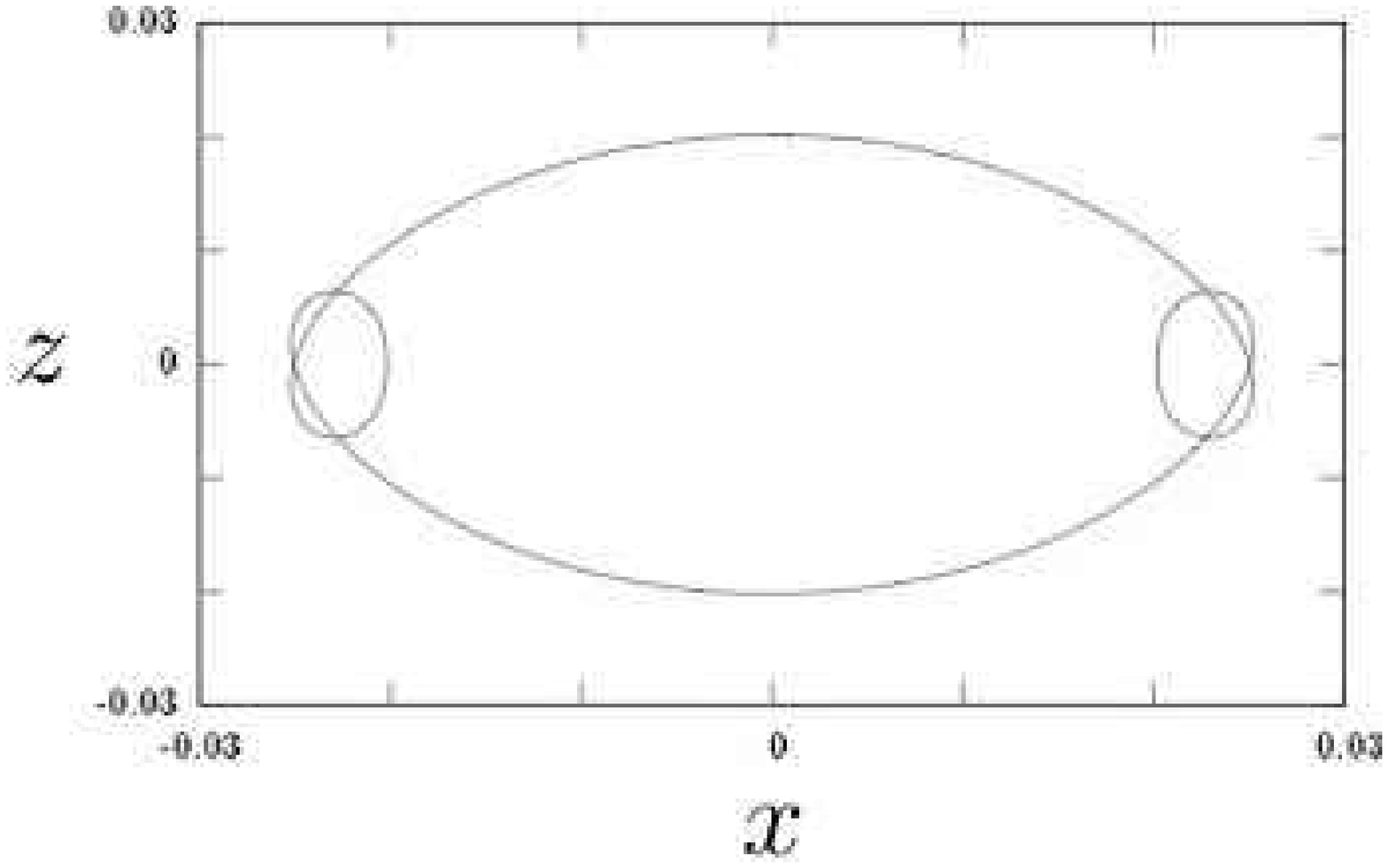}}
\subfigure[]{\label{case1-c}\includegraphics[width=2.0in,height=2.0in]
{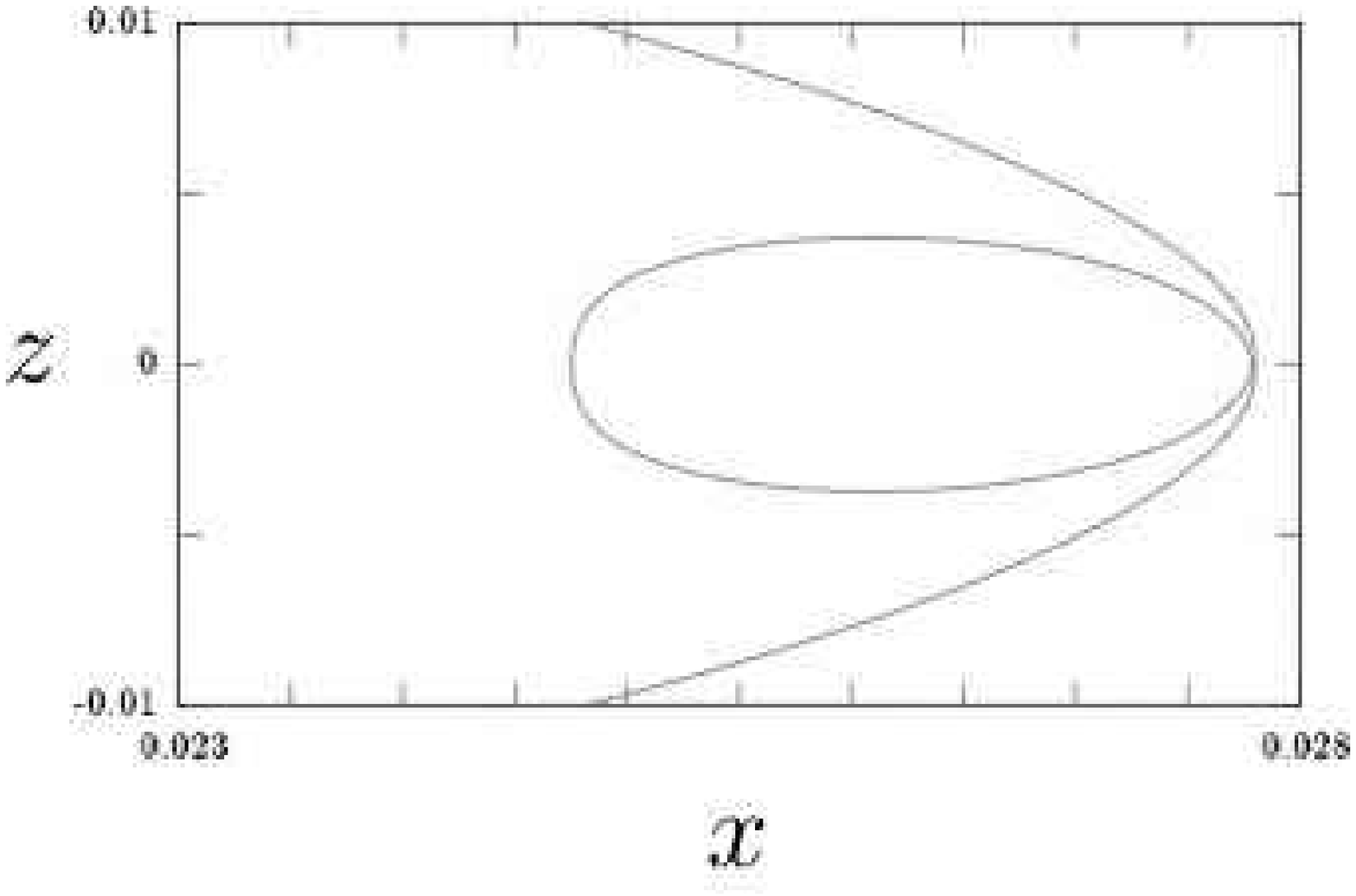}}\\
\subfigure[]{\label{case1-d}\includegraphics[width=2.0in,height=2.0in]
{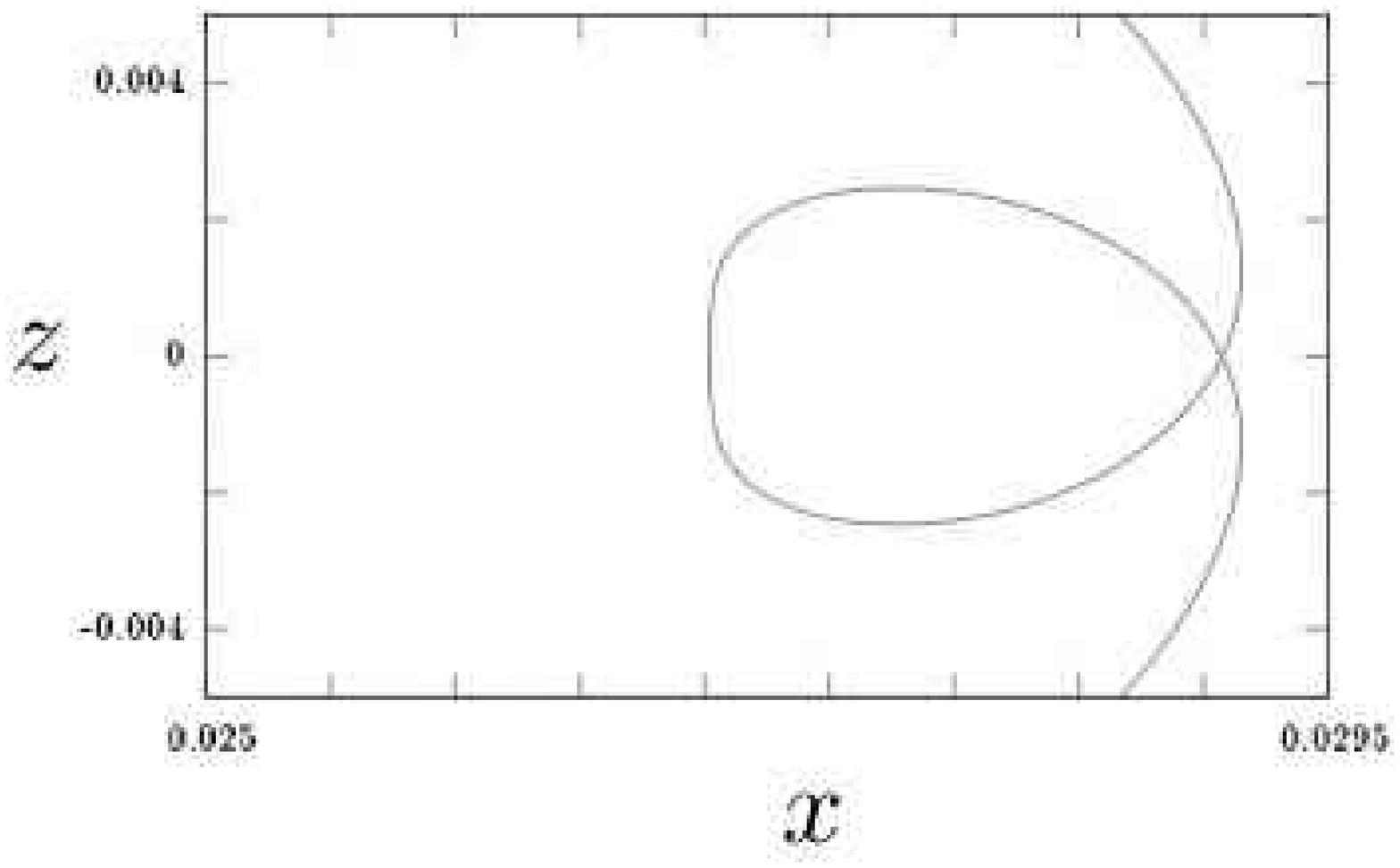}}
\subfigure[]{\label{case1-e}\includegraphics[width=2.0in,height=2.0in]
{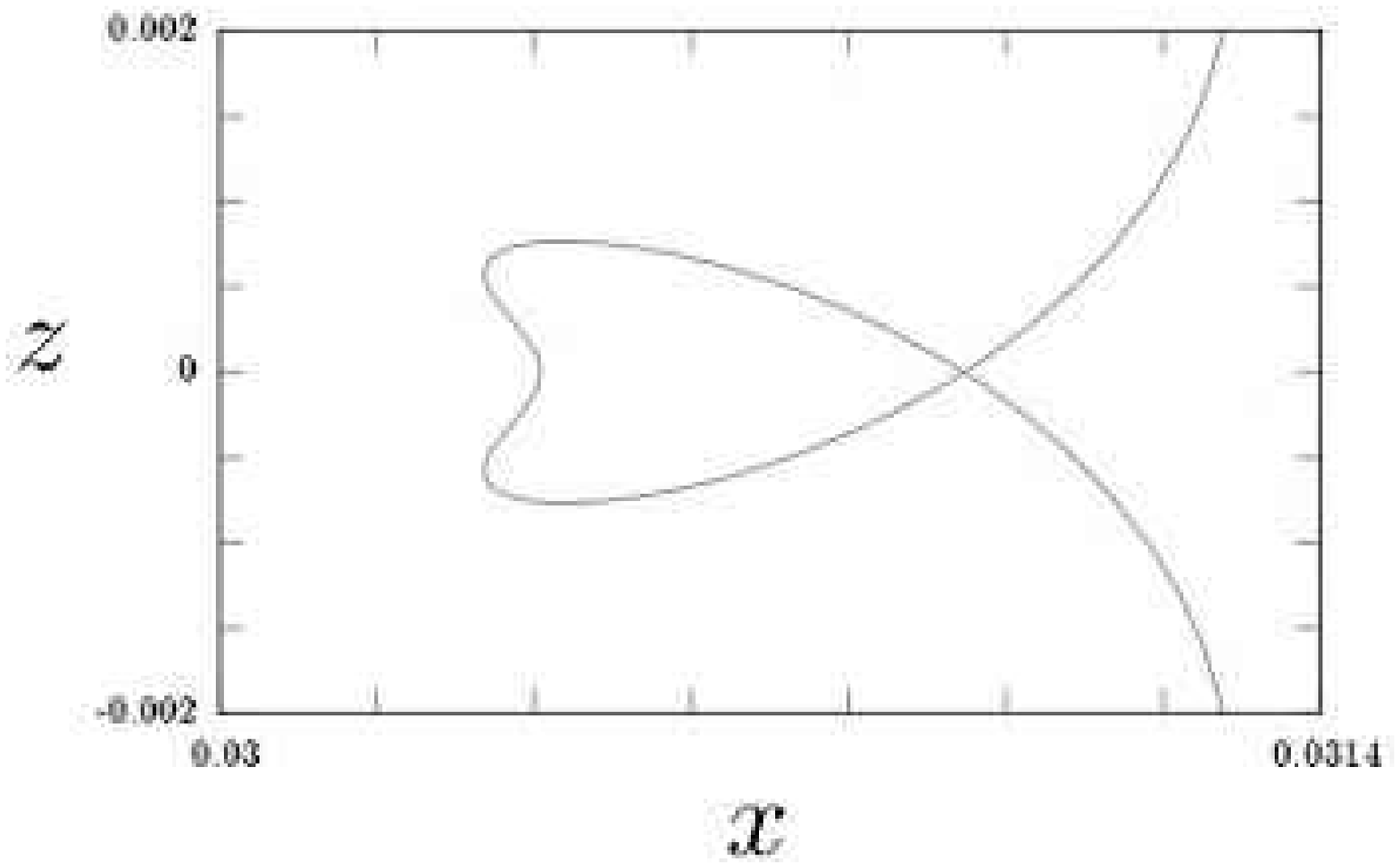}}
\subfigure[]{\label{case1-f}\includegraphics[width=2.0in,height=2.0in]
{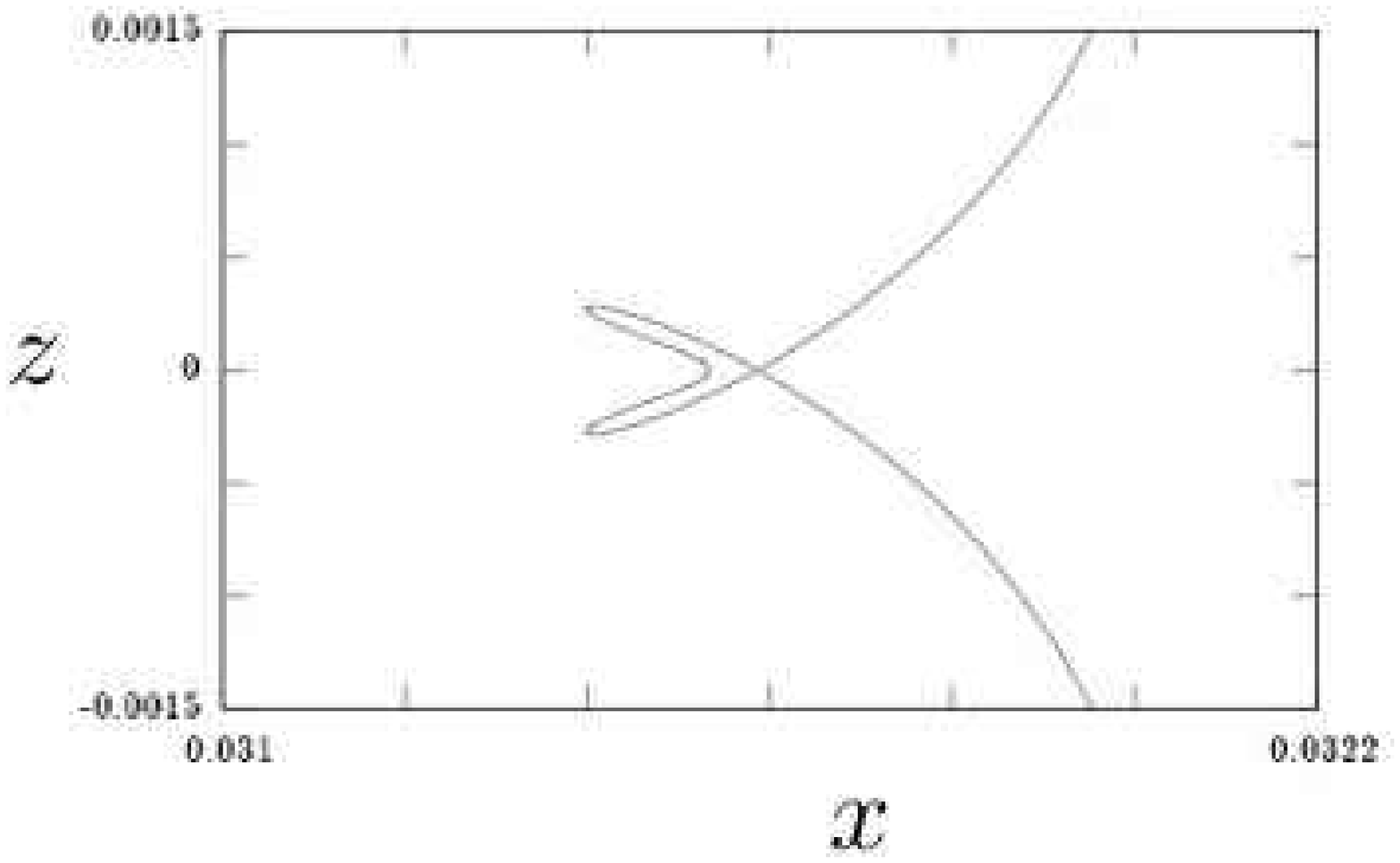}}\\
\subfigure[]{\label{case1-g}\includegraphics[width=2.0in,height=2.0in]
{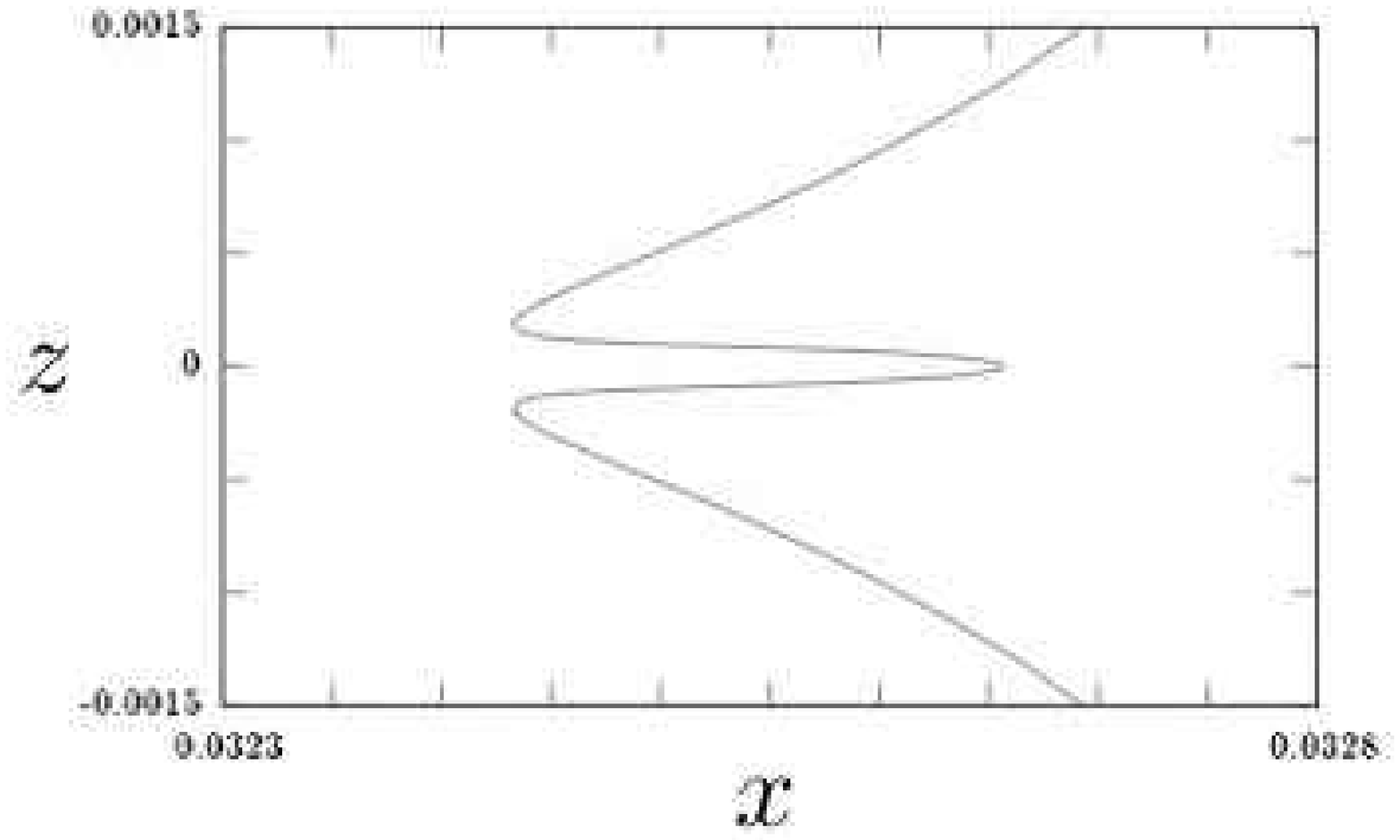}}
\subfigure[]{\label{case1-h}\includegraphics[width=2.0in,height=2.0in]
{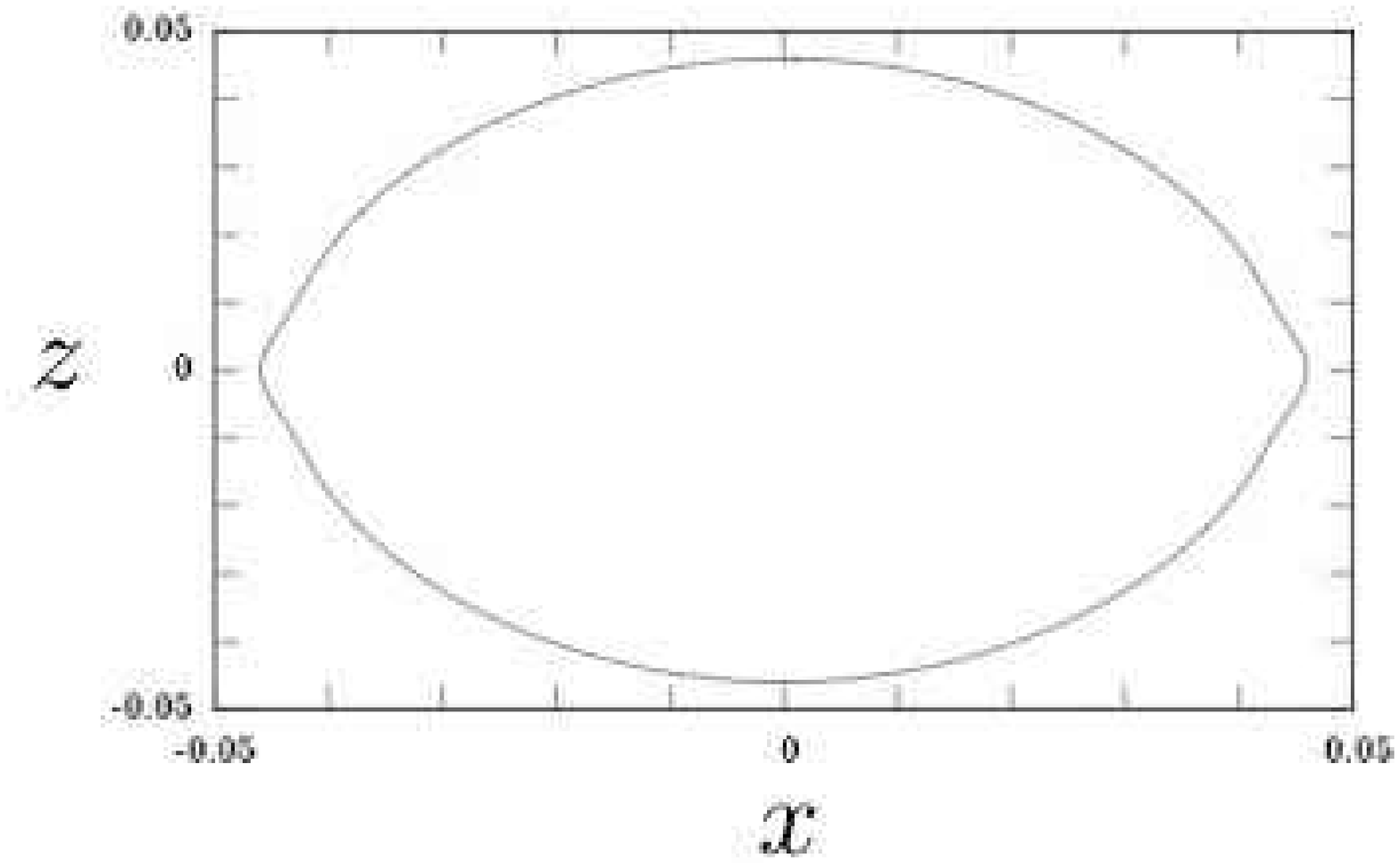}}

\caption{Continuation of the sequence of Fig.~\ref{inf}.
The shell is shown at times $t^\prime$ = 1.265, 1.269, 
1.271, 1.272, 1.273, 1.2737, 1.2742 and 1.280.
\label{case1}}
\end{figure}

\begin{figure}

\subfigure[]{\label{case2-a}\includegraphics[width=2.0in,height=2.0in]
{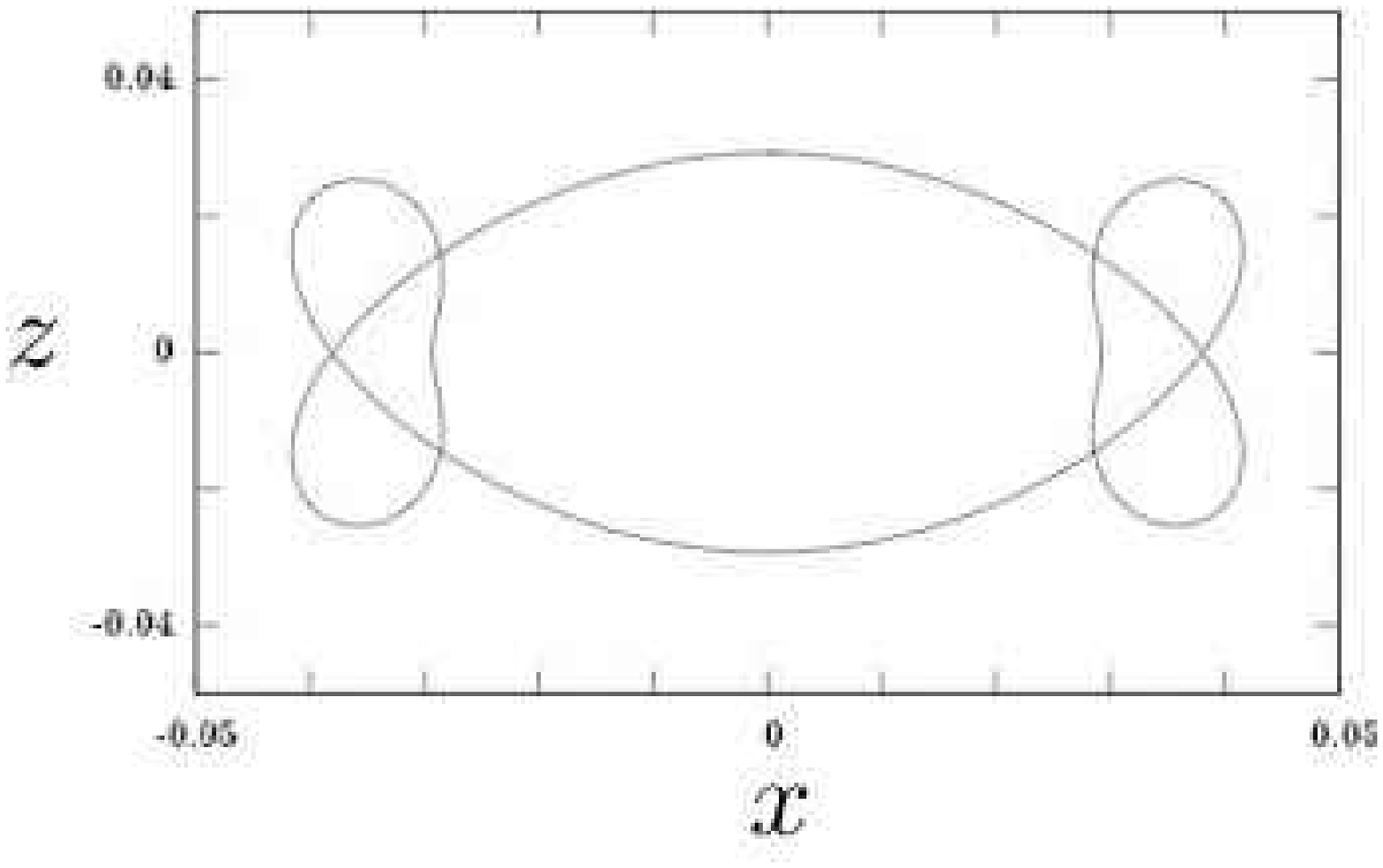}}
\subfigure[]{\label{case2-b}\includegraphics[width=2.0in,height=2.0in]
{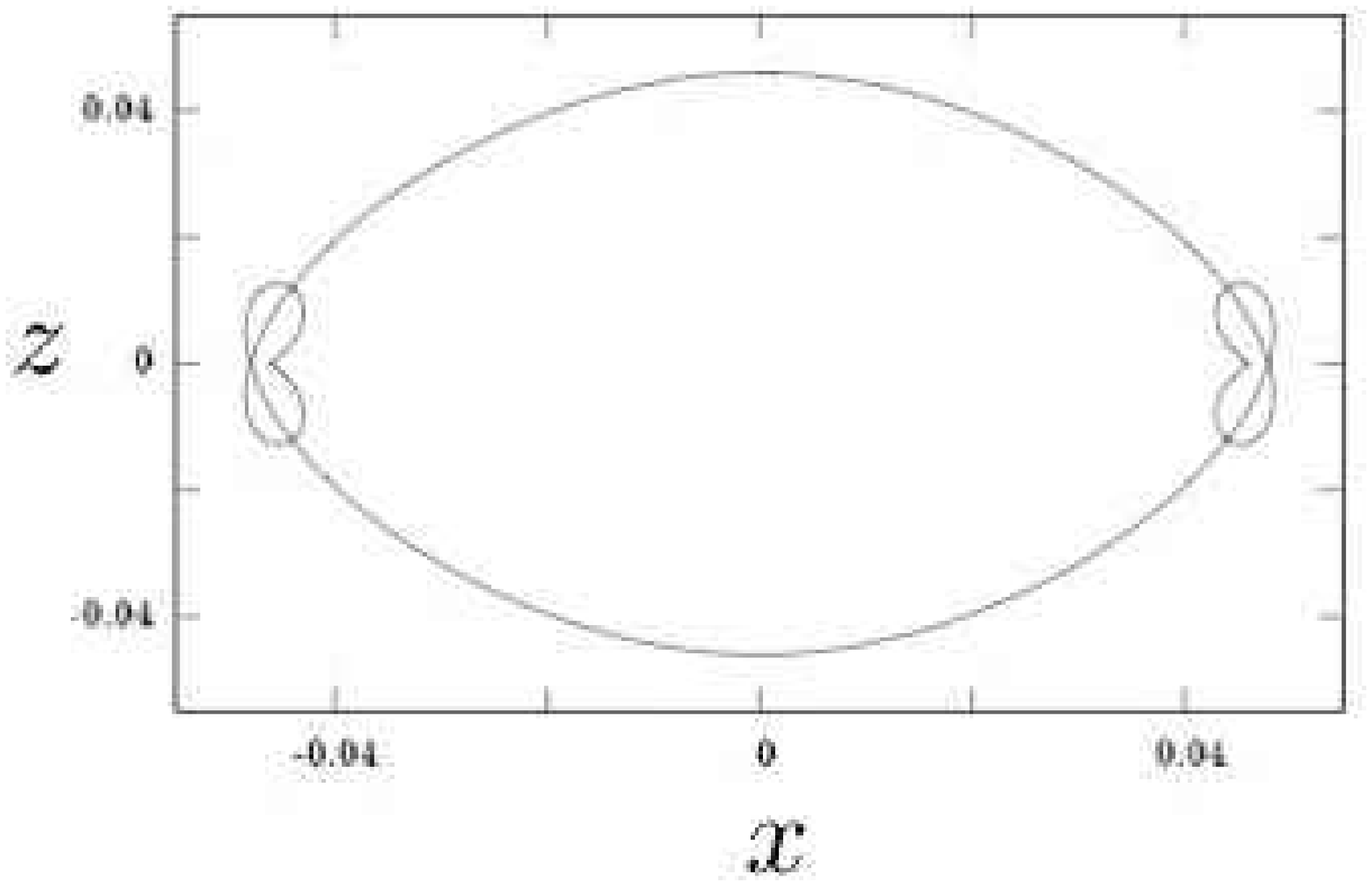}}
\subfigure[]{\label{case2-c}\includegraphics[width=2.0in,height=2.0in]
{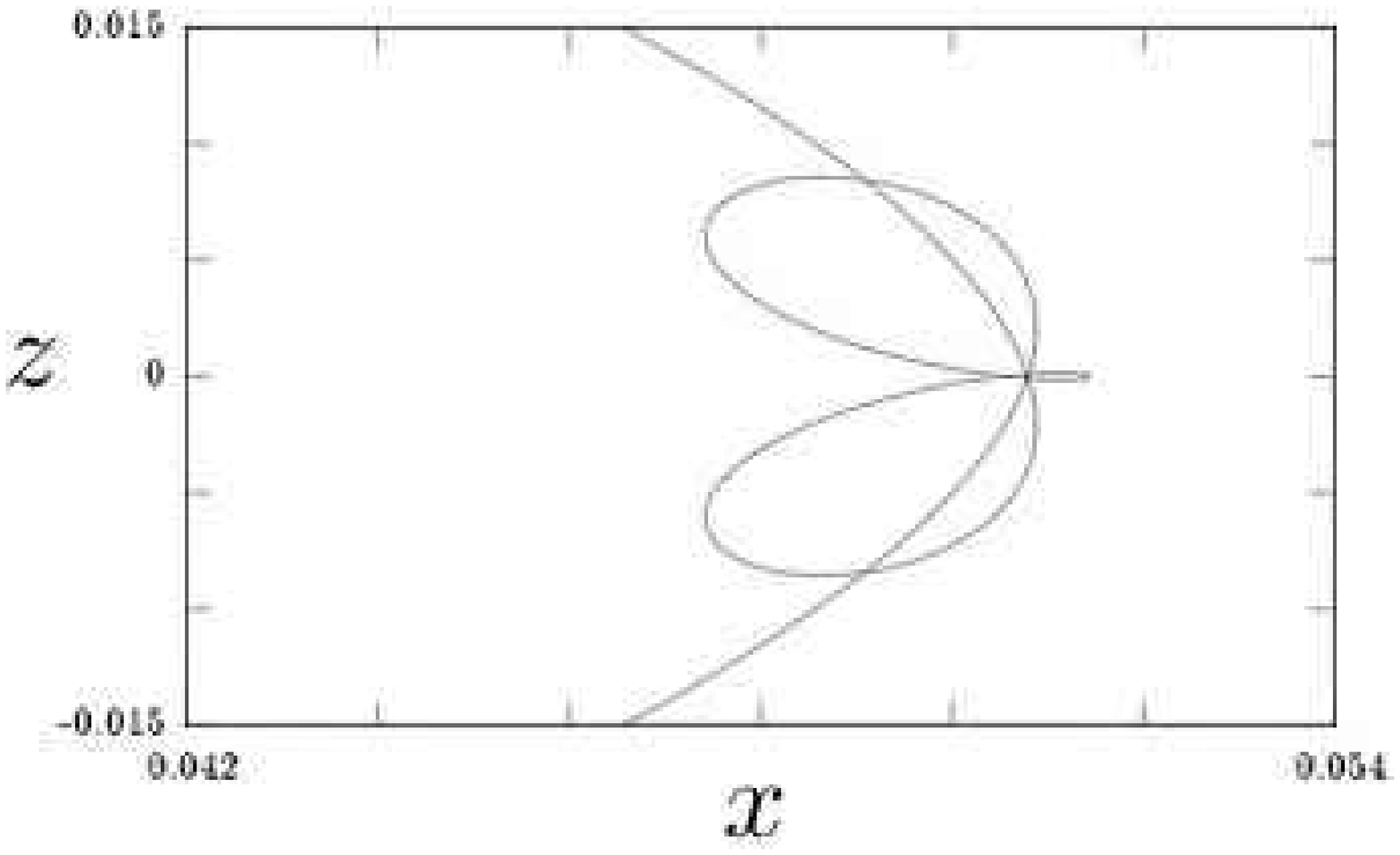}}\\
\subfigure[]{\label{case2-d}\includegraphics[width=2.0in,height=2.0in]
{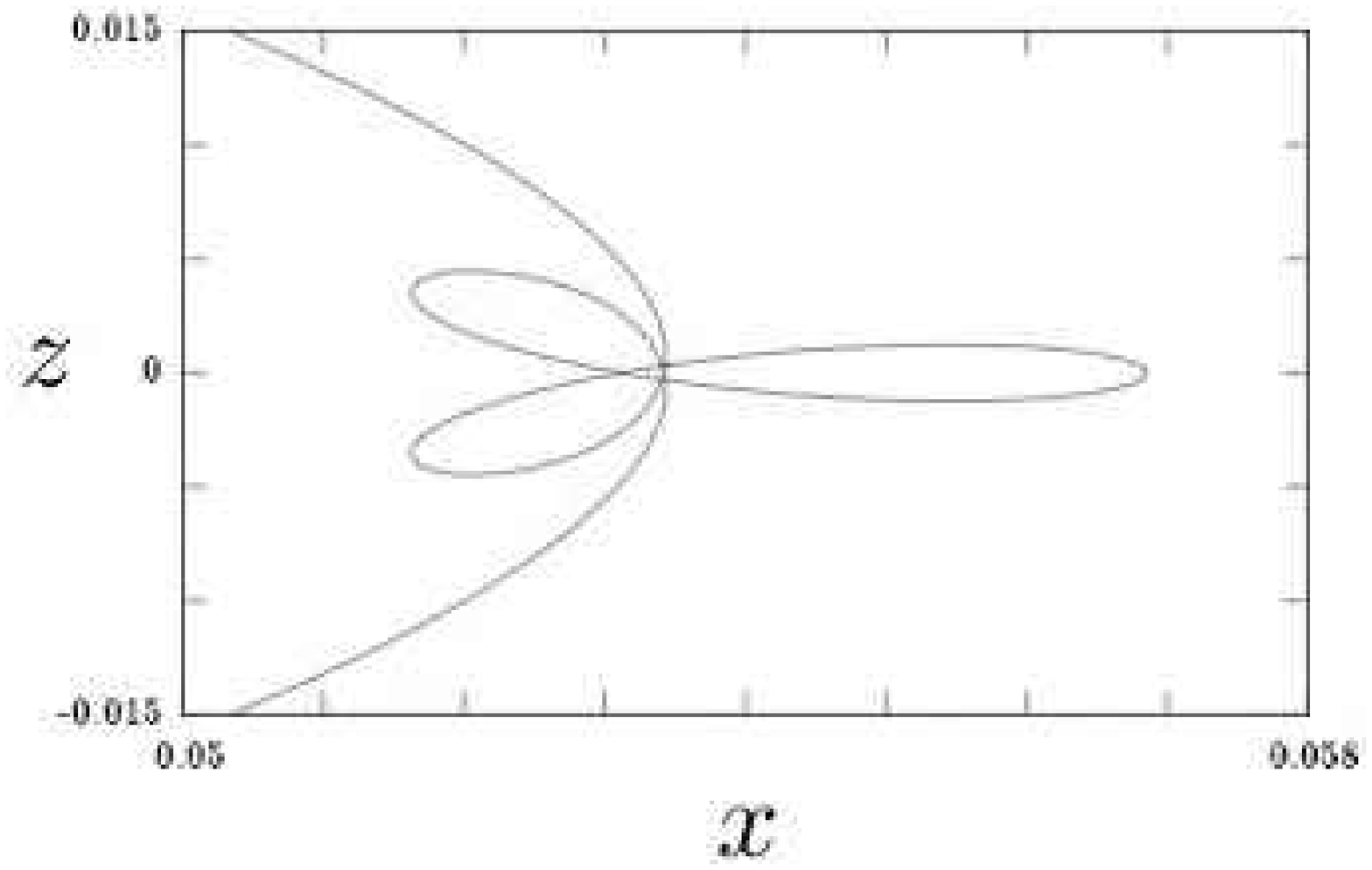}}
\subfigure[]{\label{case2-e}\includegraphics[width=2.0in,height=2.0in]
{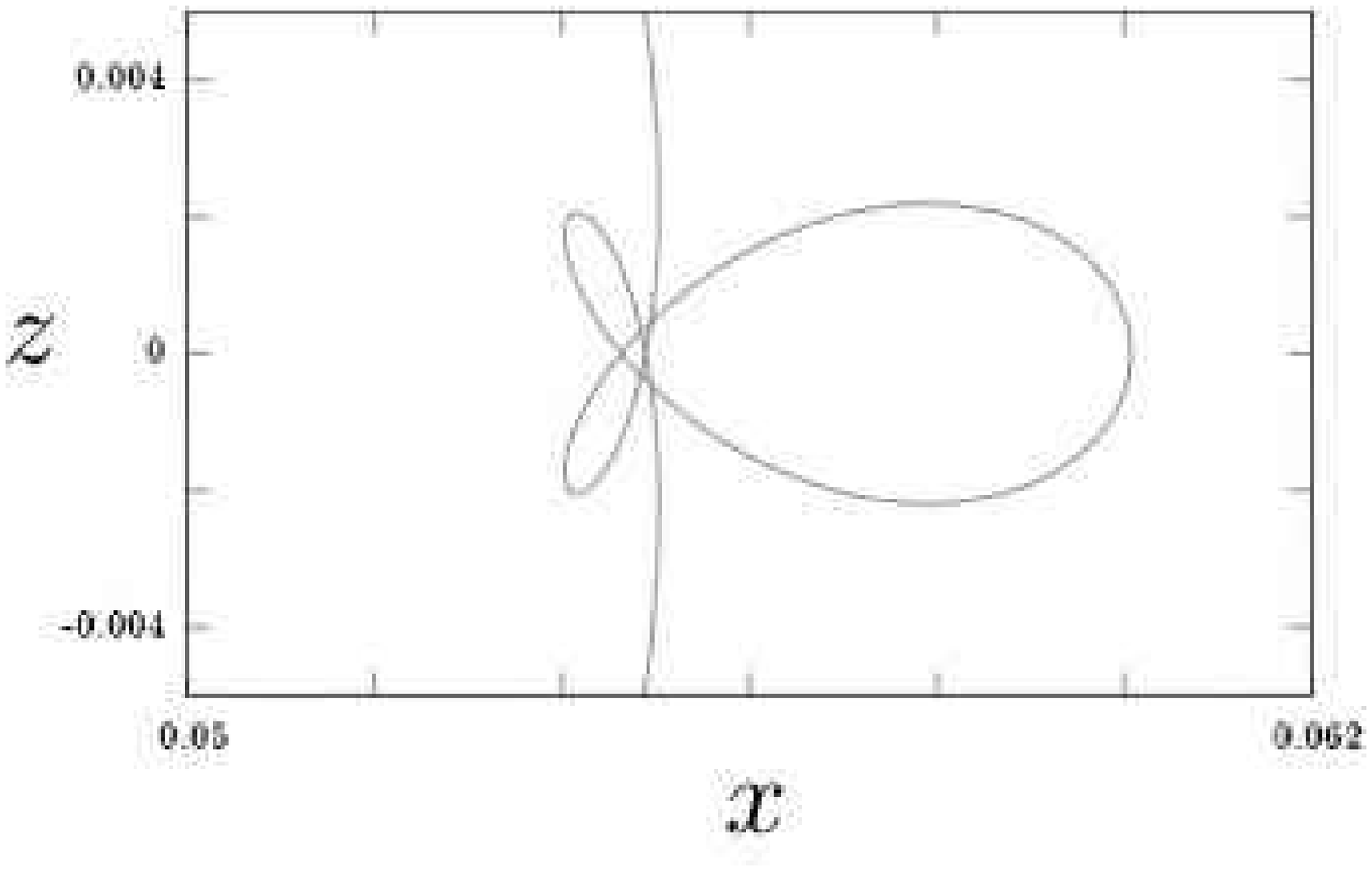}}
\subfigure[]{\label{case2-f}\includegraphics[width=2.0in,height=2.0in]
{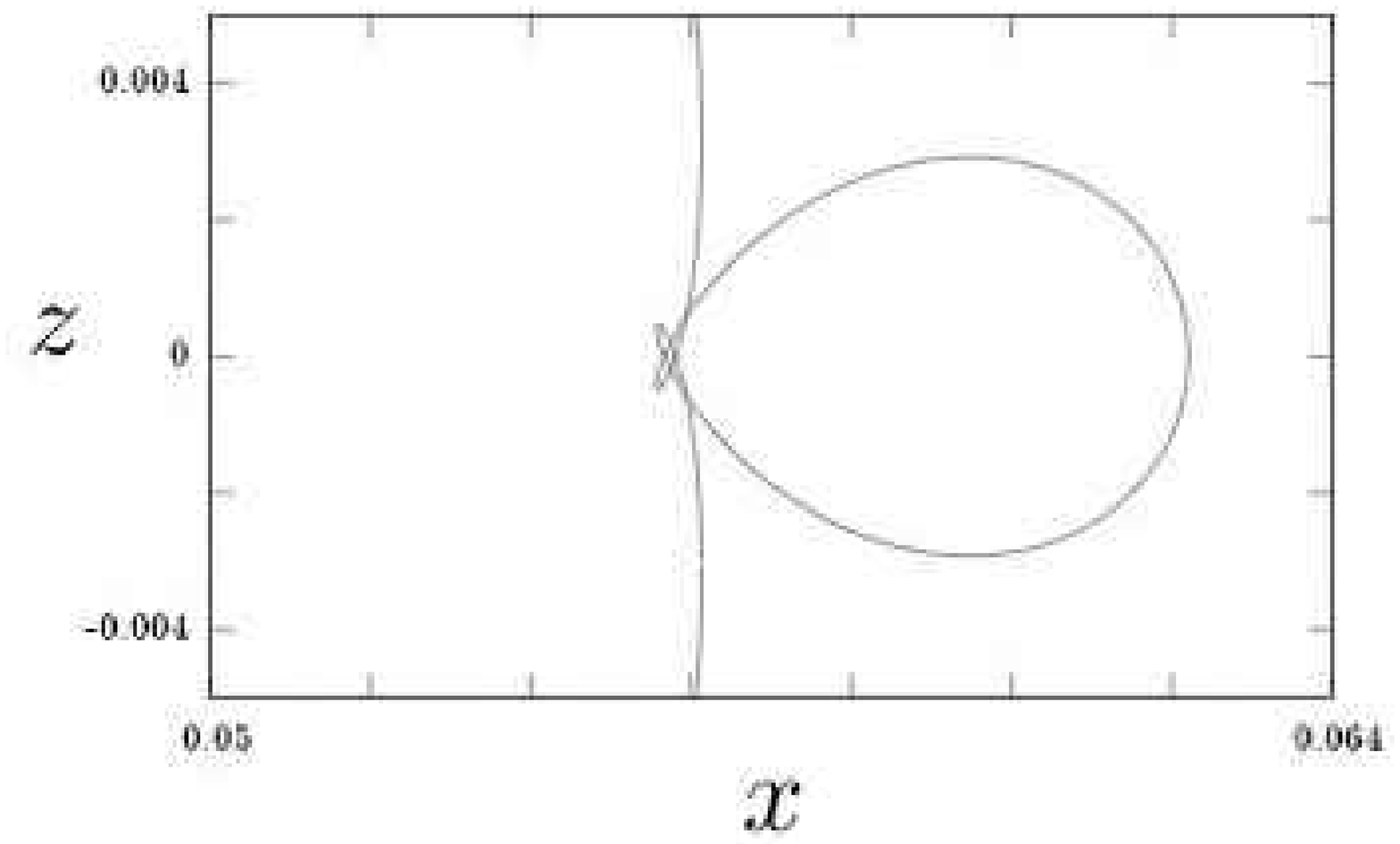}}\\
\subfigure[]{\label{case2-g}\includegraphics[width=2.0in,height=2.0in]
{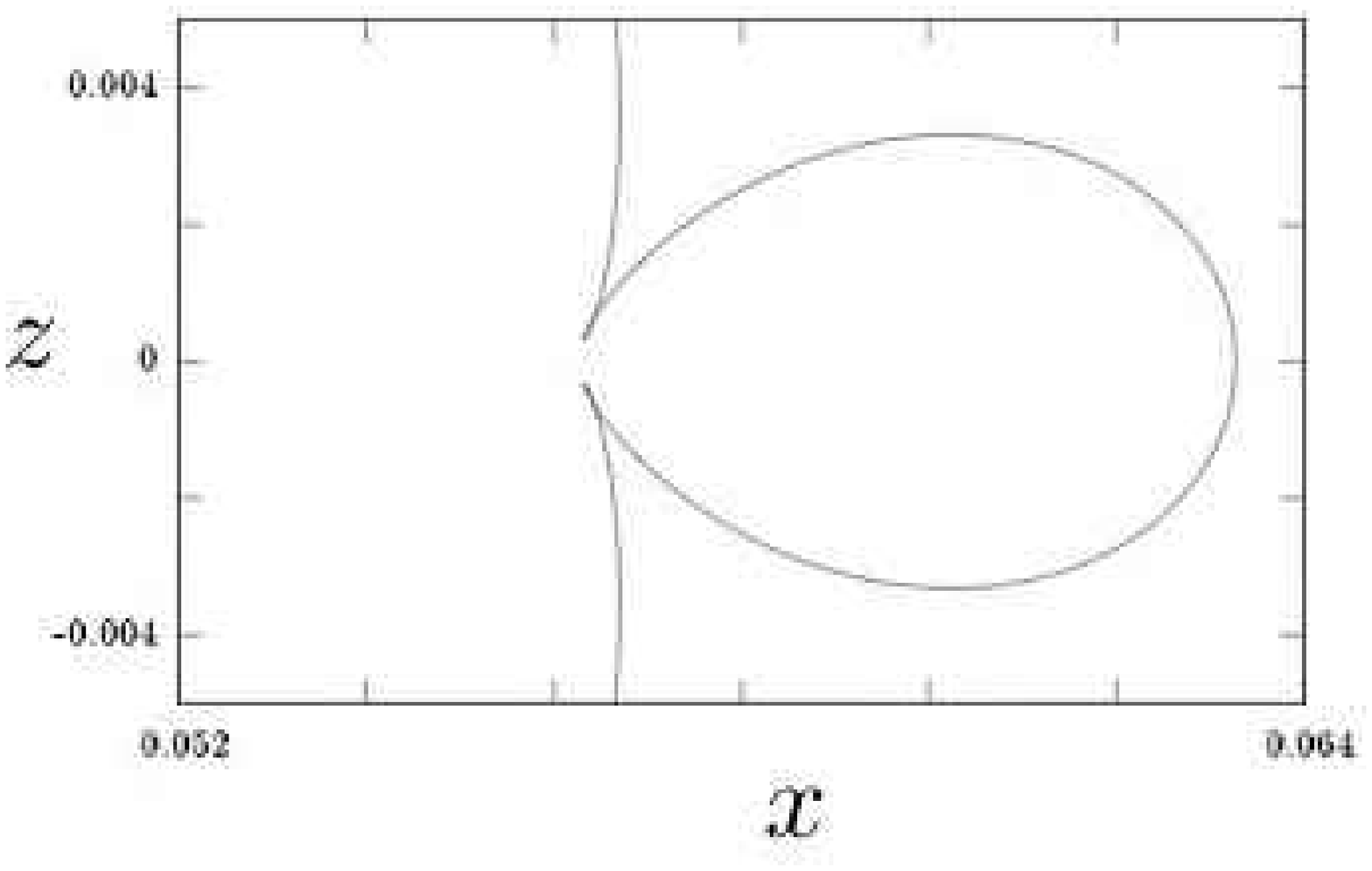}}
\subfigure[]{\label{case2-h}\includegraphics[width=2.0in,height=2.0in]
{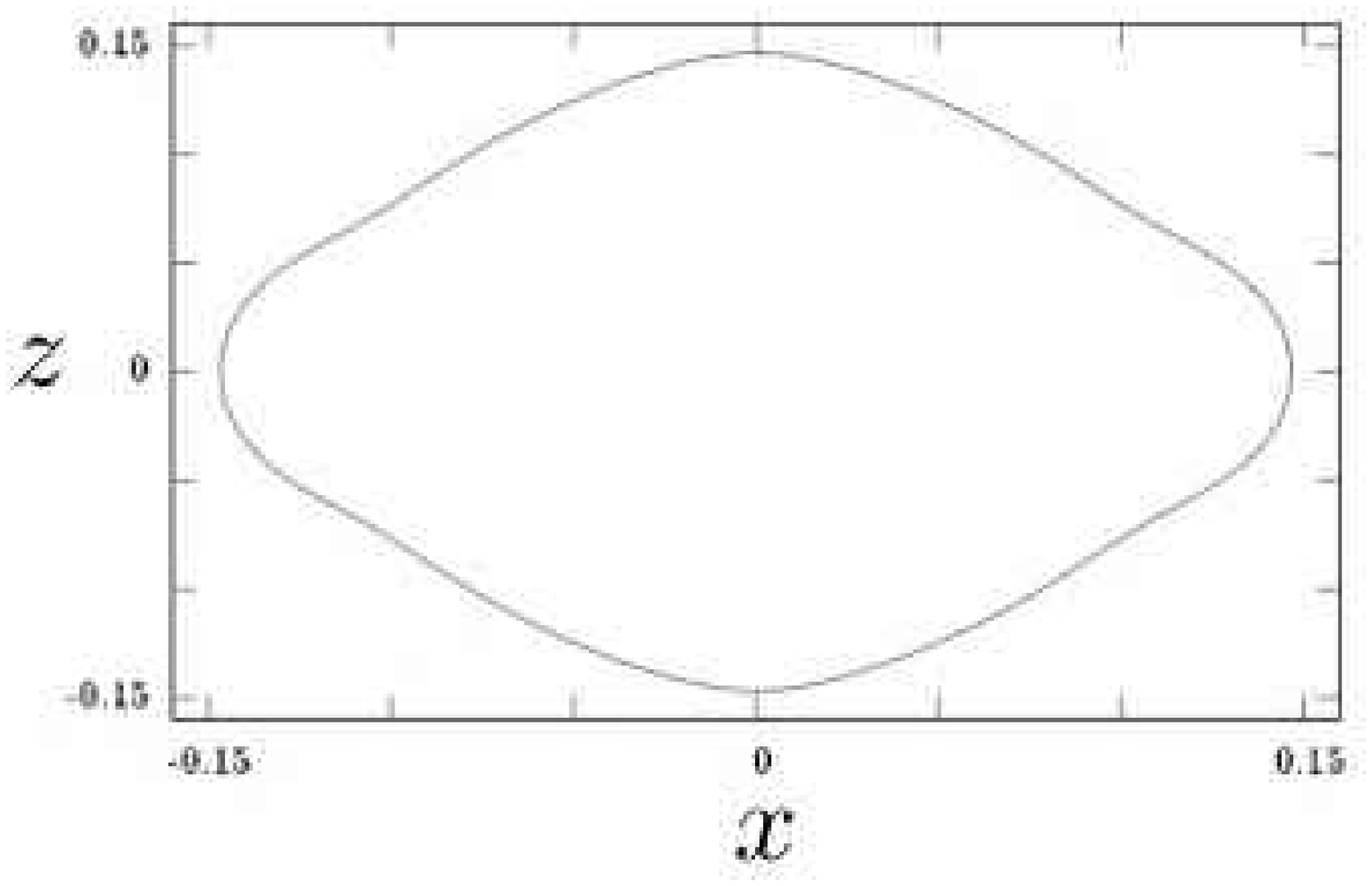}}

\caption{Same as Fig.~\ref{case1} but for the initial 
velocity field $\vec{v} = - 0.1~\sin(2\theta)~\hat{\theta}$.
The shell is shown at times $t^\prime$ = 1.273, 1.280, 1.283, 
1.285, 1.287, 1.2875, 1.2880 and 1.329.  The earlier evolution 
is qualitatively the same as in Fig. \ref{inf}.  The continuous 
infall of many shells with this initial velocity field produces 
the caustic shown in Fig. \ref{caust2-a}.\label{case2}}
\end{figure}

\clearpage

\end{document}